# Globally promoting legumes by production or use subsidies

Pablo de la Vega[1] and Wolfgang Britz[2]

## Abstract

Despite consensus on the positive economic, environmental, and health effects of expanding legume production and consumption, the share of cropland devoted to them remains relatively small, and their share of human consumption is negligible compared with other protein sources such as meat. The presence of externalities associated with legume production and consumption gives room for policy interventions. In this context, we estimate the effects of global subsidies on legume production and consumption using CGEBox, an open-source and open-access Computable General Equilibrium modelling platform, with integrated accounting for land use, land cover, and energy, and considering agricultural and non-agricultural households. The analysis offers high agri-food detail and a special focus on Ghana and Germany, two countries with different agroecological and socioeconomic conditions facing multiple food system issues. Compared to production subsidies, we find that use subsidies are more effective at reducing consumer prices of legumes and stimulating household intake of macronutrients, without displacing consumption of other foods. In contrast, production subsidies are more effective at boosting domestic supply, but also stimulate their feed use. Furthermore, subsidies drive up crop land to the detriment of pasture land and forests. The impact of these subsidies also reveals profound regional disparities, influenced by patterns of land use and of production and consumption.

[1] Institute for Food and Resource Economics, University of Bonn, Nussallee 19, D-53115 Bonn, Germany. Email: pdelaveg@uni-bonn.de. This is the corresponding author.
[2] Institute for Food and Resource Economics, University of Bonn, Nussallee 19, D-53115 Bonn, Germany. Email: wolfgang.britz@irl.uni-bonn.de

# 1. Introduction

It is widely recognized that there is an urgent need for more sustainable food systems. In this context, legumes can offer significant economic, environmental, and health benefits (Drinkwater et al., 1998; Zander et al., 2016; Willet et al., 2019) but only occupy a small share of crop land in many regions. Data from the Food and Agriculture Organization of the United Nations (FAO) suggest that legumes currently account for less than 9% of the cropped land in Europe, Central Asia, East Asia, the Middle East and North Africa, compared to 41% in Latin America.[3] Within the legume group, oilseeds account for nearly 65% of the global area devoted to legume cultivation, although there are significant heterogeneities, both by region and by income level. According to FAO (2025), legumes represent less than 10% of both dietary energy and protein supply worldwide, while cereals and livestock products account for 42.4% and 16.5% of calorie supply, respectively, and 36.1% and 35.9% of protein supply.

Promoting legume production and consumption could lead to economic, environmental, and health gains. From a production point of view, legume crops have the capacity to fix nitrogen from the air and thus reduce the need for nitrogen fertilizers and associated GHG emissions, representing significant cost savings for their cultivation and for subsequent crops (Peoples et al., 2009; Jensen et al., 2012; Bues et al., 2013). Legumes can provide further ecosystem services such as pest regulation and improving soil conditions depending on crop management practices (Willliams et al., 2014; Zander et al., 2016; Kebede, 2020; Alandia et al., 2020). From a consumption point of view, legumes are protein-rich and could thus substitute meat consumption, contributing to healthier and sustainable diets (Curran et al., 2012; Willett et al., 2019; Springmann and Freund, 2022; Kaimila et al., 2025).[4] By substituting meat, legumes could also contribute to reducing the livestock's long shadow (Steinfeld et al., 2006; Campbell et al., 2017).[5] Legumes can also be used for livestock feed as an alternative to cereal crops. Overall, the presence of externalities associated with promoting legume production and consumption gives room for policy interventions.

Several papers have analyzed different policies to incentivize increased legume production and consumption, mostly in Europe. Some of them compare different policies by using partial equilibrium models for the European agricultural sector and find that tax policies have a minor impact on domestic production due to international trade adjustments (Kuhlman et al., 2017). Various papers focus on legume production for use in feed and Europe's protein self-sufficiency (Jouan et al., 2020; Heinrichs et al., 2021; van Loon et al., 2023; Michalis et al., 2025), studying the impact of subsidies and regulations but tend to narrow the scope with regard to geographical coverage and unit-of-analysis The only reference on a global scale is Springmann and Freund (2022), who use the global computable general equilibrium (CGE) model MAGNET combined with a global risk-disease model to study options for reforming agricultural subsidies. However, the authors, like almost all papers mentioned, based their analysis on highly aggregated agricultural products. In some cases, legumes are not even present as a separate product category. There is a gap in this body of literature, which focuses primarily on Europe in its analysis, employs mainly partial-equilibrium analyses, and uses highly aggregated product lists.

[3] We consider the following crops as legumes: pulses (dried legumes or seeds), oilseed legumes (soybeans and groundnuts, excluding shelled), vegetable legumes (string beans, other green beans, green peas, and green broad and horse beans), and tree legumes (locust beans).

[4] Whilst 820 million people have insufficient food, many more consume unhealthy diets characterized by overconsumption of meat, added sugar and refined grains and underconsumption of healthy foods such as vegetables, fruits, whole grains and legumes (Willet et al., 2019). These diets generate micronutrient deficiencies and increase the incidence of obesity, coronary heart disease, stroke, and diabetes, and morbidity risk.

[5] In a highly influential report, FAO refers to the livestock's long shadow based on the overall environmental impact of the global livestock sector, being responsible for 18% of GHG emissions, occupying 26% of the planet's ice-free land area (70% of all agricultural land), consuming a massive portion of water resources and being a key driver of deforestation and biodiversity loss (Steinfeld et al. 2006).

Additionally, there are still debates on which policy mechanisms could promote the transition towards more sustainable and nutritious diets whilst minimizing distortions in agricultural markets.

We contribute to filling in this gap in the literature by estimating the likely impacts of global subsidies on the production and consumption of legumes by using CGEBox, an open-source and open-access CGE modelling platform, with integrated accounting for land use, land cover, and energy. In addition to using a global model, our study offers a high level of disaggregation of agricultural and food products and activities. This allows to accurately analyze the economic, environmental, and nutritional consequences of these policies. The analysis will specifically focus on Ghana (as one example of an LMIC) and Germany (as one example of a HIC), two countries with different agroecological and socioeconomic conditions facing multiple food system issues, and thus they will be singled out. In terms of many conditions related to agricultural structures, food supply chains, dietary patterns, and nutrition and health outcomes, Germany is somewhat typical for other countries in Europe and North America, whereas Ghana is somewhat typical for Africa

The rest of this paper is organized as follows. Section 2 performs a literature review. Section 3 describes the data, the model, and the scenario design. Section 4 presents the main results, while Section 5 discusses them and Section 6 concludes.

# 2. Literature Review

Multiple studies have shown a long-term downward trend in the production of Legumes in Europe (Helming and Kuhlman, 2015; Zander et al., 2016; Jouan et al., 2020; Heinrichs et al., 2021; Michalis et al., 2025) that now cover less than 3% of the total European arable land (Notz et al., 2023). Large amounts of legumes and products thereof are imported for feed use into Europe; the resulting low self-sufficiency in protein-rich feed raises concerns (Jouan et al., 2020; Michalis et al., 2025). Zander et al. (2016) identify factors that hinder the expansion of legumes in Europe: both market forces, such as relatively low and volatile margins compared to other crops such as wheat and rapeseed combined with low prices for imported soybean meal, and non-market forces, especially the failure to monetize their ecosystem services and health effects. Moreover, legume production not only competes with other crops such as cereals and maize, but also with livestock production, as described by the livestock's long shadow (Steinfeld et al., 2006).

Several policies have been implemented in Europe to promote legume production in Europe under the Common Agricultural Policy (CAP), which has undergone various reforms since 1978 (Kuhlman et al., 2017). In particular, since 2014, under the CAP, EU member states could introduce subsidies for legumes. Additionally, legumes can count as areas of ecological interest (EFA). These policies helped to reverse the downward trend in legume cultivation, but they were implemented unevenly across countries, as were the results achieved (Heinrichs et al. 2021). Despite these efforts, legumes are still underrepresented in Europe's crop budget. More recently, within the EU Green Deal and its Farm to Fork strategy, the EU is further promoting vegetable protein production and consumption (European Commission, 2020).

Several authors have analyzed policies to incentivize increased legume production and consumption. These papers mostly focus on Europe, motivated by the above-mentioned decline in legume production over the last decades and the concern about animal feed self-sufficiency. Helming and Kuhlman (2015) use the CAPRI model, a partial equilibrium model of the European agricultural sector, to simulate a combined additional tax on meat and a subsidy on legumes for human consumption. They find that, despite a considerable increase in legume consumption, this policy will have only a limited impact on domestic production of both products, because of international trade adjustments: net exports of meat increase and so do net imports of legumes. For the same reason, the land-use and environmental impacts are also marginal. Kuhlman et al. (2017) use the same model to simulate six policies that incentivize cultivation of legumes in Europe. They find that the best policy is a carbon tax that also covers agricultural GHG emissions for two reasons. First, ruminant meat production is highly emission-intensive and would shrink under taxed emissions, giving room for higher vegetable protein consumption, and second, cost increases for nitrogen fertilizers under a carbon tax would favor legume cultivation that requires none or little of them.

Among the papers that particularly focus on legume production for use in feed and Europe's protein self-sufficiency, Jouan et al. (2020) employ a single-region CGE model combined with a bio-economic model at the farm level to simulate an increase in demand for GMO-free animal products and an increase in coupled support for legumes in Western France. They find that production subsidies increase legume production but have a quite limited impact on protein self-sufficiency, since domestic legumes continue to be underused in feed. In contrast, when the demand for GMO-free animal products increases, legume production increases substantially, and most livestock are fed domestic legumes to replace imported GMO based feed. Heinrichs et al. (2021) use a bio-economic model representing French and German farms and find that relatively low levels of coupled support can lead to modest increases in legume production, but that more substantial changes require considerable subsidies. Instead, allowing the French farm, similar to German ones, to apply manure on legumes fosters legume production, but environmental benefits are limited. By running simulations with a crop growth simulation model, van Loon et al. (2023) suggest that a 61% yield increase or moderately expanding legume areas, i.e., to 1/12th of the cropland being occupied by legumes, could expand the supply of plant-based protein and achieve European protein self-sufficiency, which would vastly reduce cropland use and GHG emissions when substituting meat protein. Similarly, Michalis et al. (2025) use a linear programming model calibrated with farm data from Greece and find that higher yields or selling prices can be just as effective as coupled support in promoting legume cultivation for livestock. On the demand side, the push for GMO-free animal products is identified as a highly effective market lever to stimulate domestic legume-based feed production.

The only reference on a global scale is Springmann and Freund (2022), who use the global CGE model MAGNET combined with a global risk-disease model to study options for reforming agricultural subsidies (producer support estimates by the OECD). They find that repurposing up to half of agricultural subsidies to support the production of foods with beneficial health and environmental characteristics, including fruits, vegetables, and other horticultural products, and combining such repurposing with a more equal distribution (either based on GDP or population) of subsidy payments globally could significantly reduce global diet-related mortality and agricultural GHG emissions without sacrificing economic welfare. But the study used highly aggregated product groups and did not analyze legumes separately; they are partly included in their fruits and vegetable aggregate and partly aggregated in their vegetable oils and sugar aggregate.

Following from this, we see a gap in this body of literature, which focuses primarily on Europe in its analysis, employs supply-side or partial-equilibrium analyses, or uses highly aggregated product lists. We contribute to this literature by estimating the likely impacts of global subsidies on the production and consumption of legumes by using CGEBox, an open-source and open-access global CGE model, with a high level of disaggregation of agricultural and food products and activities, which allows us to more accurately analyze the economic, environmental, and nutritional consequences of these policies.

# 3. Data and Methodology

Given their multi-sector and global perspective, CGE models are a useful tool for ex-ante estimation of changes in policies such as subsidies on production and consumption at the global level. Particularly, global CGE models allow for a rigorous capture of interregional interactions by modeling bilateral trade flows and ensuring the simultaneous equilibrium of all markets in response to relative price or policy shocks that occur in one or more economies and are transmitted to the rest of the world. We use CGEBox, an open-source CGE modelling platform, with integrated accounting for land use, land cover, and energy. Below, we describe the database used in our study and the model's configuration.

## 3.1 Database

The database used in this study builds on the GTAP v12 Power (Aguiar et al., 2025) and GTAP_AEZ databases (ref), which represent the global economy in 2023, including data on power generation, emissions, air pollution, and land use and land cover (LULC). To improve computational efficiency, the GTAP v12 Power Database is partly aggregated for this study. Electricity products have been consolidated, and labor types have been grouped into skilled and unskilled. The GTAP data does not provide the necessary detail on crop production and diets; thus, following Britz (2022a), agricultural and food-related production products are split into 42 and 39 categories, respectively, and irrigation water is introduced as a production factor, allowing differentiation between irrigated and rainfed cropland production activities as described in Britz (2022b). This results in 115 agricultural and food-related production activities. We also split households into agricultural and non-agricultural ones.

Regions have been grouped based on ten geographical world regions that are disaggregated further by income level according to the World Bank classification into Low Income (LI), Low and Middle Income (LMI), Upper Middle Income (UMI), and High Income (HI) countries. In addition, the analysis will specifically focus on Ghana, as one example of an LMI country, and Germany, as one example of a HIC country. These two countries, with different agroecological and socioeconomic conditions, face multiple food system issues, and thus they will be singled out. We do not claim that these countries are representative. HI and LMI countries are simplistic categories, as the heterogeneity within these two groups of countries is at least as large as the heterogeneity between them. However, in terms of many conditions related to agricultural structures, food supply chains, dietary patterns, and nutrition and health outcomes, Germany is somewhat typical for other countries in Europe and North America, whereas Ghana is somewhat typical for Africa. The final regional aggregation is included in Table A1 in Appendix A. The resulting data encompasses 27 countries or regions, 174 production activities, 130 products, and 6 production factors (Land, Natural Resources, Water, Capital, Unskilled Labor, and Skilled Labor). The final product and sectoral aggregations are included in Table A2 and Table A3 in Appendix A.

In Table 1, we include a description of base-year consumption and production of Legumes. Production of soybeans is mainly concentrated in Latin America UMI (43.7%), North America (35.6%), and East Asia HI (12%). Groundnut and other Leguminosae production are mainly concentrated in East Asia HI (48.8% and 40.5%, respectively), South Asia LMI (22.2% and 18.4%, respectively), and North America (6.2% and 7%, respectively). In terms of consumption, soybean is mainly demanded by East Asia HI (46.7%), North America (17%), and Latin America UMI (13.4%); whilst Groundnut and other Leguminosae are mainly demanded by East Asia HI (51% and 41.9%, respectively) and South Asia LMI (22.6% and 21%, respectively).

In summary, the largest producer and consumer regions include East Asia HI, Latin America UMI, North America, and South Asia LMI. However, it is important to note that: (i) East Asia HI is a net importer of soybeans and self-sufficient in Groundnut and other Leguminosae; (ii) Latin America UMI and North America are net exporters of soybeans and self-sufficient in Groundnut and other Leguminosae; and (iii) South Asia LMI is a net importer of other Leguminosae and self-sufficient in soybeans and groundnut. Germany and Ghana account for a negligible share of soya production and consumption, apart from small quantities consumed in Germany (1.2% of global demand) and small quantities of other pulses produced and consumed (less than 1% of the global total). Ghana is self-sufficient in other pulses, whilst Germany is a net importer.

**Table 1. Legume Consumption and Production, Base Description for All Regions (% World)**

| Region | Output | | | Domestic Demand (Final and Intermediate) | | |
|---|---|---|---|---|---|---|
| | Soybean | Groundnut | Other Leguminosae | Soybean | Groundnut | Other Leguminosae |
| EU27, HI | 1.1 | 0.0 | 4.3 | 3.7 | 0.3 | 5.5 |
| East Asia, HI | 12.0 | 48.8 | 40.5 | 46.8 | 51.0 | 41.9 |
| East Asia, UMI | 0.1 | 0.1 | 0.4 | 4.2 | 0.2 | 0.9 |
| Germany | 0.0 | 0.0 | 0.4 | 1.2 | 0.1 | 0.6 |
| Ghana | 0.0 | 0.9 | 0.3 | 0.0 | 0.7 | 0.3 |
| Latin America, HI | 0.3 | 0.0 | 0.3 | 0.2 | 0.0 | 0.5 |
| Latin America, LMI | 0.6 | 0.1 | 0.4 | 0.5 | 0.1 | 0.4 |
| Latin America, UMI | 43.7 | 4.6 | 4.4 | 13.4 | 4.6 | 3.8 |
| MENA, HI | 0.0 | 0.2 | 0.1 | 0.3 | 0.1 | 1.1 |
| MENA, LI | 0.0 | 0.2 | 0.3 | 0.0 | 0.2 | 0.4 |
| MENA, LMI | 0.5 | 2.0 | 1.1 | 1.3 | 1.9 | 1.3 |
| MENA, UMI | 0.1 | 0.8 | 2.4 | 2.2 | 0.7 | 4.1 |
| North America, HI | 35.6 | 6.2 | 7.0 | 17.0 | 5.4 | 4.2 |
| Oceania, HI | 0.0 | 0.0 | 3.7 | 0.0 | 0.0 | 0.7 |
| Oceania, UMI | 0.0 | 0.0 | 0.0 | 0.0 | 0.0 | 0.0 |
| RoW, HI | 0.8 | 0.0 | 3.4 | 1.3 | 0.5 | 2.2 |
| RoW, LMI | 0.0 | 0.2 | 0.2 | 0.0 | 0.2 | 0.1 |
| RoW, UMI | 0.6 | 0.0 | 0.6 | 0.3 | 0.0 | 0.4 |
| SSA, LI | 0.2 | 5.6 | 6.5 | 0.1 | 3.2 | 5.7 |
| SSA, LMI | 0.1 | 6.6 | 2.6 | 0.1 | 6.7 | 2.5 |
| SSA, UMI | 0.7 | 0.2 | 0.1 | 0.7 | 0.1 | 0.1 |
| South Asia, LI | 0.0 | 0.0 | 0.0 | 0.0 | 0.0 | 0.1 |
| South Asia, LMI | 3.4 | 22.2 | 18.4 | 3.7 | 22.6 | 21.0 |
| South Asia, UMI | 0.0 | 0.3 | 0.0 | 0.0 | 0.3 | 0.3 |
| South-East Asia, HI | 0.0 | 0.0 | 0.0 | 0.0 | 0.0 | 0.1 |
| South-East Asia, LMI | 0.0 | 0.6 | 0.8 | 0.0 | 0.3 | 0.0 |
| South-East Asia, UMI | 0.0 | 0.7 | 1.5 | 3.0 | 0.7 | 1.8 |
| **World** | **168.89** | **46.38** | **76.90** | **180.99** | **46.52** | **78.10** |

Note: Own elaboration.

In Figure 1, we show the land under legumes as a share of cropland (red bars) and as a share of cropland plus pasture land (blue bars). As discussed before, legumes account for less than 10% of the cropped land in most of the World, with some exceptions in Latin America and North America. From this figure, it is also evident that there is competition between legumes and not only other crops but also with livestock production. If these shares are taken as indicators, competition is intensive in various regions such as Latin America, North America, East Asia (HI), MENA, Oceania (HI), RoW, and SSA. Globally, legume share on cropland is 11.1% compared to just 3.7% on cropland plus pastureland.

**Figure 1. Legume Share of Land Cover (%), by Model Region**

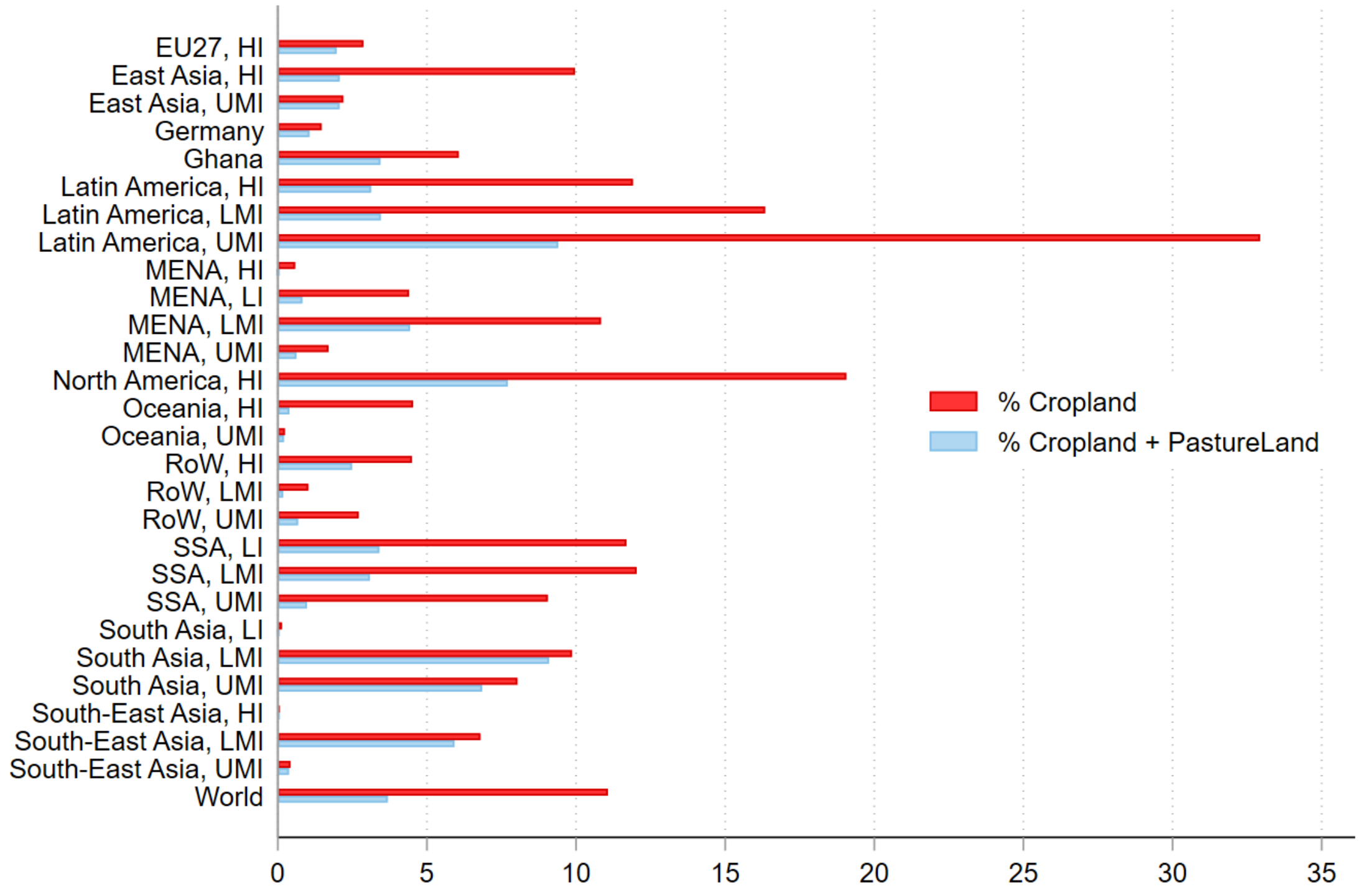

Note: Own elaboration.

## 2.2 Methodology

### 2.2.1 Model configuration

The CGE analysis is implemented in the open-source modelling platform CGEBox (Britz and van der Mensbrugghe, 2018), with G-RDEM (Britz and Roson, 2019; Roson and Britz, 2021) building on the GTAP standard model (van der Mensbrugghe, 2018) at its core. The CDE (Constant Differences in Elasticities) demand system in GTAP is not flexible enough to capture differentiated cross-price effects. We therefore employ a demand structure which bundles nests comprising more closely substitutable food products: cereals, oilseeds, vegetable oils, fruits and vegetables, livestock products (fish, meat, dairy) and introduces further sub-nests for meat and dairy; see Figure 2 for household demands. These nests are also active for government and investment demand as well as intermediate demand.

**Figure 2. CES-sub nets relating to agri-food used for private, government and investment demand**

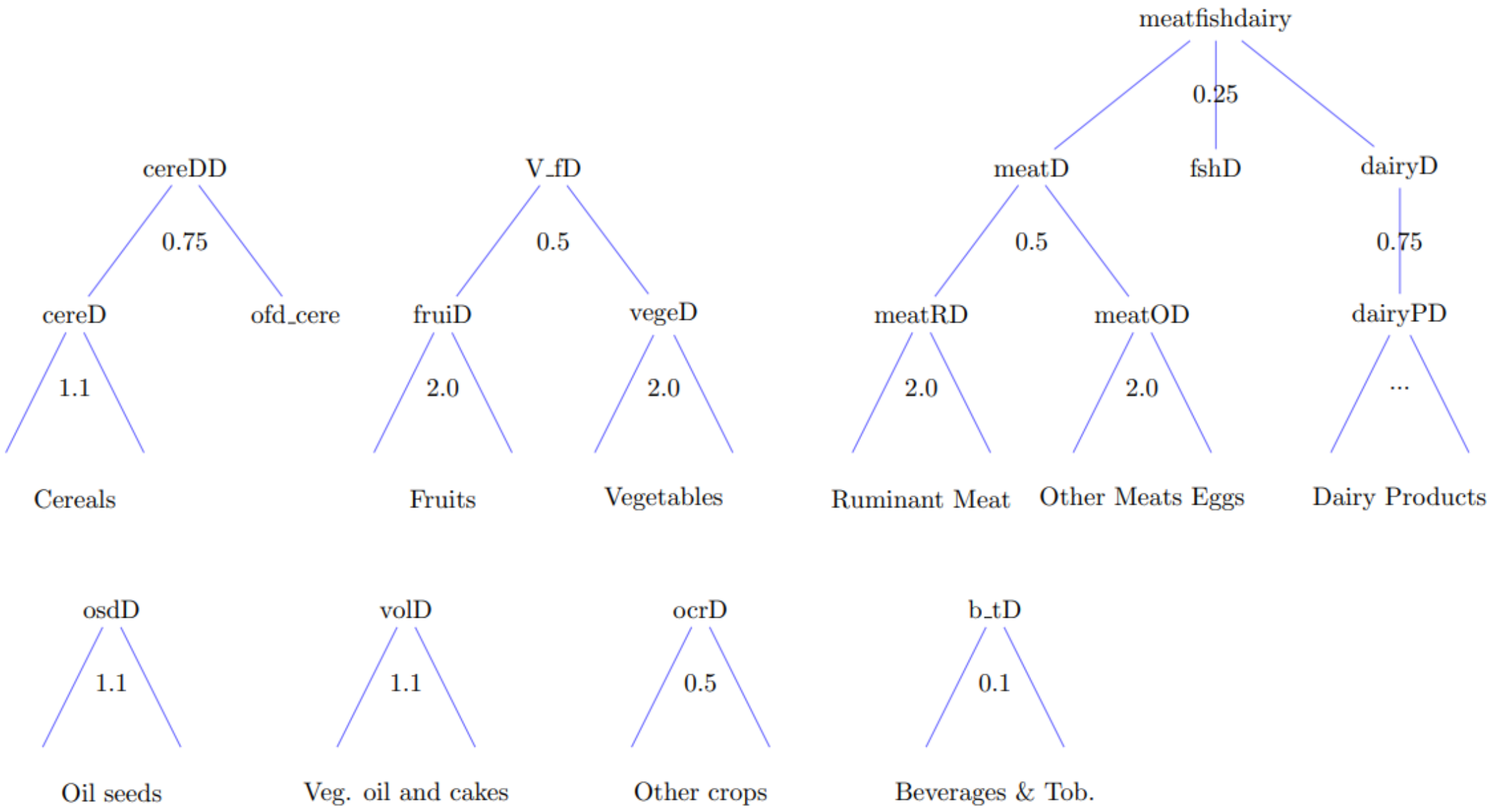

Notes: Own elaboration.

The focus on the agricultural sector requires detail in land use, realized based on the GTAP-AEZ (Agro-Ecological Zones) model (Lee et al., 2009). The AEZ module enhances LULC representation by disaggregating land demand and supply at the regional level into 18 AEZs with similar biophysical properties and by improving substitution among land uses. Land is immobile across these units. Compared to the original GTAP-AEZ model, CGEBox introduces three key improvements (see Figure 3): (1) an additive constant elasticity of transformation land supply function to preserve physical area balances (van der Mensbrugghe, Peters, 2016); (2) a land nesting structure distinguishing annual from perennial crops; and (3) statistically estimated cropland conversion patterns from grassland, managed forest, or unmanaged land (Britz, 2022b).

**Figure 32. CET nesting in the GTAP-AEZ model, extensions by ACET, and additional CET nesting in the GTAP-AEZ model**

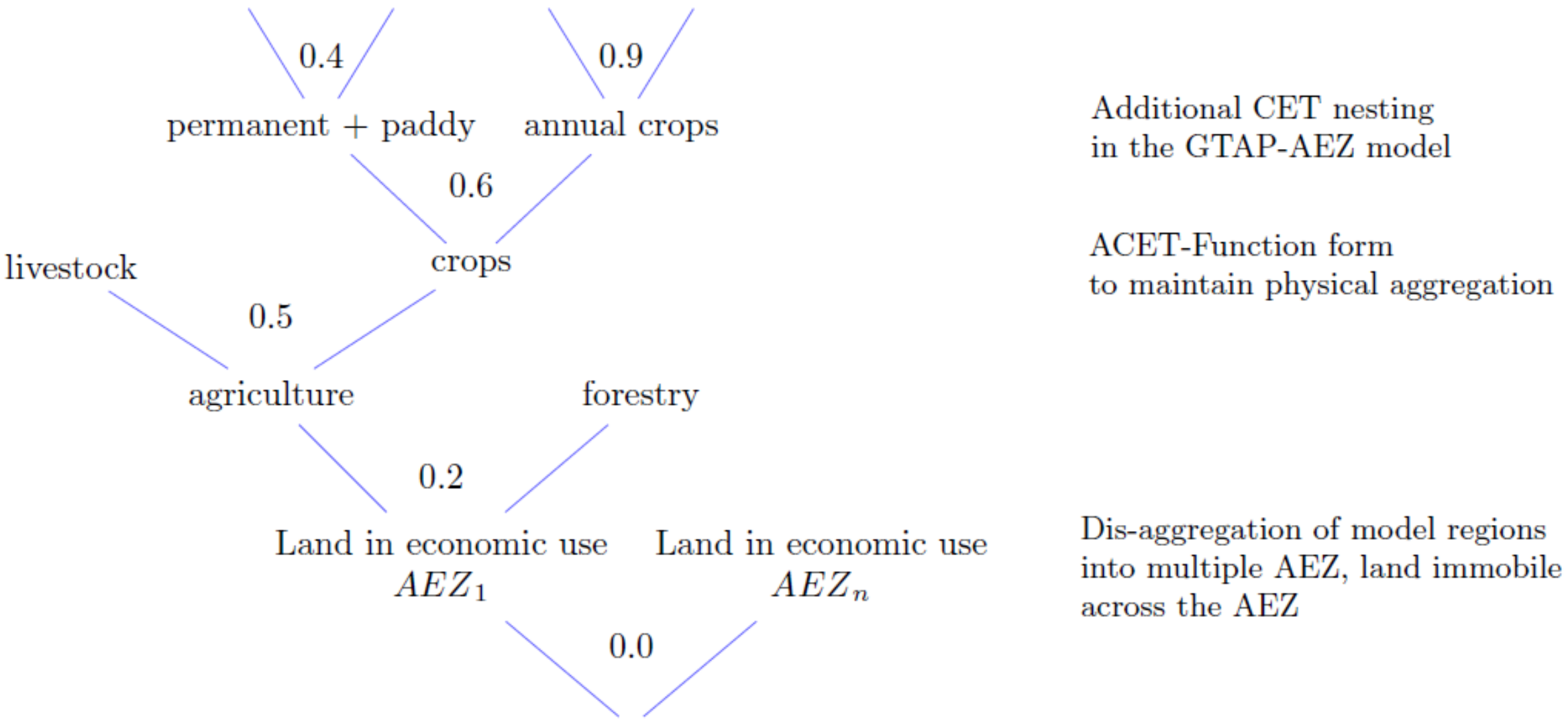

Notes: Own elaboration.

Additional nestings in the production function provide a more realistic depiction of substitution possibilities in agricultural production, drawing on a similar layout in the ENVISAGE model (van der Mensbrughe, 2008) (see Figure 4). For livestock production activities, feed from pastureland and concentrates are treated as imperfect substitutes. Sub-nests describe substitution elasticities in the feed composition (see Figure 5).

**Figure 4. CES-Nestings for crop production activities**

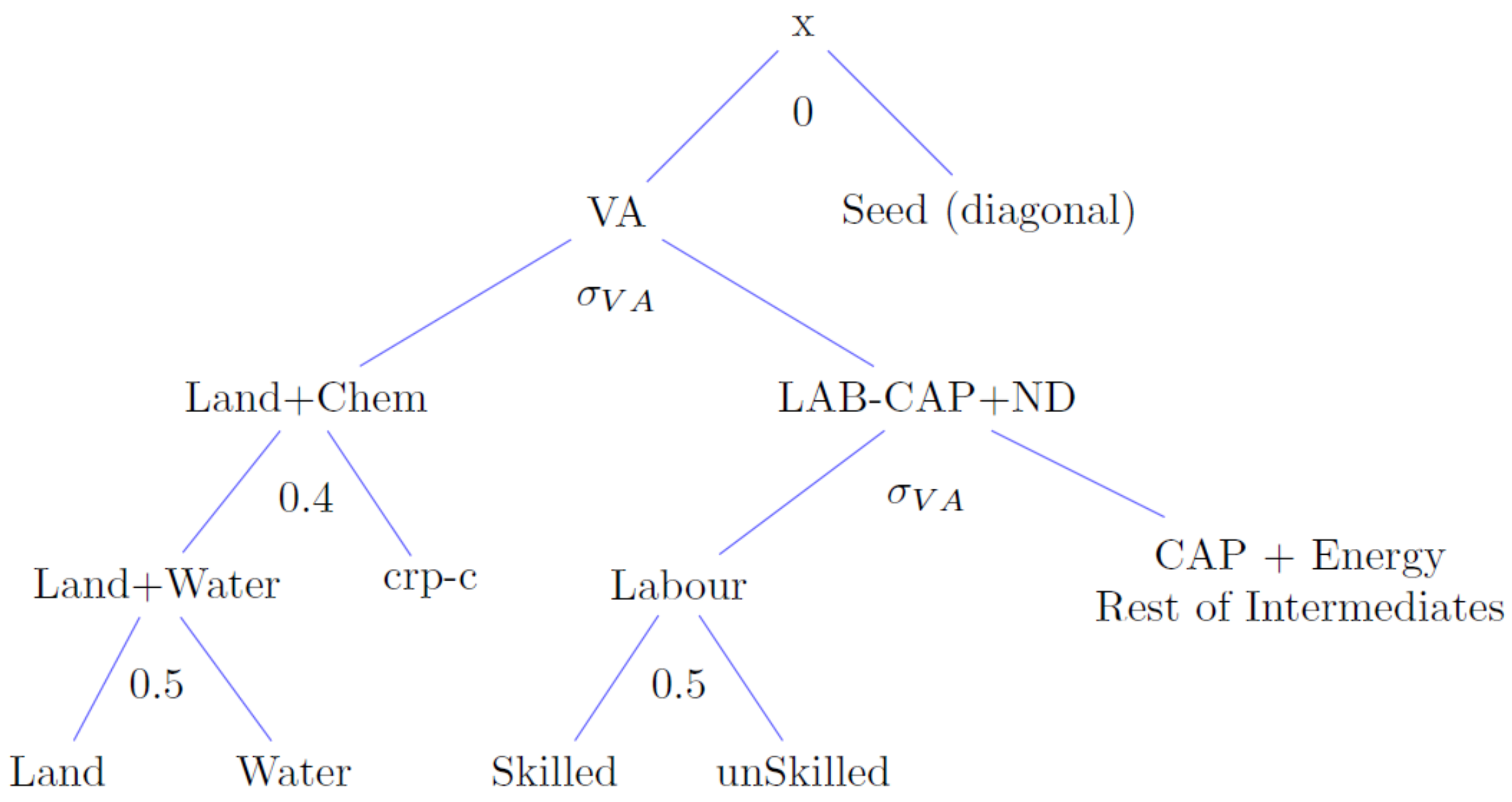

Notes: Own elaboration.

**Figure 5. CES-Nestings for livestock production activities**

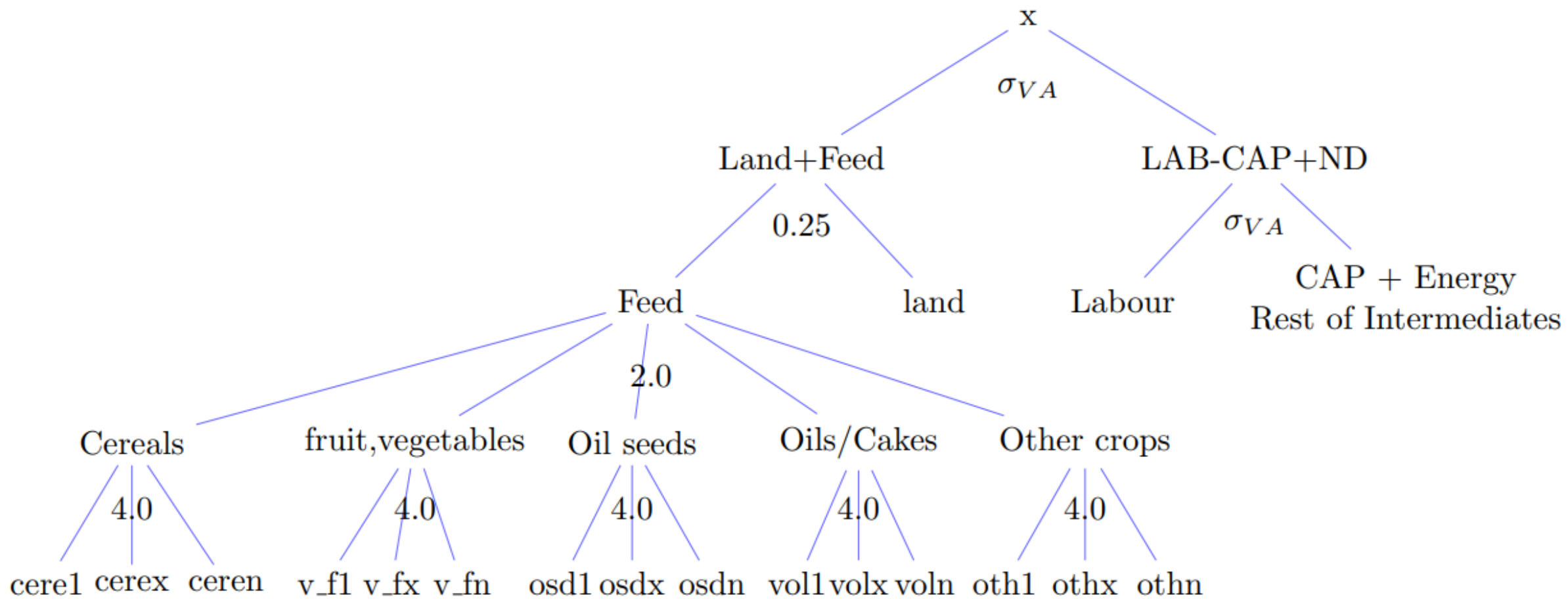

Notes: Own elaboration.

The energy module, derived from GTAP-E (McDougall and Golub, 2007), supports detailed substitution among energy carriers and between energy and capital and represents energy quantities in million tonnes of oil equivalent (MTOE); see Figure 6. CO2 and non-CO2 emission accounting modules track and tax climate-relevant emissions using GTAP-related data (Chepeliev, 2020; 2022).

**Figure 6. CES-Nesting for energy and capital and different energy carriers**

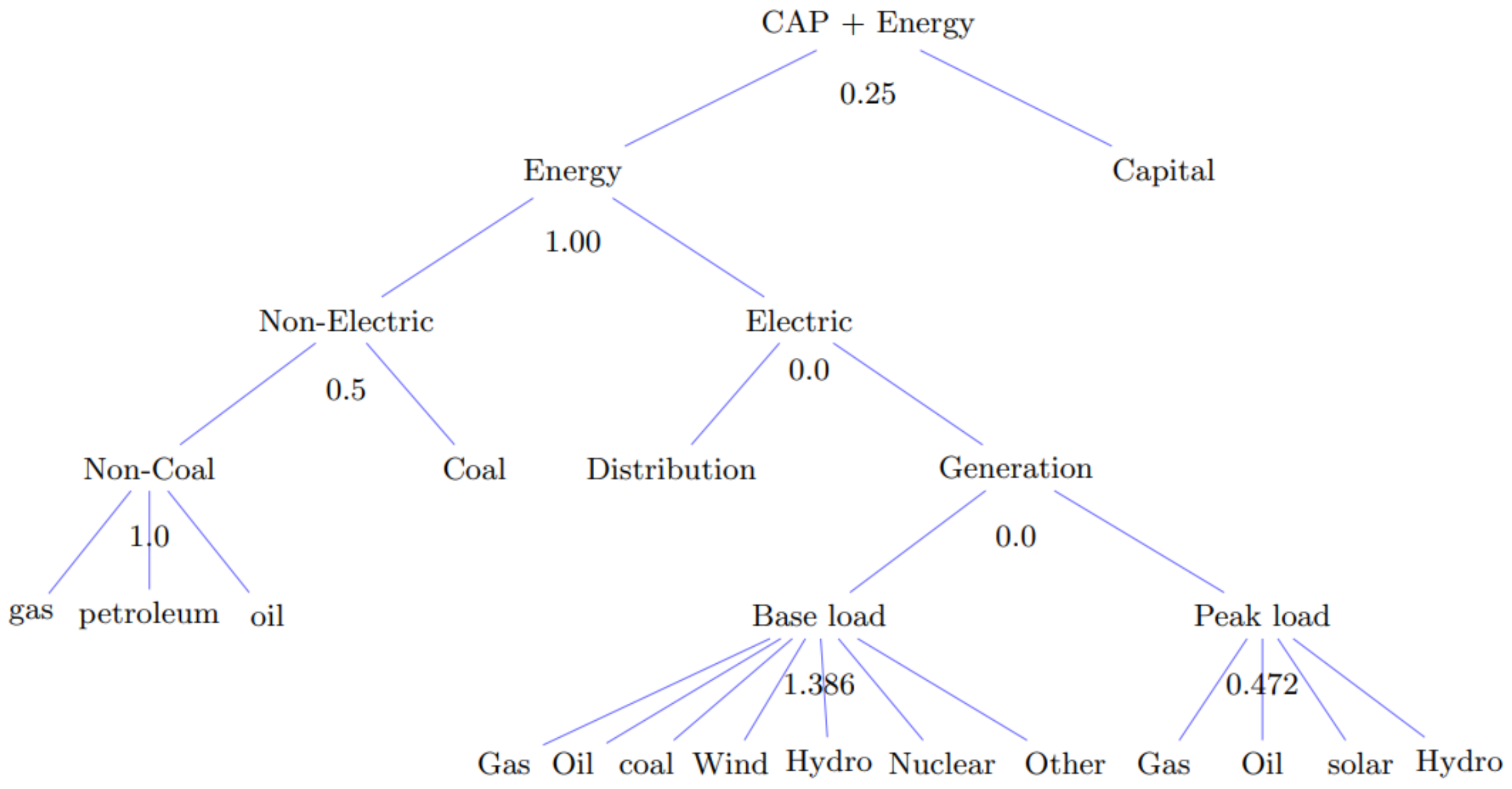

Notes: Own elaboration.

The sub-national module disaggregates production functions and factor markets to sub-national levels (Britz, 2020a). Lastly, of the different possibilities to depict international trade, the simplest version available in CGEBox is used, which employs a two-stage Armington presentation with identical domestic and import shares for all agents, differently from the GTAP Standard model, where these shares are allowed to vary across demanders. This Armington representation is usual in single-country CGE models, but also found, for instance, in the multi-regional GLOBE CGE model (McDoland and Thierfelder, 2014). The parameters for the substitution between importer regions are taken from Fontagne et al. (2019), which estimate elasticities at HS6 tariff line level. These estimates are aggregated based on trade weights to the product detail of the database. Substitution elasticities between the imported and domestic origin use half of the ones between individual importers in the lower nest.

### 2.2.2 Scenario design

The analysis is comparative-static, comparing a benchmark situation with a new equilibrium after adjustments provoked by our shock scenarios. These scenarios introduce either production subsidies or use subsidies. In order to make them comparable, they imply the same cost to the taxpayer at benchmark data. Specifically, in order to calculate equivalent subsidy rates to production, we first calculate the monetary cost of subsidizing the household legume consumption bundle by either 10% or 20% at the benchmark data, and then support their production by an ad valorem subsidy that imposes the same cost to the taxpayer.

In our model, legumes related to soybeans, groundnuts, and other Leguminosae. Since they differ in terms of their production and demand structures, as well as their nutritional value, we will simulate four scenarios for each subsidy rate in which the legumes eligible for the subsidy are only soybeans, groundnuts, or other Leguminosae, or all three together. In the latter case, since we are calculating the monetary cost of subsidizing consumption, the same ad valorem subsidy will be applied to all three. This could imply that, for instance, despite very little household consumption of soy, soy production could still receive a sizeable subsidy[6]. We call this a cross-subsidization effect.

[6] In one extreme case, for example, suppose a country imports all the soybeans it consumes; we are using the budget generated by the demand for Leguminosae to subsidize the non-existent production of soybeans.

Each scenario first calculates the monetary cost of subsidizing the consumption of households of one type of legume or of all by a certain percentage in the benchmark data, and derives from the estimated costs the subsidy rate to production. Similar to what is stated in footnote 8, suppose a country imports all the soybeans it consumes, and we analyze the case of subsidizing soybeans only. In this case, the budget required for the consumption subsidy cannot be mapped to a production subsidy, as no domestic production is observed. If production is small compared to consumption, we could generate extremely large per-unit production subsidies. For this reason, we limit production subsidies in each country or model region to a maximum of 75%.

Subsidizing final use or production will have different implications and challenges. While final-use subsidies target household consumption, production subsidies promote both intermediate (e.g., feed use) and final use, as well as exports. In addition, from an implementation perspective, it might be easier to subsidize production, especially in a setting where a lot of fresh produce is marketed locally. In contrast, our use subsidy targets household consumption and requires a well-established tax system at the point of sale, such that a subsidy could be implemented, for instance, as a change in the VAT tax level applied.

In summary, we will simulate 16 scenarios: two subsidy types (use and production), two subsidy rates (10 % and 20%), and four legume targets (soybeans, groundnuts, other Leguminosae, and all three types of legumes).

# 4. Results

In this section, we present simulation results compared to the benchmark for soybeans, groundnuts, and other Leguminosae.[7] To ease visualization, the figures shown in this section correspond to scenarios with 20% subsidies on all legumes. They also exclude cases with very small benchmark values that vary widely under the shock scenarios. Particularly, the figures are limited to cases with deviations lower than 25% in absolute value. In Appendix C, we include Table C1 for selected results (output, household demand, and nutritional intake) and figures with the complete distribution of results for all regions. As discussed above, we also pay special attention to Ghana and Germany. In Tables B1-B3 in Appendix B, we summarize the results for Germany, Ghana, and the World.

## 4.1. Prices

The direct effect of the shock scenarios is evident in consumer and producer prices. With use subsidies, consumer prices for legumes go down and producer prices up (see Figure 7 and Figure 8 and also Figure C1 in Appendix C). This reflects that the higher demand for legumes stimulated by the use subsidy implies increased per-unit production costs. With production subsidies, firms' unit costs are decreasing, implying both lower producer and consumer prices. Use subsidies to households are more effective in reducing consumer prices, as they are more targeted: production subsidies benefit all uses, including intermediate demand such as feed and exports. In terms of regional heterogeneities, subsidizing in producer regions like East Asia HI, Latin America UMI, and North America leads to relatively small producer price variations. Regions like Southeast Asia and MENA, with large producer price variations, tend to consume and produce small amounts of legumes (see Figure 7 and Figure 8).

As shown in Table B1, at the global level, this becomes clear: use subsidies decrease consumer prices of soybeans between 1.6% and 3.4%, of groundnuts between 9.1% and 18.6%, and of other Leguminosae between 7.1% and 14.3%. In contrast, production subsidies reduce consumer prices of soybeans between 0.9% and 1.9%, of groundnuts between 5.6% and 12.3%, and of other Leguminosae between 3.6% and 7.3%. Use subsidies increase producer prices between 0.1% and 0.4% for soybeans, between 0.5% and 1.4% for groundnuts, and between 3.3% and 7.5% for other Leguminosae, with a similar reduction also for consumers. With production subsidies, there are differences between the scenarios in which single legumes are subsidized separately and the scenario in which all are subsidized, suggesting the presence of the aforementioned cross-subsidization effect. At the global level, this effect benefits groundnuts, which experience much larger reductions in producer price when all legumes are subsidized, at the expense of soybeans and other Leguminosae, indicating that there is competition among these goods on the production side.

[7] We have also analyzed results for meat and livestock products, which represent an alternative important source of macronutrients. However, they are not shown here, as the deviations are very small. As a consequence of the small price responses, livestock and meat output is virtually unaffected. Those results are available upon request.

Price effects in Germany and Ghana are larger. In Germany, use subsidies reduce soybean prices by 7.1% and 14.6%, and Leguminosae prices by 6.7% to 13.3%; whilst production subsidies reduce soybean prices between 4.8% and 16%, and Leguminosae prices between 2.3% and 6.7%. As there is no production of groundnuts in Germany, there are no producer price responses, but consumer prices do slightly increase under use subsidies. Groundnut consumer prices decrease under production subsidies because imports are now cheaper. It is interesting to note that cross-subsidies in Germany are bad for all legumes, i.e., the price reduction is lower when all legumes are subsidized together compared to when each legume is subsidized separately. This suggests that the expansion of legumes in Germany is constrained by land availability. In Ghana, use subsidies reduce soybean consumer prices between 5.4% and 11.5%, groundnut prices between 9.4% and 19.3%, and Leguminosae prices between 6% and 12.4%, whilst they increase producer prices between 0.3% and 11.4% for soybeans, between 0.4% and 1.6% for groundnuts, and between 4.3% and 9.9% for other Leguminosae. Meanwhile, production subsidies reduce soybean prices between 4.9% and 9.9%, groundnut prices between 0.8% and 6.7%, and other Leguminosae prices between 3.5% and 7%. Cross-subsidies effects are large and benefit groundnut over soybeans and other Leguminosae. For example, Leguminosae consumer prices decrease by 7% when only subsidizing Leguminosae, but by 2.6% when subsidizing all legumes, while groundnut prices decrease by 1.6% when only subsidizing groundnut, but by 6.7% when subsidizing all legumes.

## Figure 7. Producer Prices, By Regions

### Panel A: Soybeans

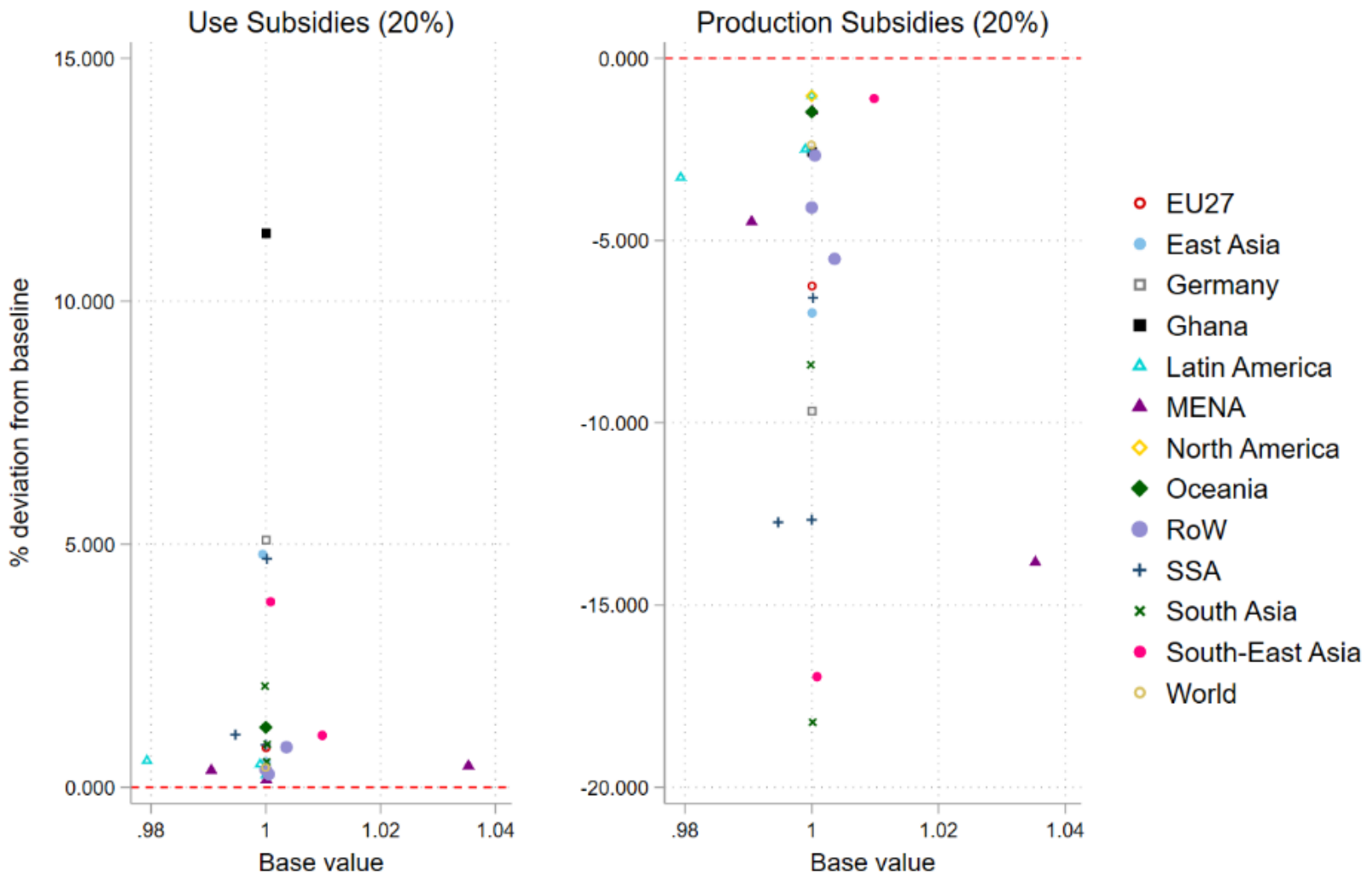

### Panel B: Groundnuts

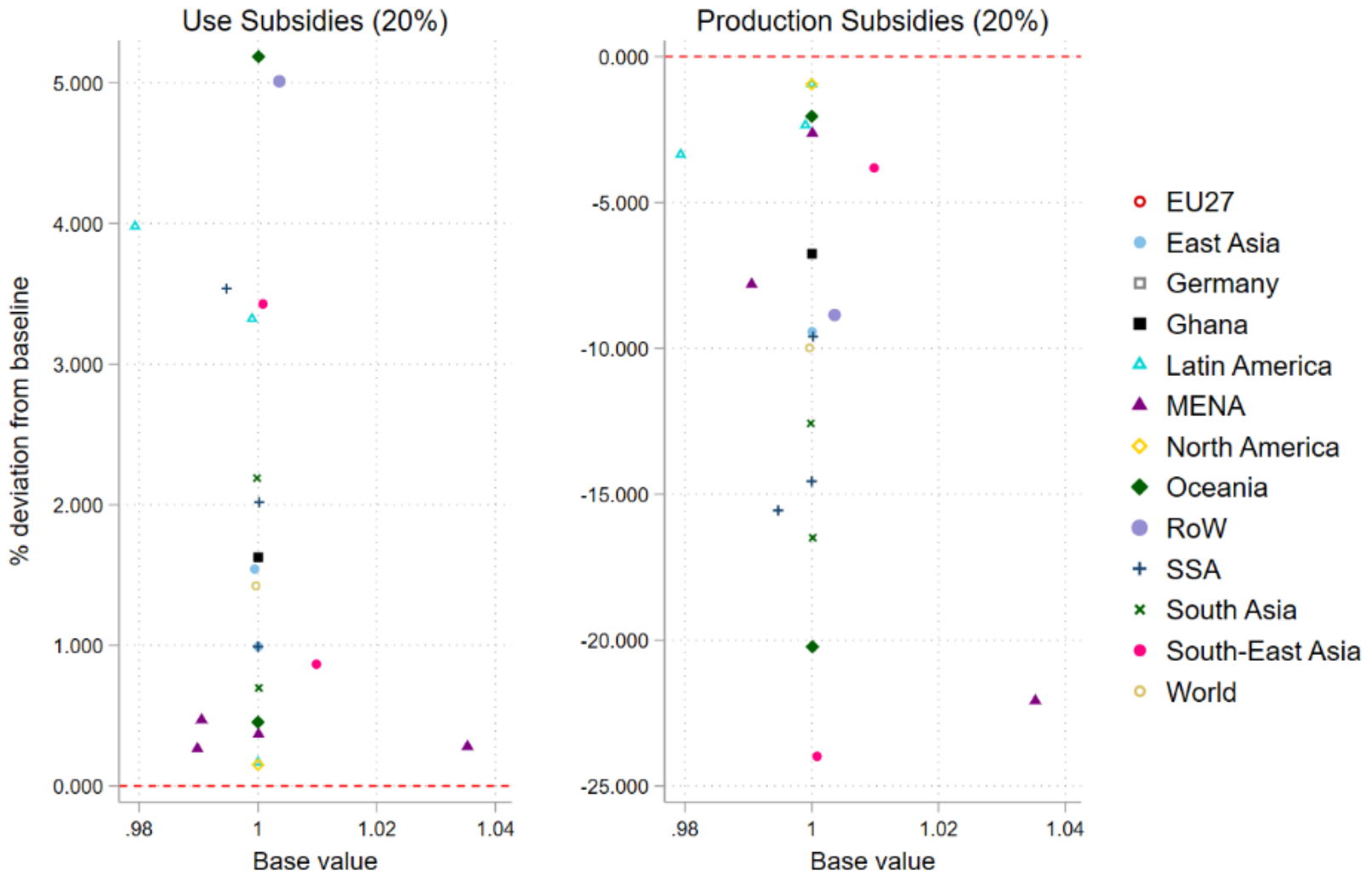

### Panel C: Other Leguminosae

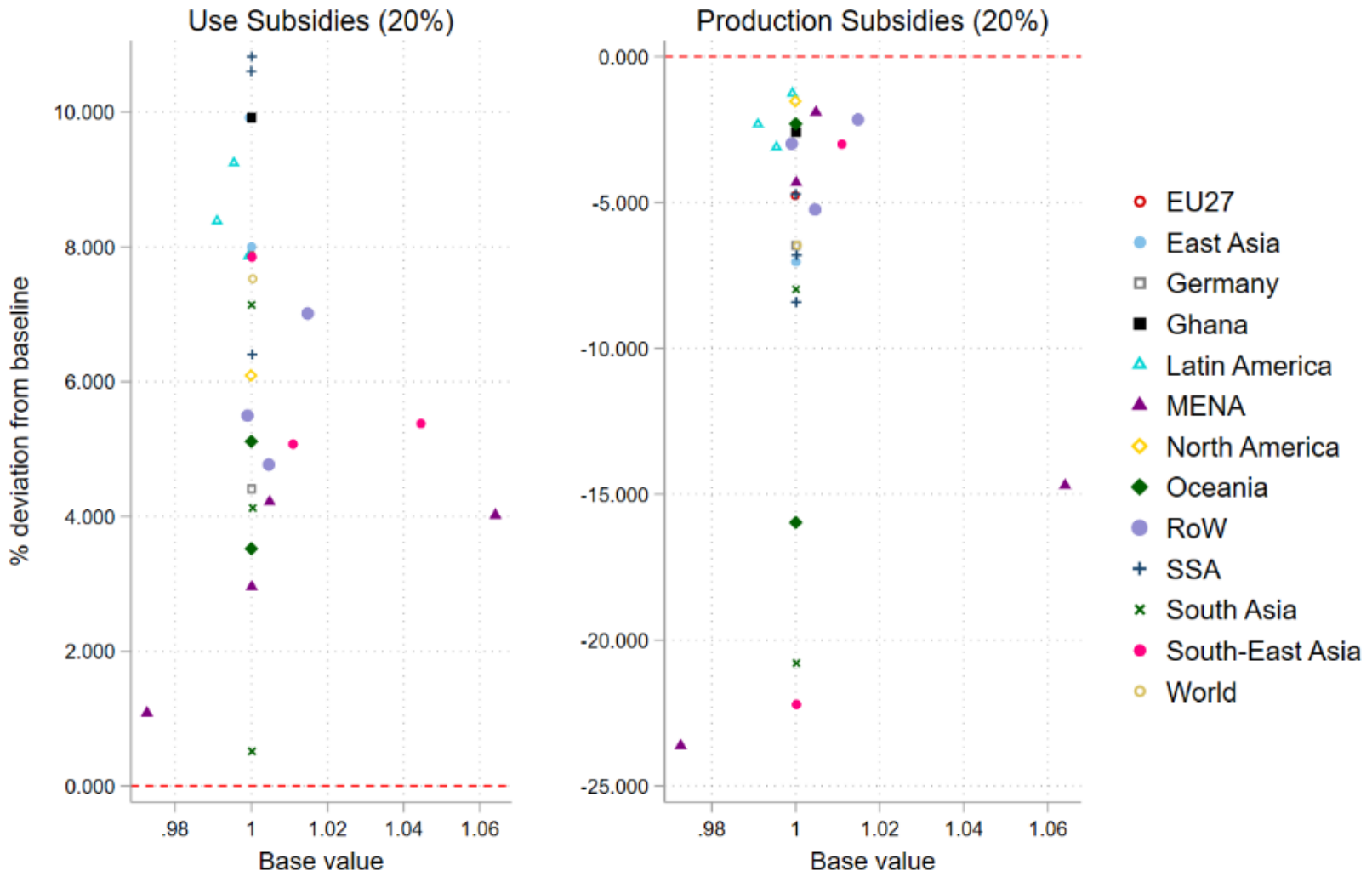

Notes: Own elaboration. The figure shows the results for scenarios with 20% subsidies on all legumes.

**Figure 8. Consumer Prices, By Region**

**Panel A: Soybeans**

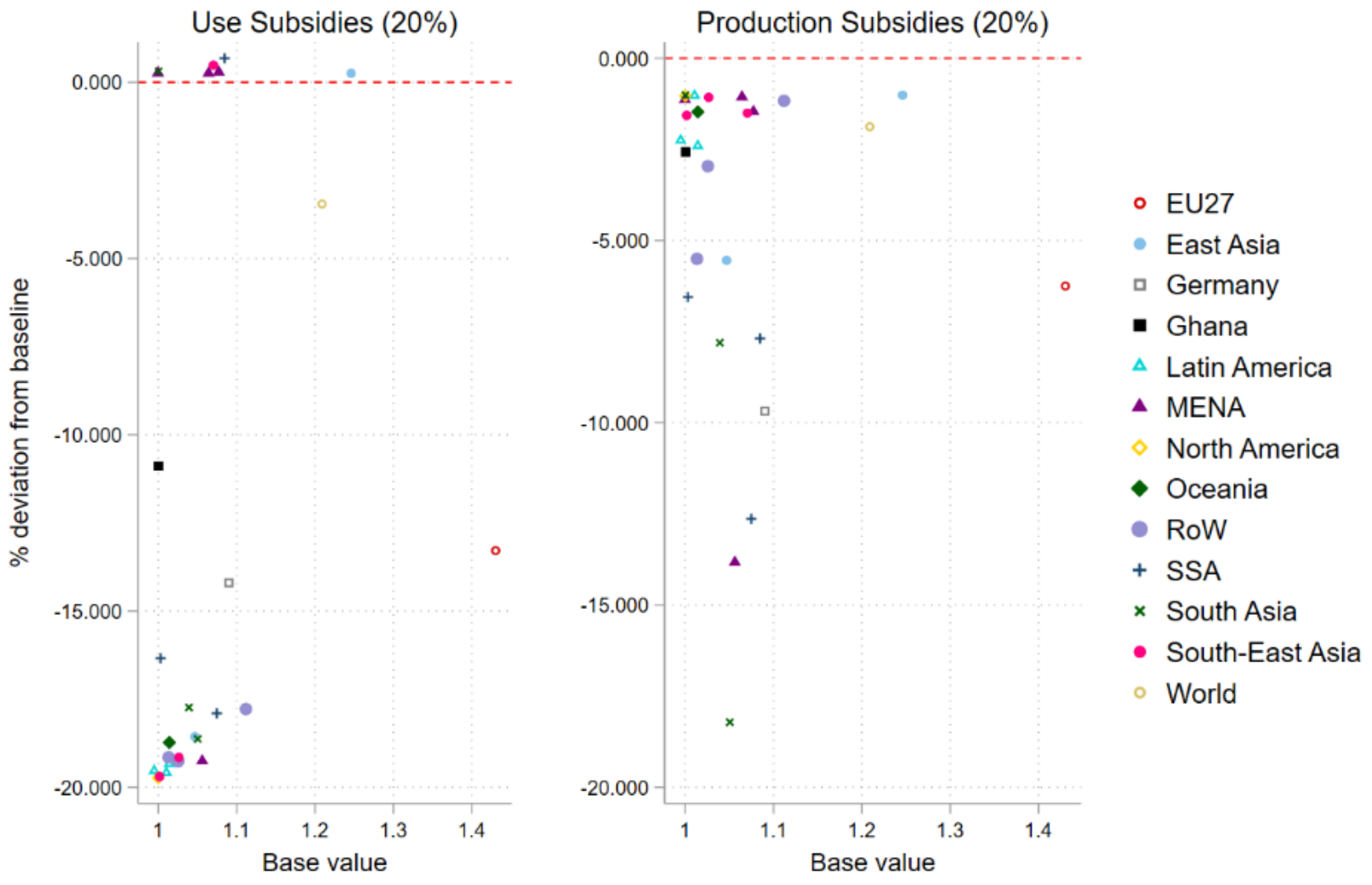

**Panel B: Groundnuts**

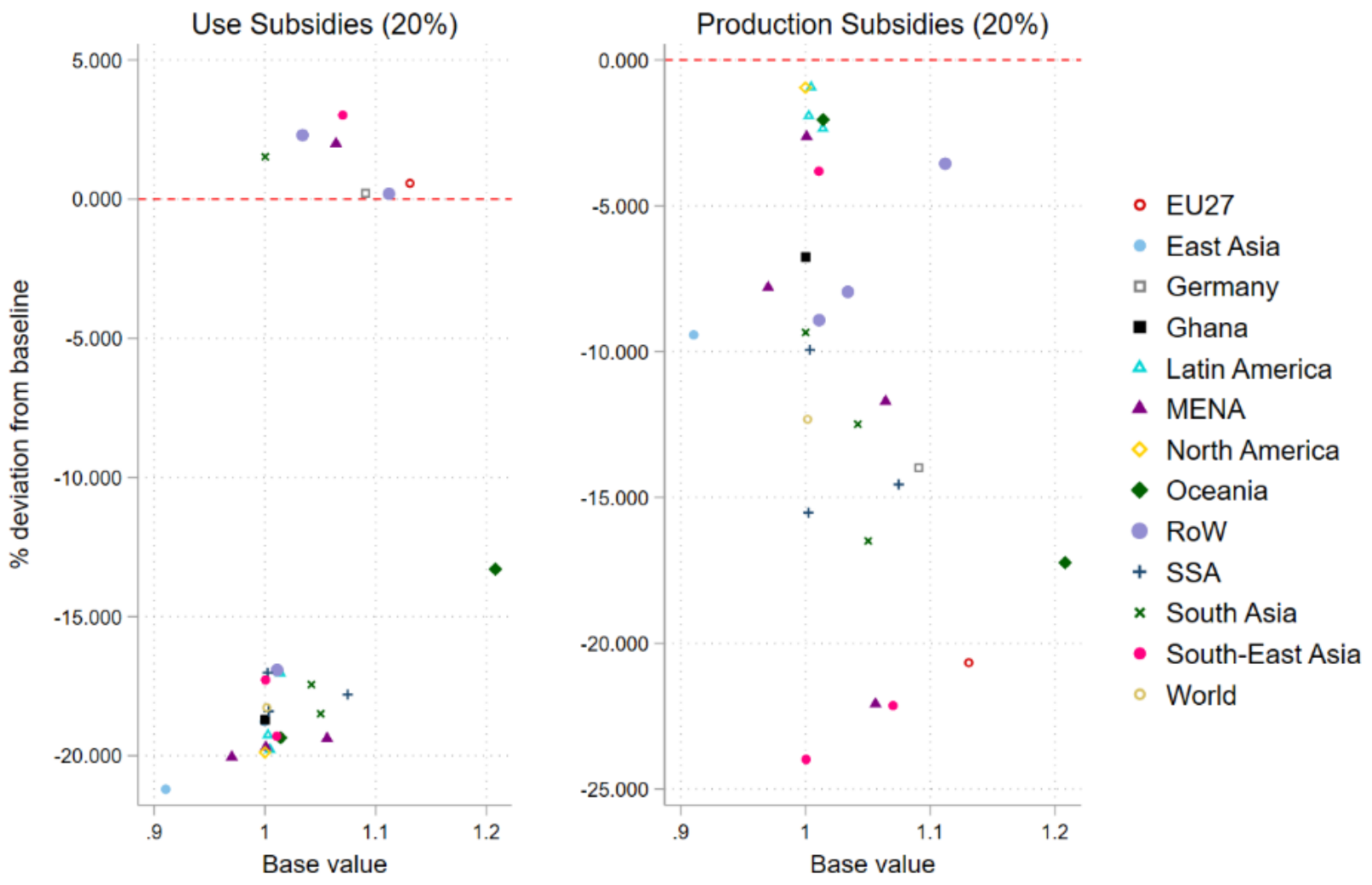

**Panel C: Other Leguminosae**

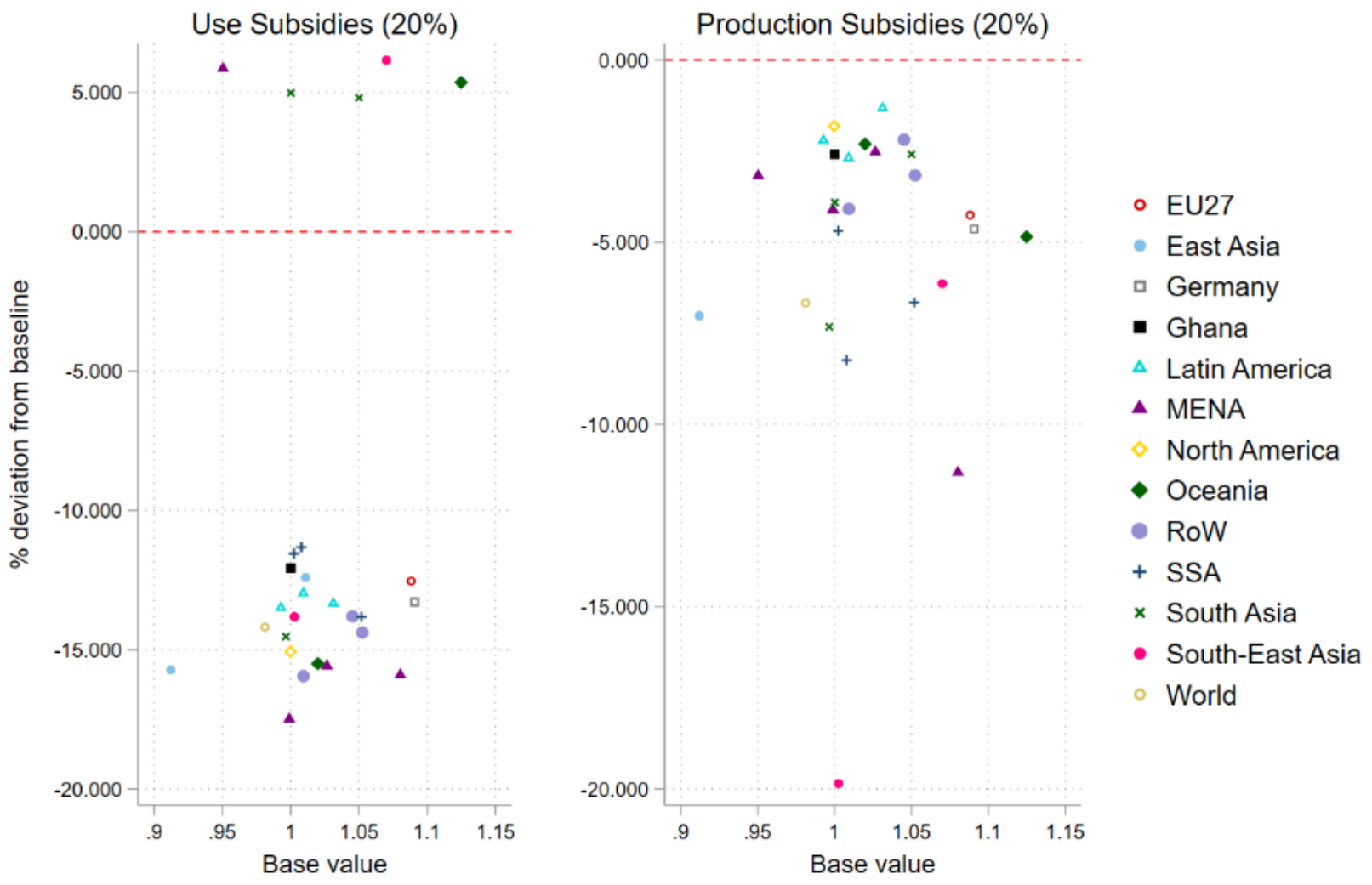

Notes: Own elaboration. The figure shows the results for scenarios with 20% subsidies on all legumes.

## 4.2. Production

In response to the price signals, production of legumes increases (see Figure 9 and Figure C2), especially when subsidizing production in the case of soybeans, but similarly for Leguminosae. Logically, production subsidies are more effective than use subsidies in increasing domestic production. There are significant regional heterogeneities. In particular, subsidizing in producer regions, such as East Asia HI, Latin America UMI, and North America, leads to small price variations, which in turn reduces the incentive to increase production. In fact, in the case of Latin America UMI and North America, this tends to even reduce output (see Figure 9). In contrast, subsidies in East Asia HI are highly effective in increasing output for all legumes, which is related to the initial low share of cropland devoted to legumes in that region. In contrast, in Latin America and North America, they already occupy a large share of cropland, suggesting there is less room for their expansion (see Figure 1). Regions that experience large output variations, like MENA, Southeast Asia, and South Asia, produce negligible amounts of legumes.

At the global level (see Table B1 in Appendix B), soybean output increases slightly under final-use subsidies, while it could rise by up to 3.2% under 20% production subsidies. Groundnut output increases between 2.2% and 5.2% under use subsidies, whilst between 2.4% and 10.4% under production subsidies. On the contrary, for other Leguminosae, output responses are larger under use subsidies: production increases between 4.6% and 10.4% under final-use subsidies and between 3.6% and 7.7% under production subsidies. This reflects the cross-subsidy effects already discussed and that other Leguminosae are mostly produced for final consumption.

Output responses in Ghana and Germany are somewhat larger, although the benchmark values are rather low for both economies. In Germany, use subsidies increase soybean production by between 6.1% and 13.2%, whilst production subsidies increase it by between 4% and 14.6%. Germany does not produce groundnuts. Leguminosae output increases by between 2.7% and 6.2% under use subsidies, whilst by 4% and 10.9% when using production subsidies. In this case, cross-subsidies benefit other Leguminosae. For example, Other Leguminosae production increases by 8.4% when only subsidizing other Leguminosae, but by 10.9% when subsidizing all legumes, while soybean production increases by 14.6% when only subsidizing soybeans, but by 8.3% when subsidizing all legumes. In Ghana, use subsidies increase soybean production by between 4.1% and 9%, whilst production subsidies increase it by between 1.2% and 9.4%. Leguminosae output increases by between 4.8% and 10.6% under use subsidies, whilst by 3.7% and 7.9% when using production subsidies. Cross-subsidies benefit groundnut over soybean and other Leguminosae in Ghana. For example, groundnut production decreases by 2.4% when only subsidizing groundnut (global supply increase reduces groundnut world price and Ghana’s competitiveness), but increases by 1.2% when subsidizing all legumes, while soybean production increases by 9.4% when only subsidizing soybeans, but by 2.4% when subsidizing all legumes.

# Figure 9. Production, By Region

## Panel A: Soybeans

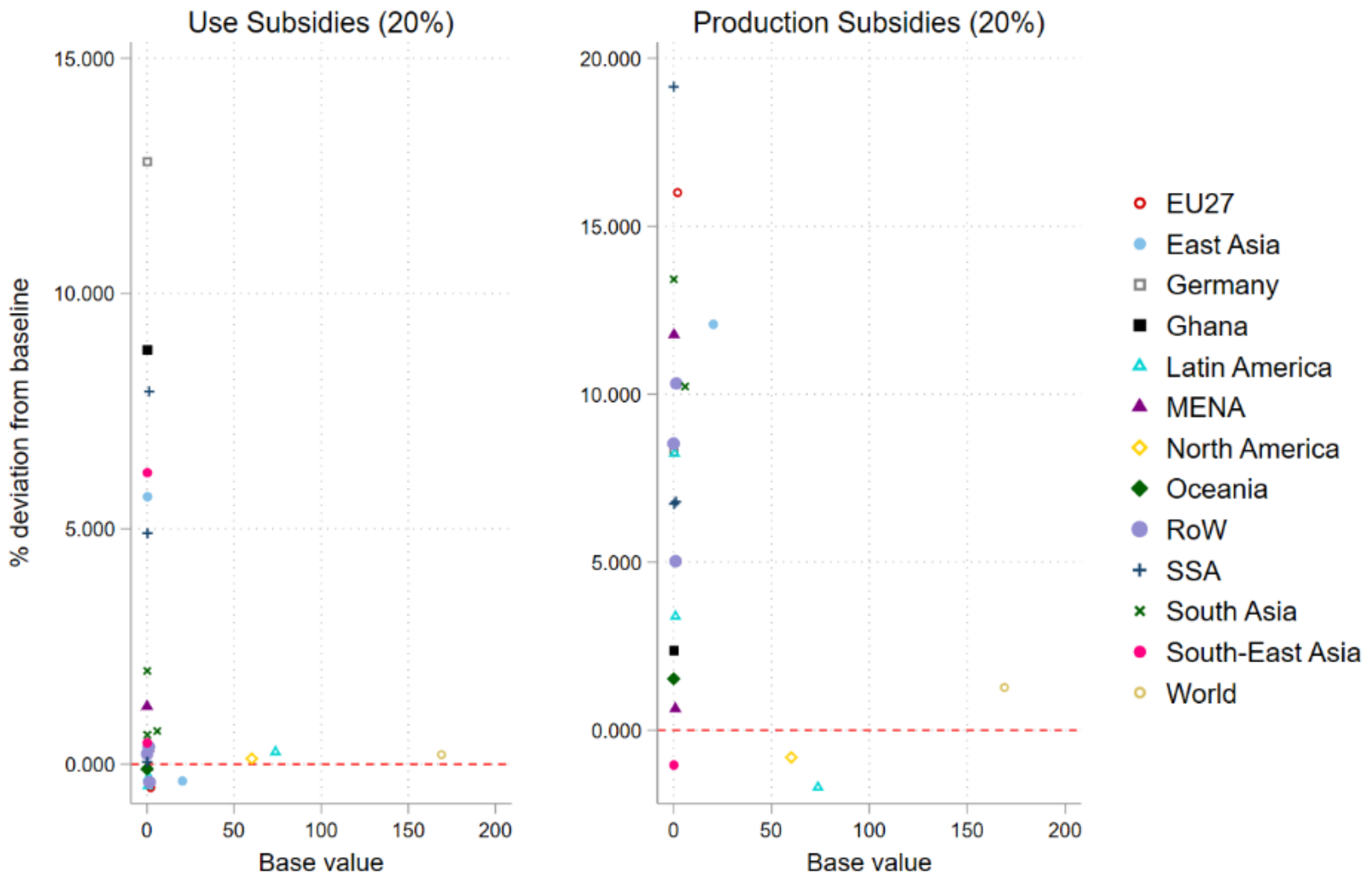

## Panel B: Groundnuts

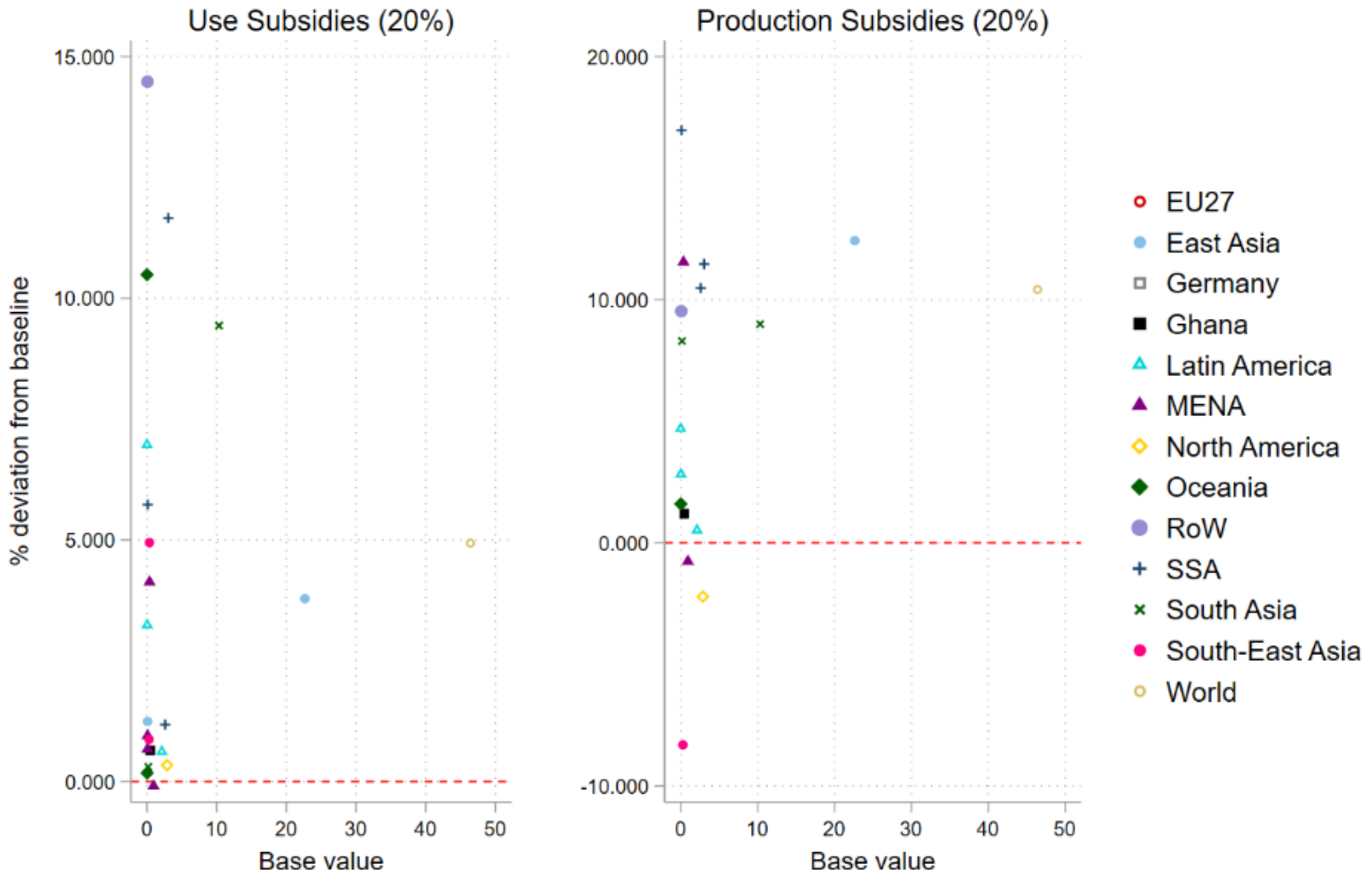

## Panel C: Other Leguminosae

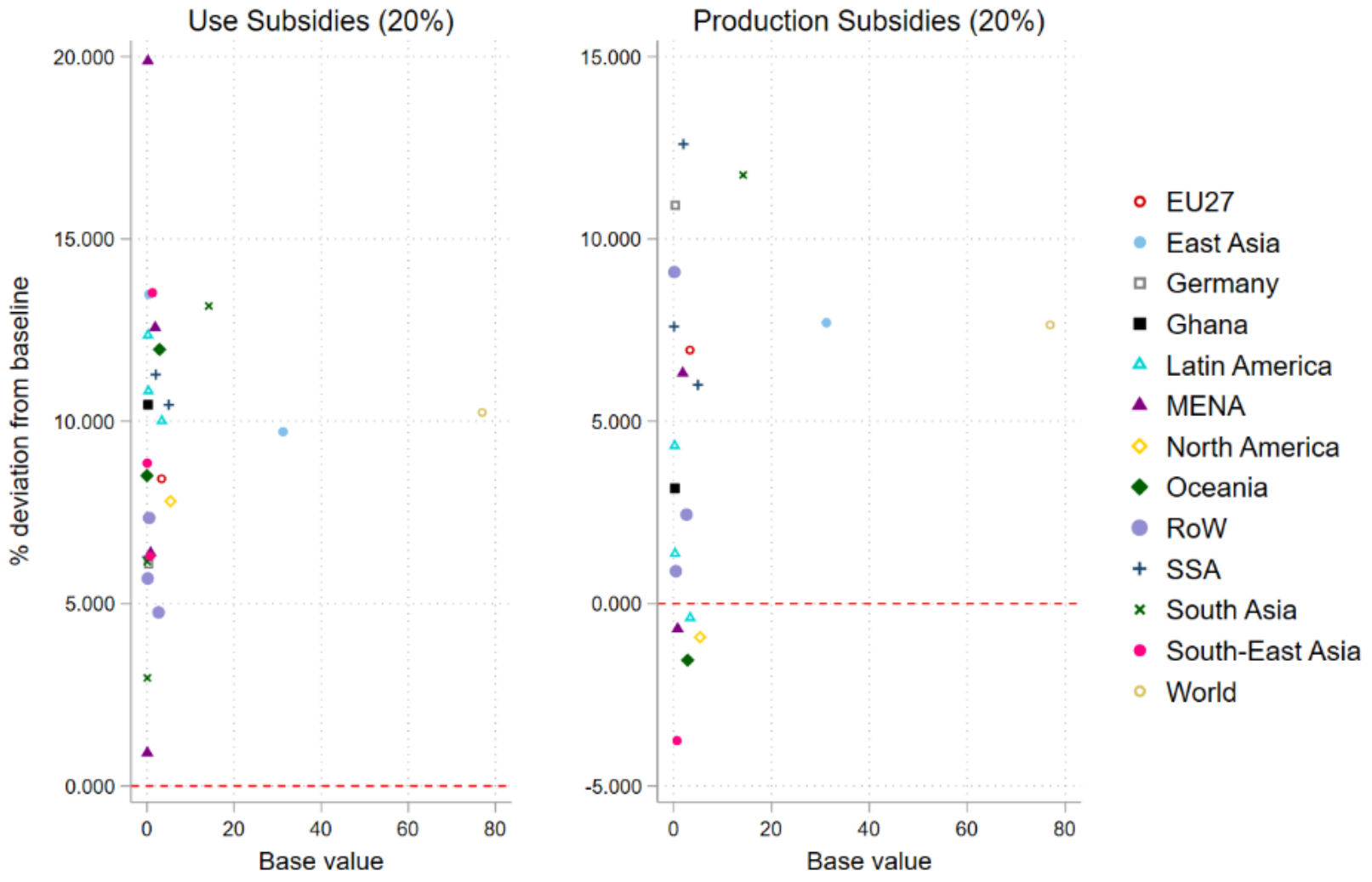

Notes: Own elaboration. The figure shows the results for scenarios with 20% subsidies on all legumes.

## 4.3 Land Cover and Land Use

As production of legumes increases, so does the area of land for their cultivation. The global share of cropland under legumes increases slightly at the expense of livestock area (see Table B2). At the regional level, this effect is also small in magnitude (see Figures 10 and D4). Land use changes are larger (see Figures 11 and D3). At the global level, land use changes for other Leguminosae are very similar under production subsidies. Subsidizing other Leguminosae could increase land use both for irrigated and rainfed variants by between 3.7% and 10.9% and between 2.8% and 10.9%, respectively. Under use subsidies, land use changes for soybeans are relatively modest, ranging between 0.2% and 0.8%. On the contrary, subsidizing production could significantly increase land use for soybeans up to 13.1% for the irrigated variant and 1.4% for the rainfed variant, under a 20% subsidy policy. Groundnut land use increases between 2% and 4.4% under use subsidies, whilst by between 1.6 and 7.7% under production subsidies. The impact of cross-subsidies benefiting groundnut is also evident here: When subsidizing all legumes, groundnut land use increases by 7.7%, compared with the 3.5% increase when subsidizing only groundnut. Meanwhile, for example, the increase in irrigated soybean land use decreases from 13.1% to 6.9%.

In Germany, there is no production of irrigated soybeans or groundnut of any variant, and the production of irrigated other Leguminosae is very small. For the rainfed variants, production subsidies are more effective in increasing their land use area. A targeted production subsidy package of 20% could increase other Leguminosae land use up to 7.5%, while a use subsidy of the same size only increases it by 5.5%. The cross-subsidies affecting soybean production in Germany are clear here too: rainfed soybean production increases by 12.5% when only subsidizing soybean, but only 6.7% when subsidizing all legumes. There is no production of irrigated other Leguminosae in Ghana. Use subsidies are relatively more effective in increasing rainfed legume production. Under production subsidies, cross-subsidy effects benefiting groundnuts are also considerable in Ghana. Subsidizing only soybeans, for example, could increase land use by 8.8% for the irrigated variant and 12.3% for the rainfed variant. However, when subsidizing all legumes, these variations are reduced to 2.2% and 2.8%, respectively.

## Figure 10. Land Cover changes, By Region

### Panel A: Forest

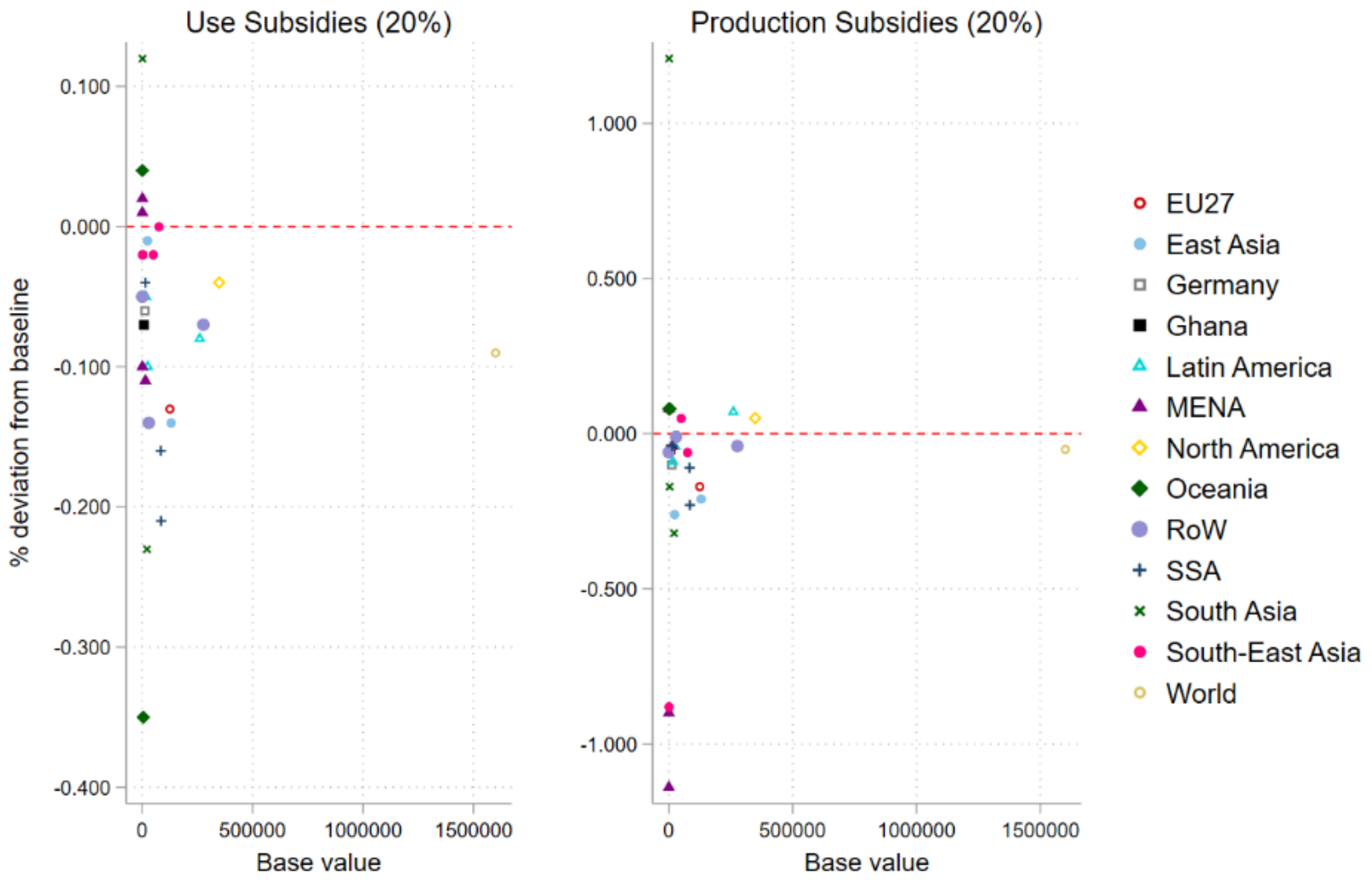

### Panel B: Pastures

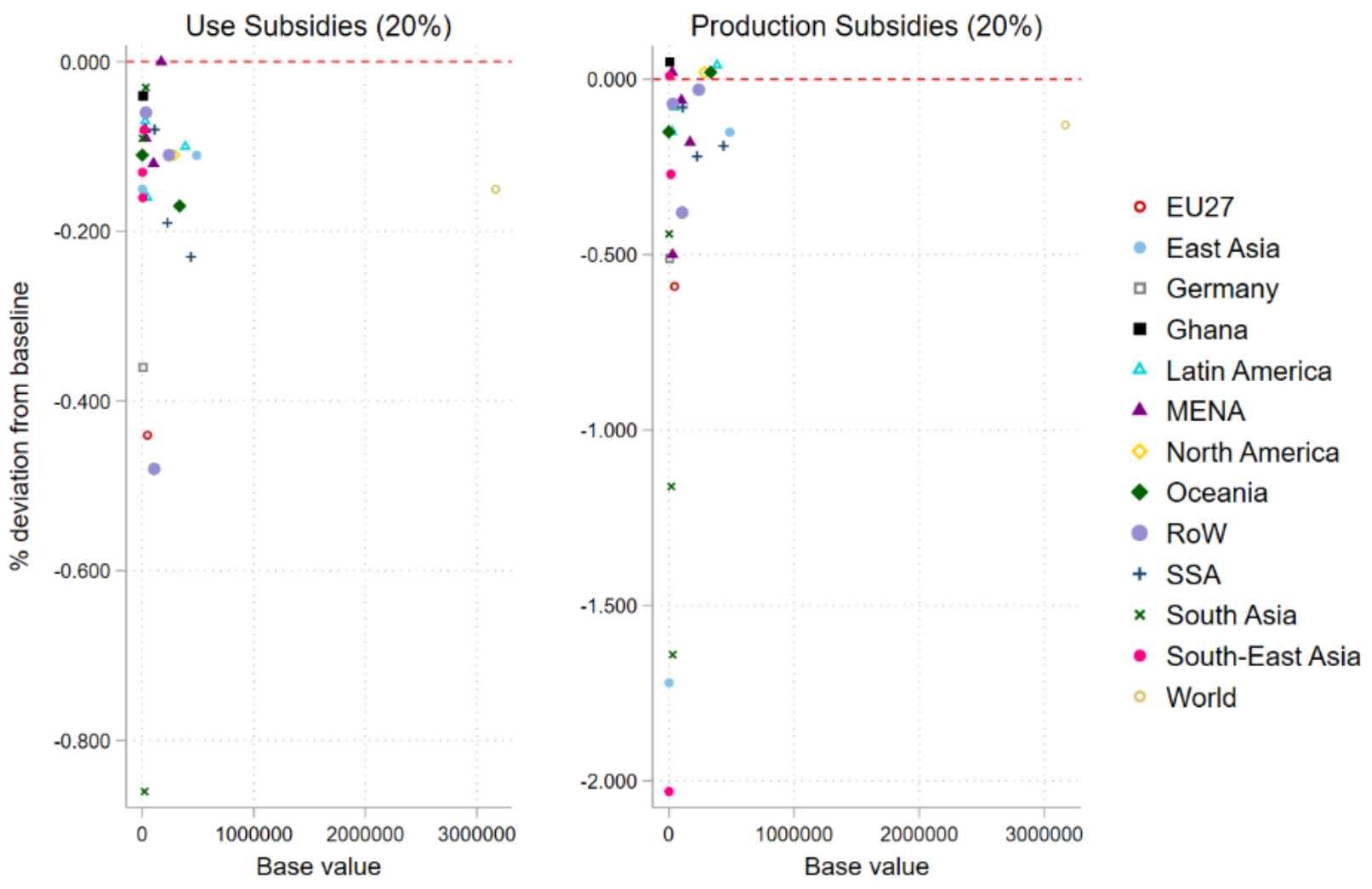

### Panel C: Cropland

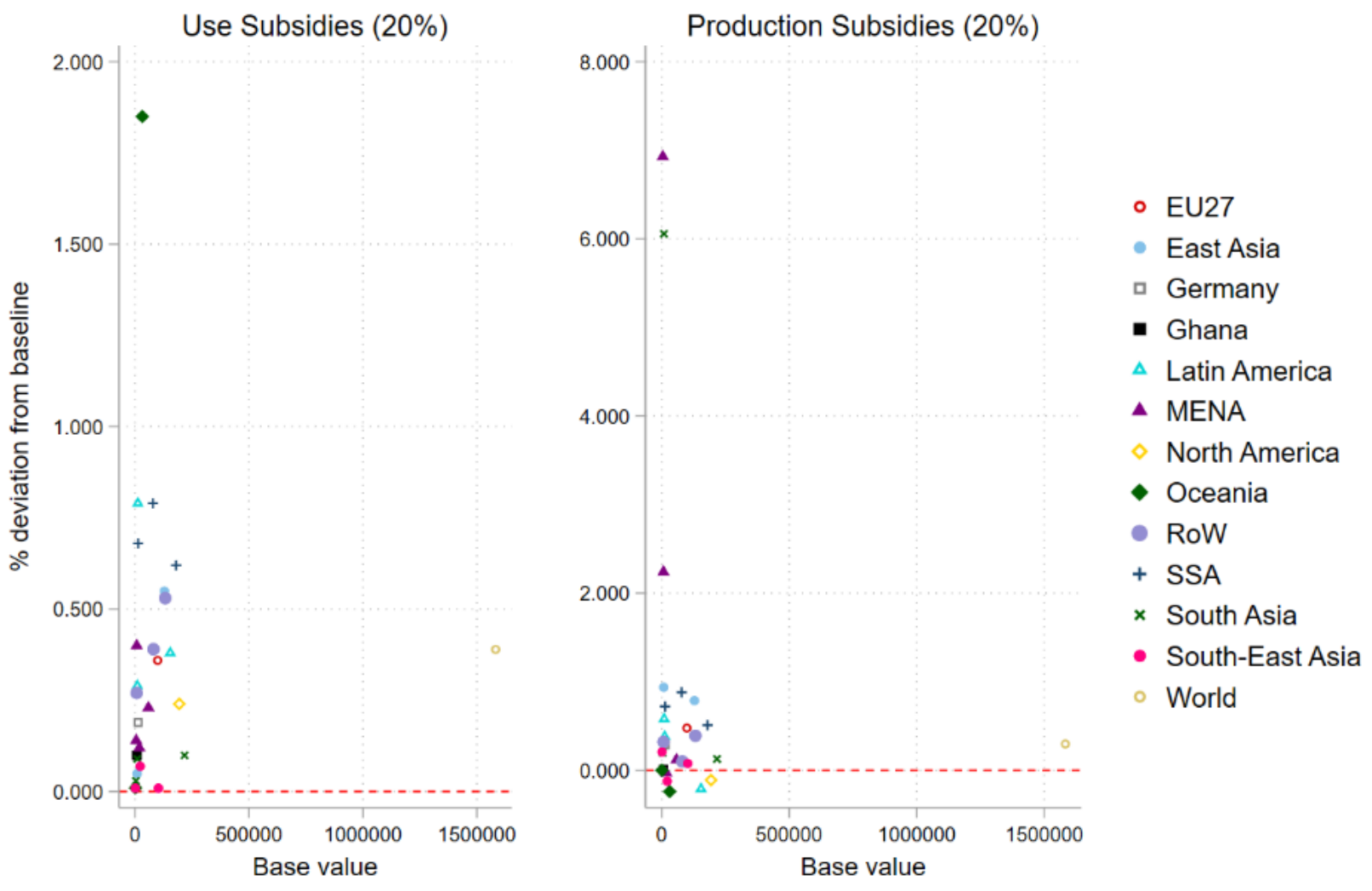

Notes: Own elaboration. The figure shows the results for scenarios with 20% subsidies on all legumes.

## Figure 11. Land Use for Legume Cultivation (million hectares), By Region

**Panel A: Soybeans, irrigated (left) and rainfed (right)**

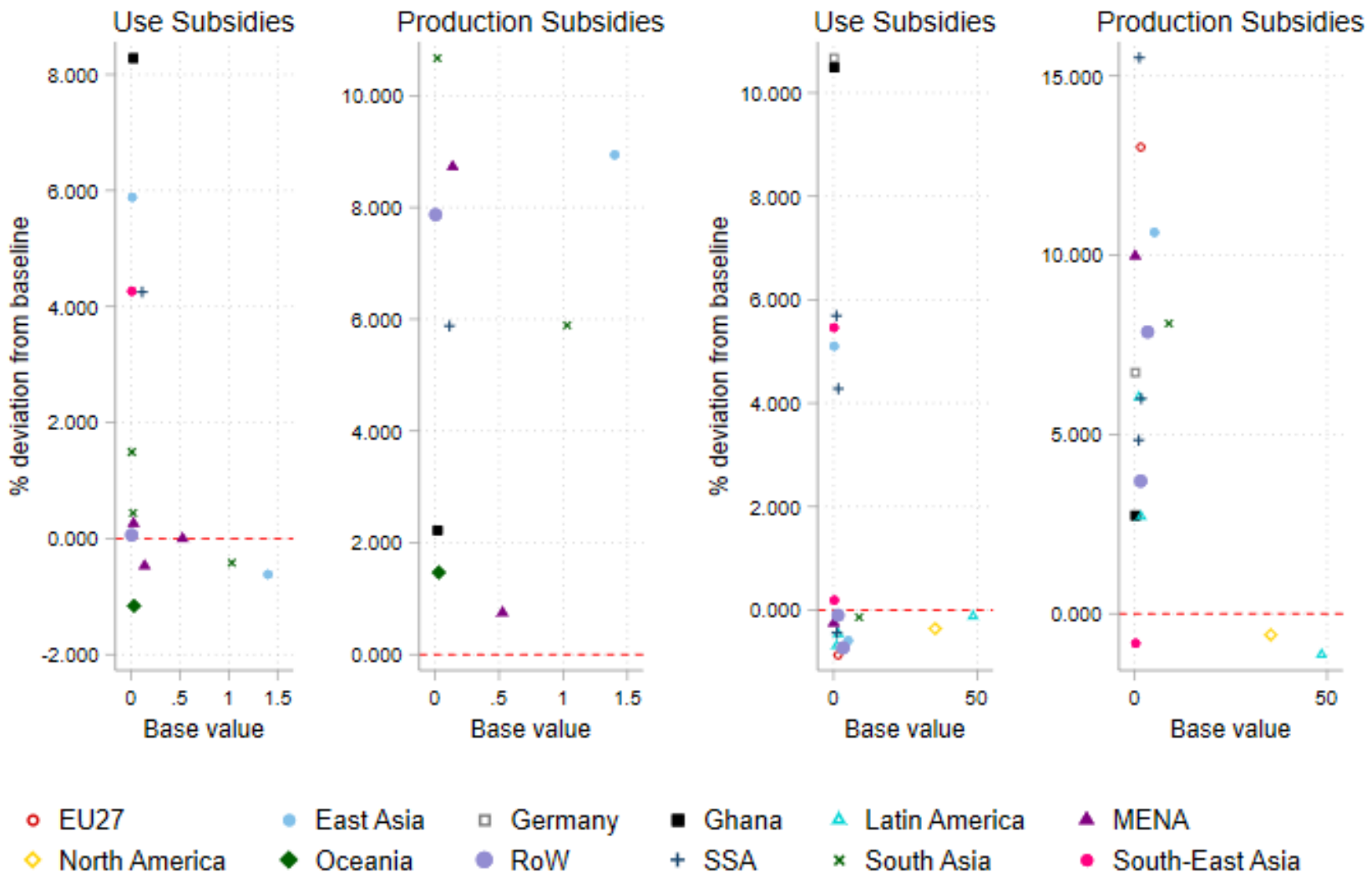

**Panel B: Groundnuts, irrigated (left) and rainfed (right)**

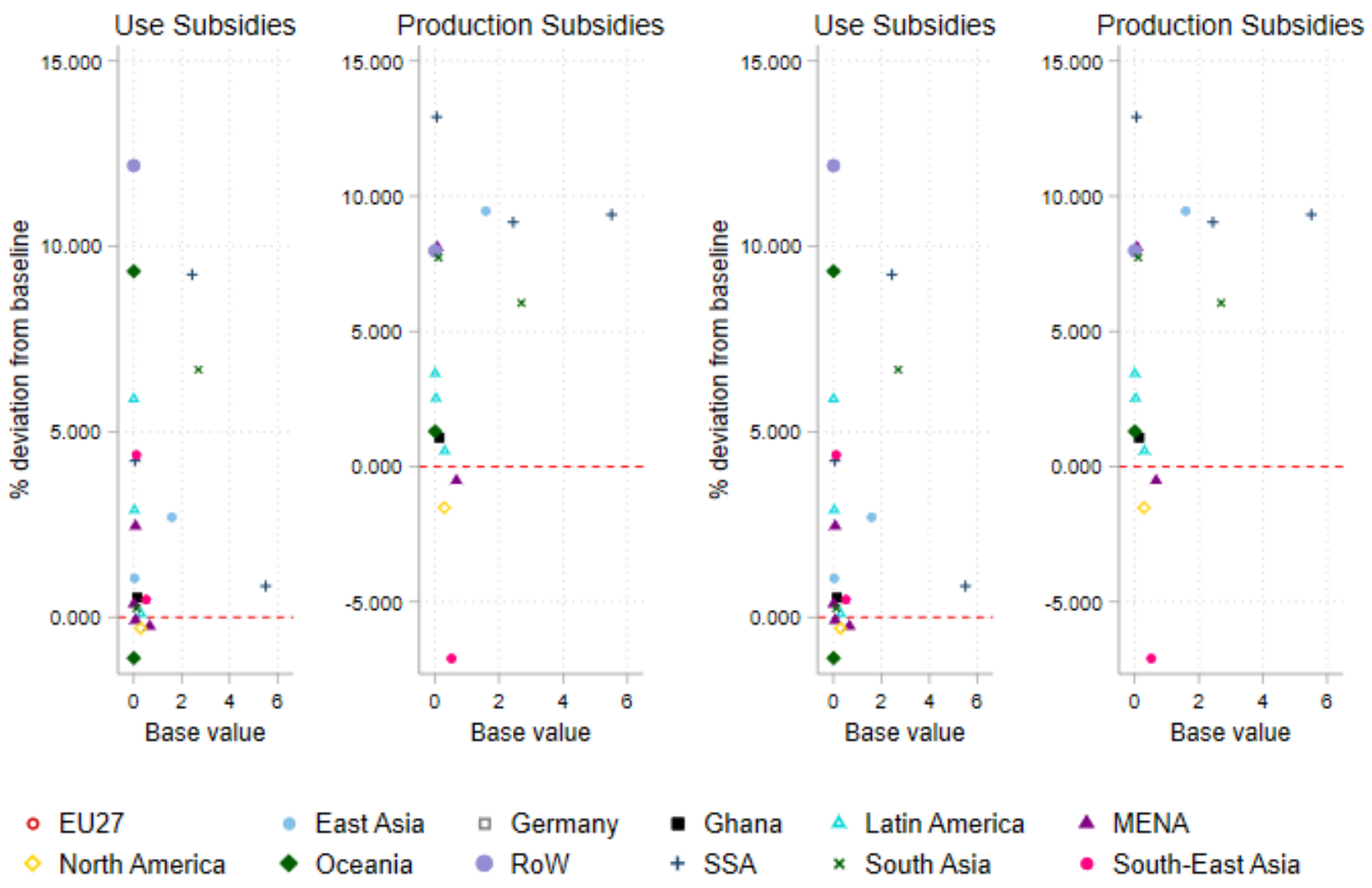

**Panel C: Other Leguminosae, irrigated (left) and rainfed (right)**

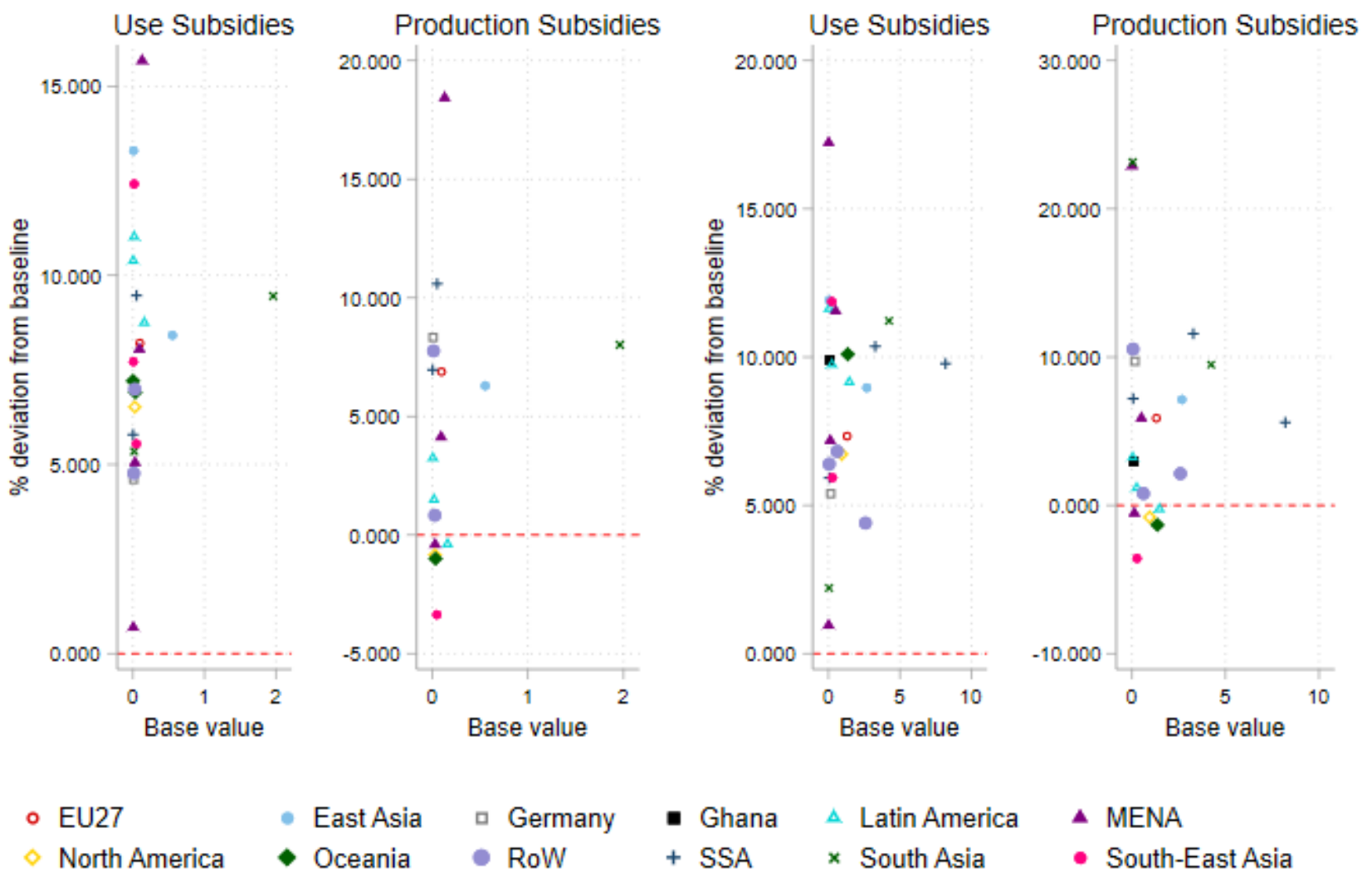

Notes: Own elaboration. The figure shows the results for scenarios with 20% subsidies on all legumes.

## 4.4. Domestic Demand

Household consumption of legumes increases in every scenario (see Figure 12 and Figure C5). Again, as expected, use subsidies are more effective than production subsidies in increasing legume household demand, since production subsidies promote both intermediate and final use as well as exports. At the global level (see Table B1 in Appendix B), soybean household consumption increases between 0.7% and 2.8% when subsidizing final use, and between 0.3% and 1.2% when subsidizing production. Groundnut consumption increases between 7.3% and 17.1% under use subsidies, and by between 8.5% and 9.9% under production subsidies. The consumption of other Leguminosae by households, in turn, increases by between 9.4% and 21.1% when subsidizing use, and by 4.2% and 9.8% when subsidizing production.

Variations in Germany and Ghana are larger, although the base values are rather low for both economies. In Germany, the increase in household consumption of soybeans under use subsidies is 5.9% and 12.8%, whilst under production subsidies it is between 4% and 14.6%. Meanwhile, the increase in household consumption of other Leguminosae under use subsidies is between 9.1% and 20.2%, whilst under production subsidies it is between 3% and 9.2%. In Ghana, the increase in household consumption of soybeans under use subsidies is between 4.8% and 10.5%, whilst between 1.3% and 9% under production subsidies. Meanwhile, the increase in household consumption of other Leguminosae under use subsidies is between 8.1% and 18.1%, whilst between 2% and 9.8% under production subsidies. Ghana's households do not consume groundnut.

When subsidizing legume final consumption, its intermediate demand decreases, particularly for other Leguminosae (see Figure 13 and Figure C6). At the global level (see Table B1 in Appendix B), such reduction ranges between 1.6% and 3.4%. As shown in Figure C7, when subsidizing legume use, all economic sectors reduce their demand for legumes, except for the aggregate grain crops sector, which includes those sectors that produce legumes. The intermediate demand for legumes in Ghana is negligible, except for groundnuts. Production subsidies increase the intermediate demand for groundnuts in Ghana by between 1.4% and 2.7%. In Germany, the intermediate demand for other Leguminosae under use subsidies is somewhat larger, ranging between 2.2% and 4.8%. As shown in Figure C7, this is mainly related to reduced use as animal feed. When subsidizing legume production, all sectors increase their demand for legumes, particularly eggs and livestock and meat products. Globally, the intermediate demand increases for Other Leguminosae under production subsidies range between 2.6% and 6.2%, between 0.6% and 3.2% for soybeans, and by 1.7% and 10.5% for groundnut.

Subsidizing production of other Leguminosae in consumer regions such as East Asia HI and South Asia LMI increases their domestic demand substantially (see Figures 12 and 13). In the case of soybean, East Asia HI is a net importer and final demand is less responsive to domestic production. North America and Latin America, other legume consumer regions, show a minor increase in legume domestic demand.

## Figure 12. Household Demand, By Region

### Panel A: Soybeans

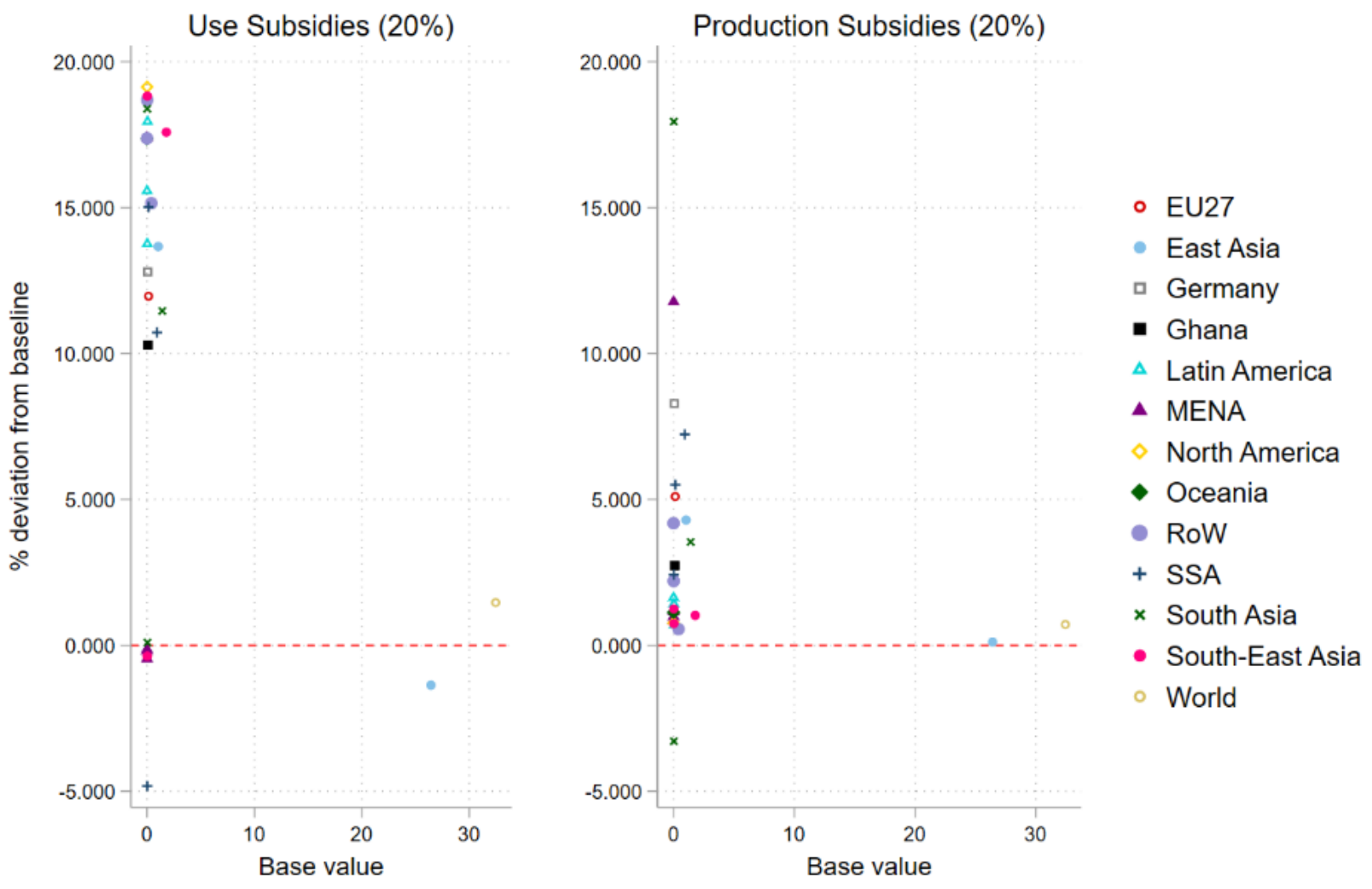

### Panel B: Groundnuts

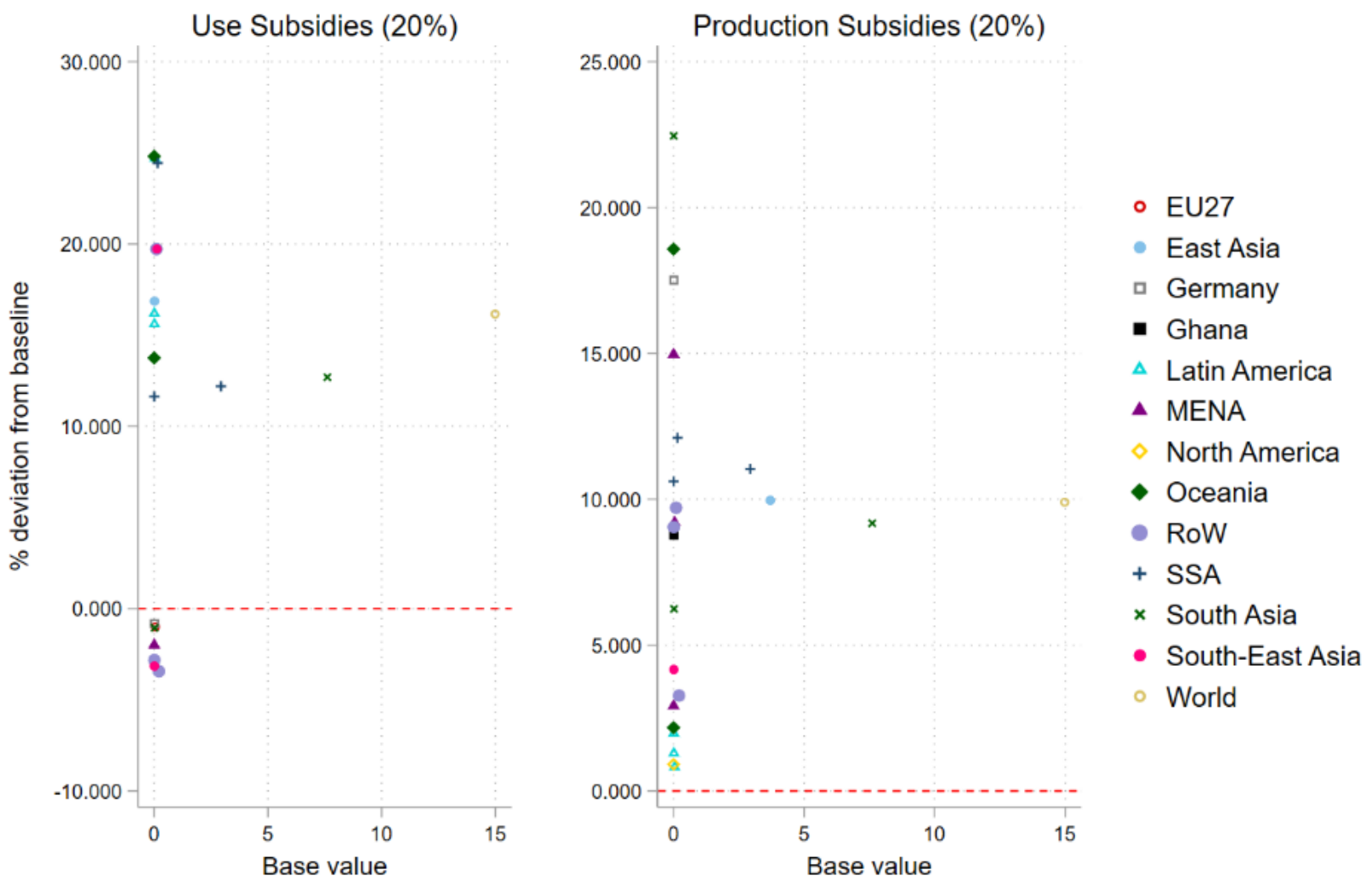

### Panel C: Other Leguminosae

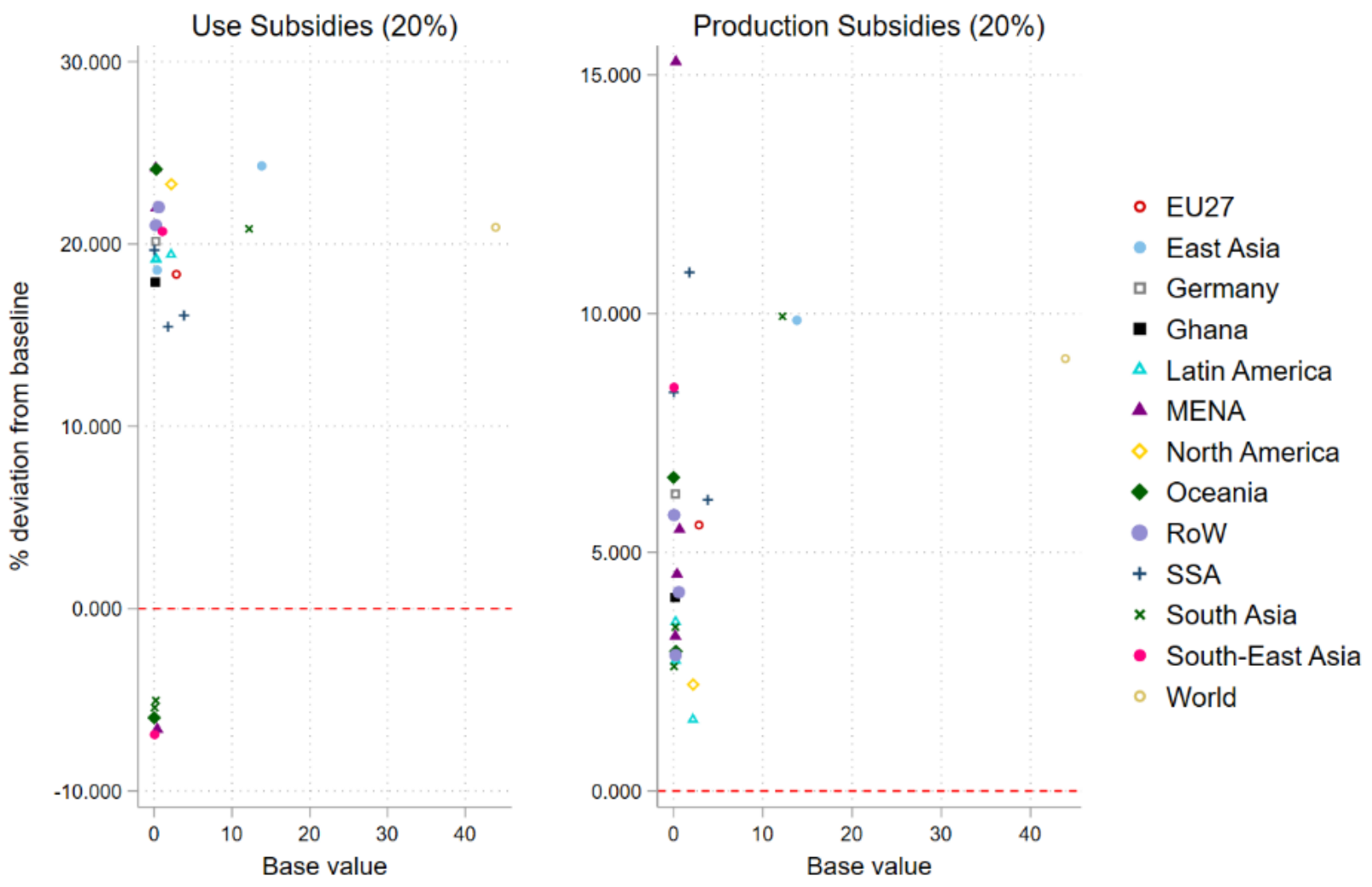

Notes: Own elaboration. The figure shows the results for scenarios with 20% subsidies on all legumes.

## Figure 13. Intermediate Demand, By Region

### Panel A: Soybeans

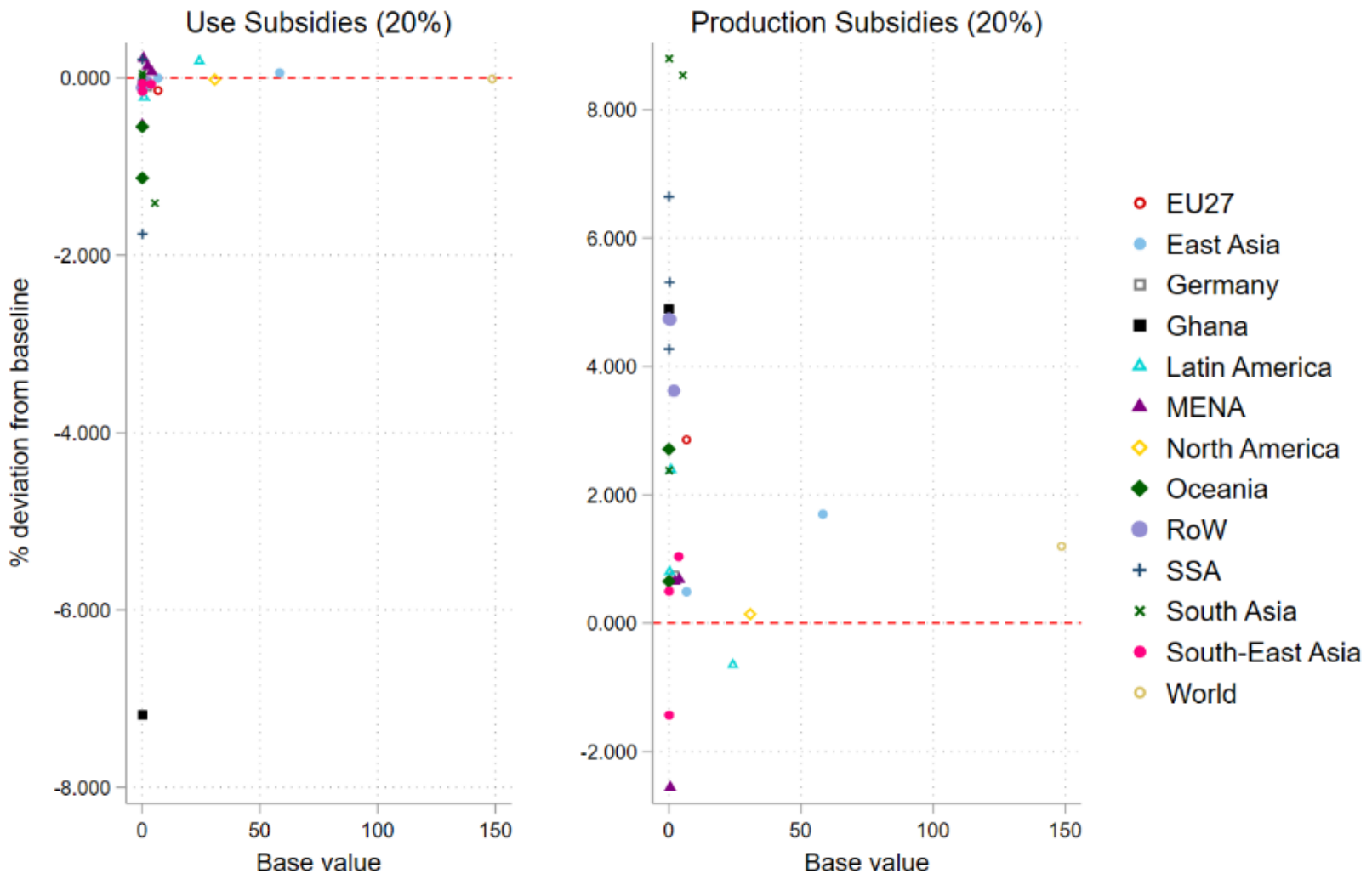

### Panel B: Groundnuts

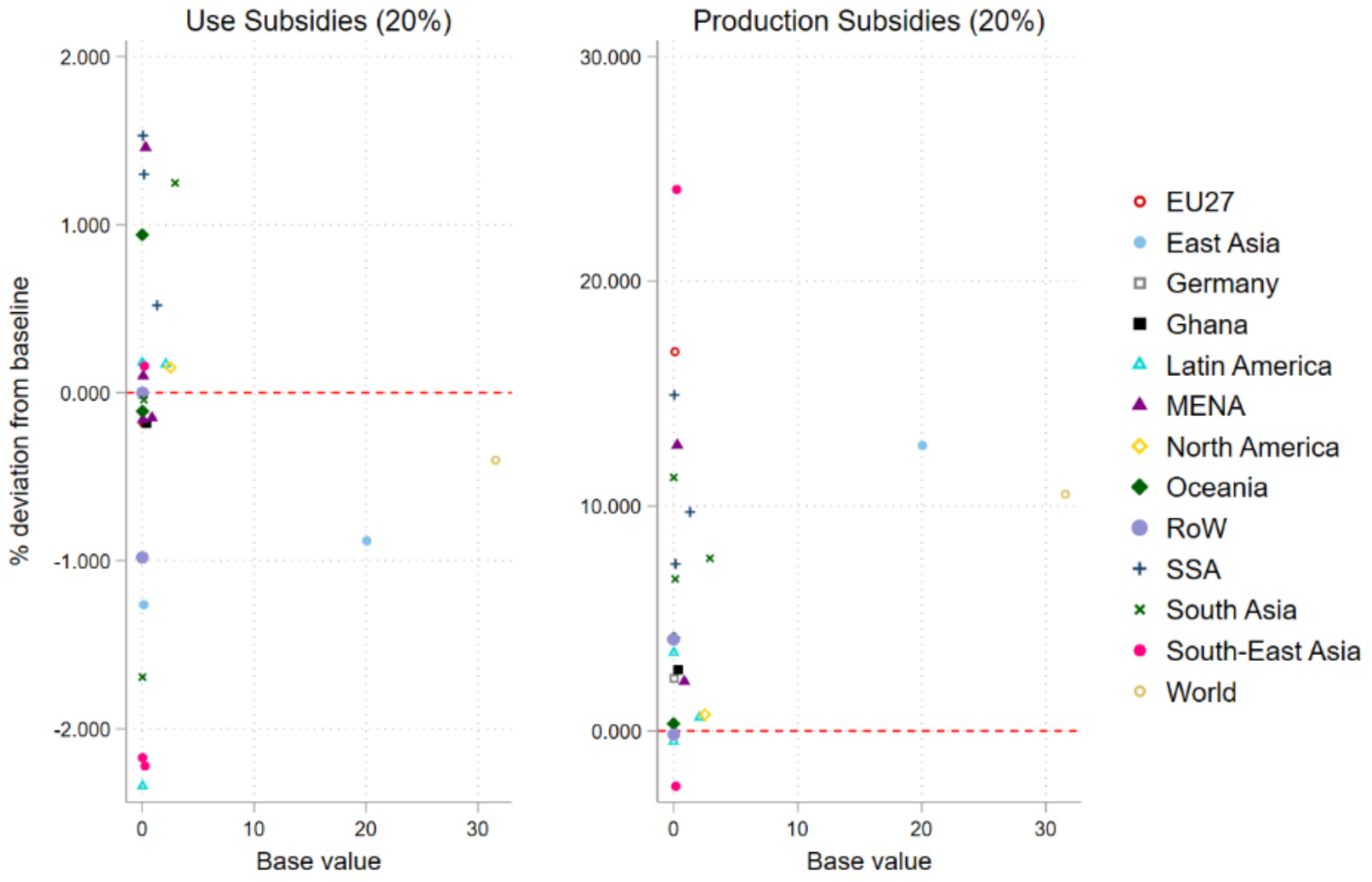

### Panel C: Other Leguminosae

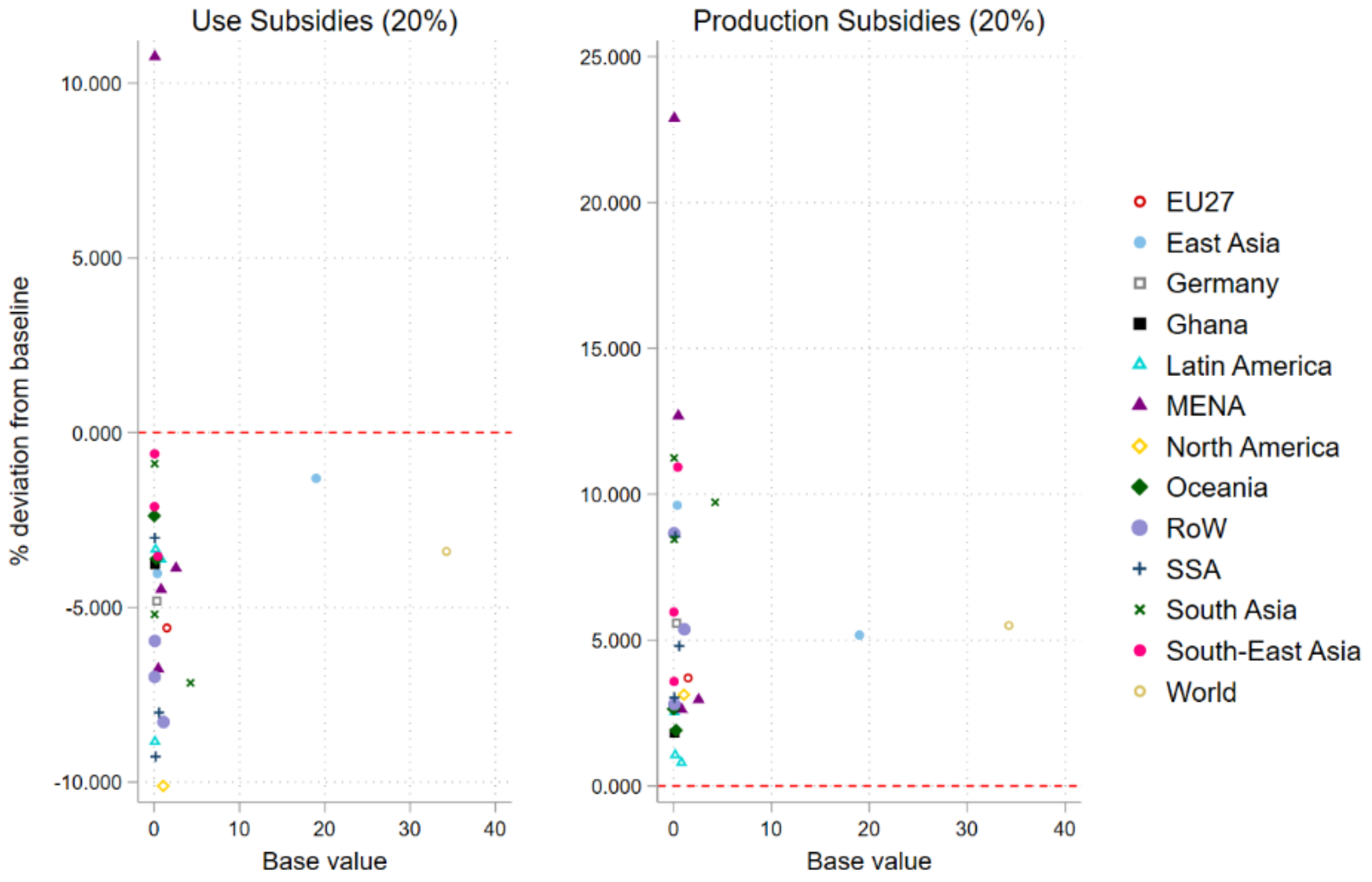

Notes: Own elaboration. The figure shows the results for scenarios with 20% subsidies on all legumes.

## 4.5 International Trade

Subsidies also alter international trade, with considerable heterogeneity of impacts (see Figures 14, D8, and D9). Use subsidies tend to increase legume imports, as expected, but also exports in some cases, driven by import demand increases in other countries. Production subsidies tend to increase both exports and imports in most cases. At the global level (see Table B1 in Appendix B), both use and production subsidies tend to increase the international trade of legumes, particularly for other Leguminosae and groundnuts. For soybeans, the variations are rather small, and production subsidies tend to decrease their international trade.

Subsidies in East Asia HI, a major consumer and producer region, tend to reduce international trade, particularly under production subsidies. In the case of major soybean net exporters such as Latin America UMI and North America, production subsidies reduce exports, while use subsidies tend to slightly increase them. Meanwhile, in the case of other Leguminosae and groundnuts, in which these regions are self-sufficient, production subsidies increase imports.

Germany is a net importer of legumes, but base export and import values are negligible, except for imports of soybeans, which are somewhat larger. Their imports almost do not change under use subsidies because they are only demanded for intermediate use. In Ghana, both exports and imports of legumes are very low. In line with the output results already discussed, there is a reduction in groundnut exports and an increase in imports in Ghana under production subsidies.

## Figure 14. Exports and Imports, By Region

### Panel A: Soybeans, exports (left) and imports (right)

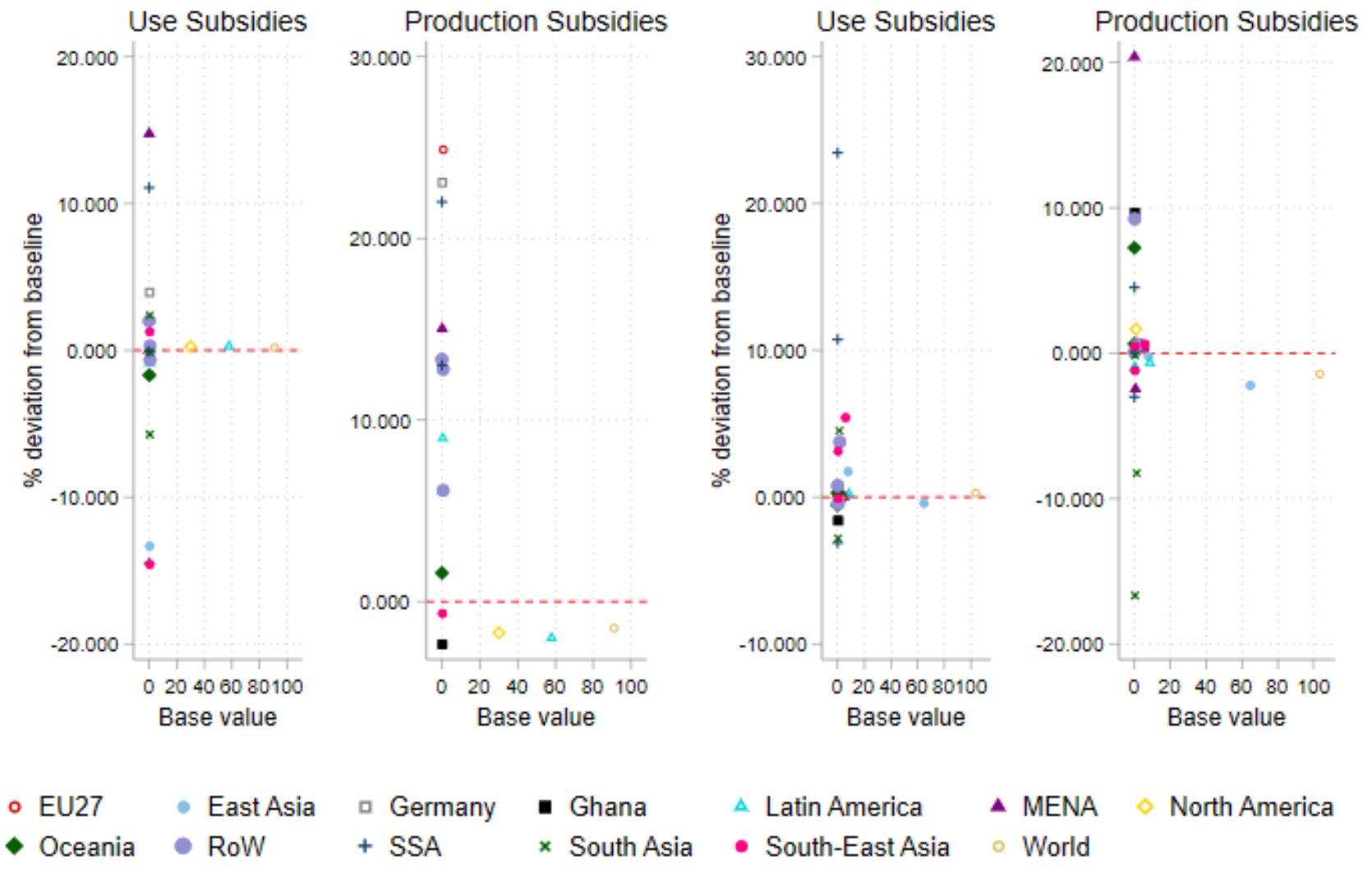

### Panel B: Groundnuts, exports (left) and imports (right)

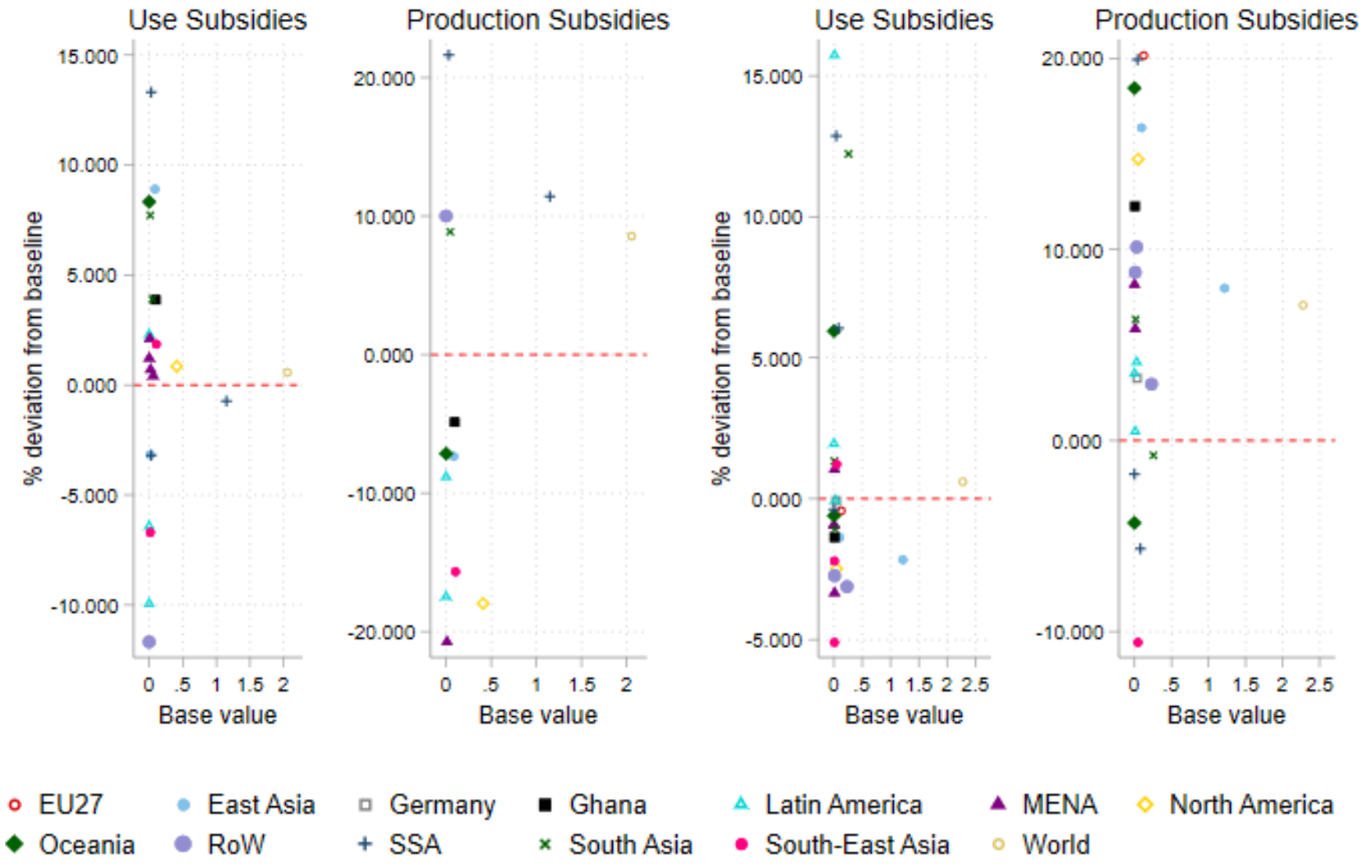

### Panel C: Other Leguminosae, exports (left) and imports (right)

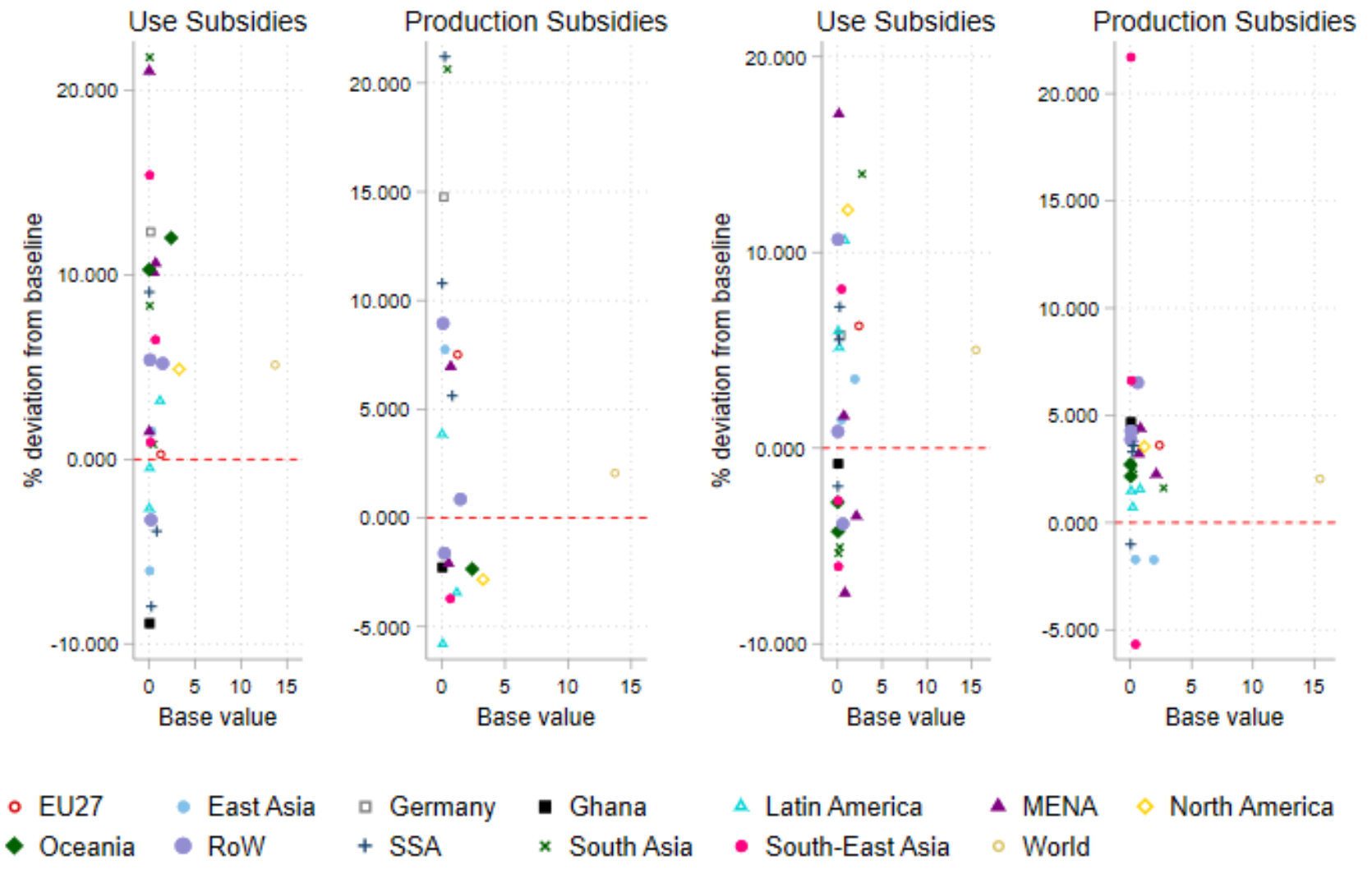

Notes: Own elaboration. The figure shows the results for scenarios with 20% subsidies on all legumes.

### 4.6 Nutritional Intake

In line with household demand changes, nutritional intake from legumes increases, but more under use subsidies, and particularly for other Leguminosae (see Figure C10 and Table C1 in Appendix C). At the global level, nutrient intake from legumes increases, as does total nutrient intake, suggesting that increased consumption of legumes does not displace the consumption of other foods with higher content of these nutrients (see Table B1 in Appendix B). This effect is more pronounced in the case of protein, which highlights the high content of this nutrient in legumes. In Germany, subsidizing legume production could increase all macronutrient intake (slightly displacing the intake of nutrients from livestock products under use subsidies), without affecting overall nutritional intake. In Ghana, production subsidies contribute to considerable increases in macronutrient intake from legumes.

## Figure 15. Total Nutrient Intake, By Region

### Panel A: Calories

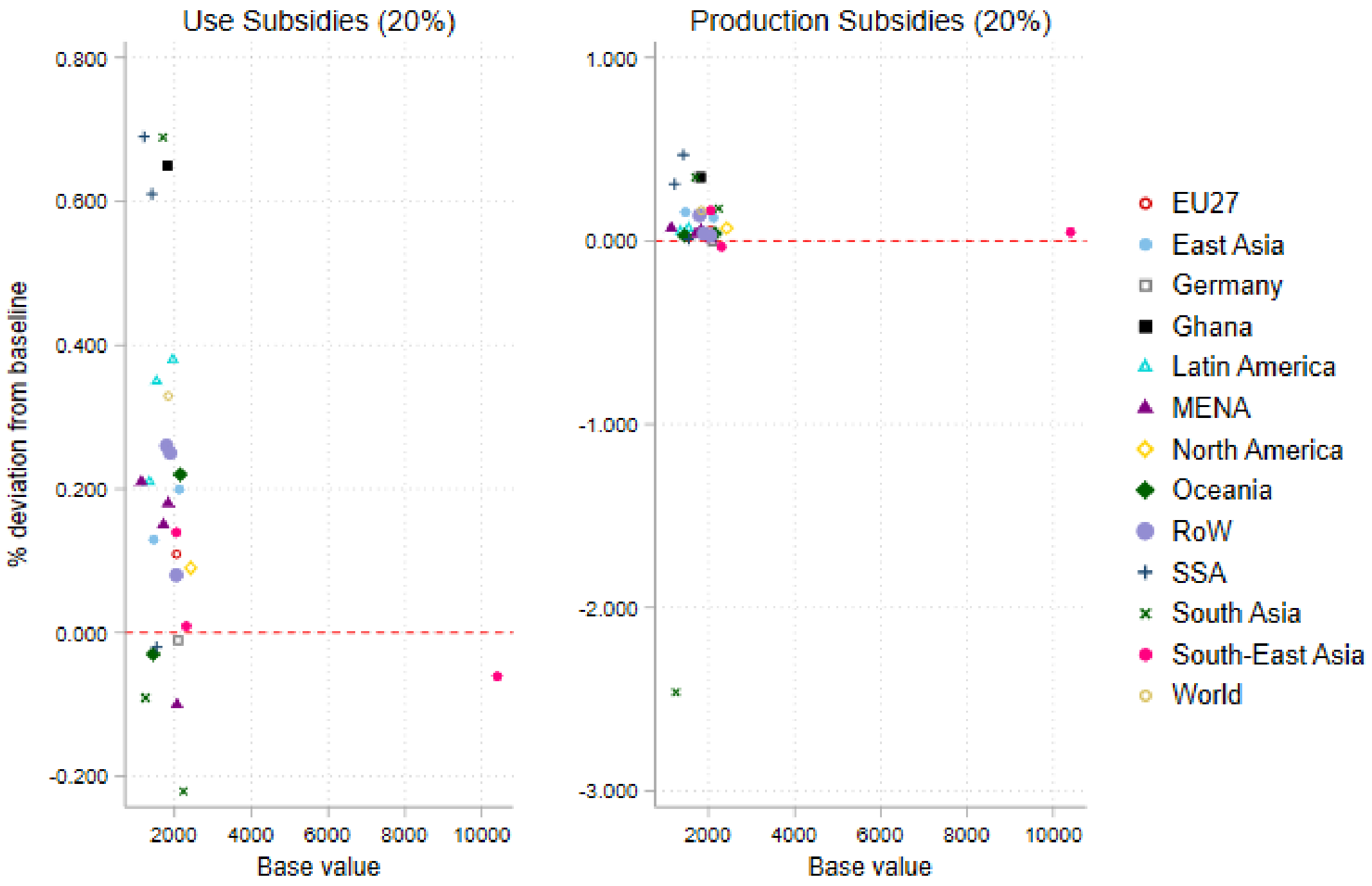

### Panel B: Protein

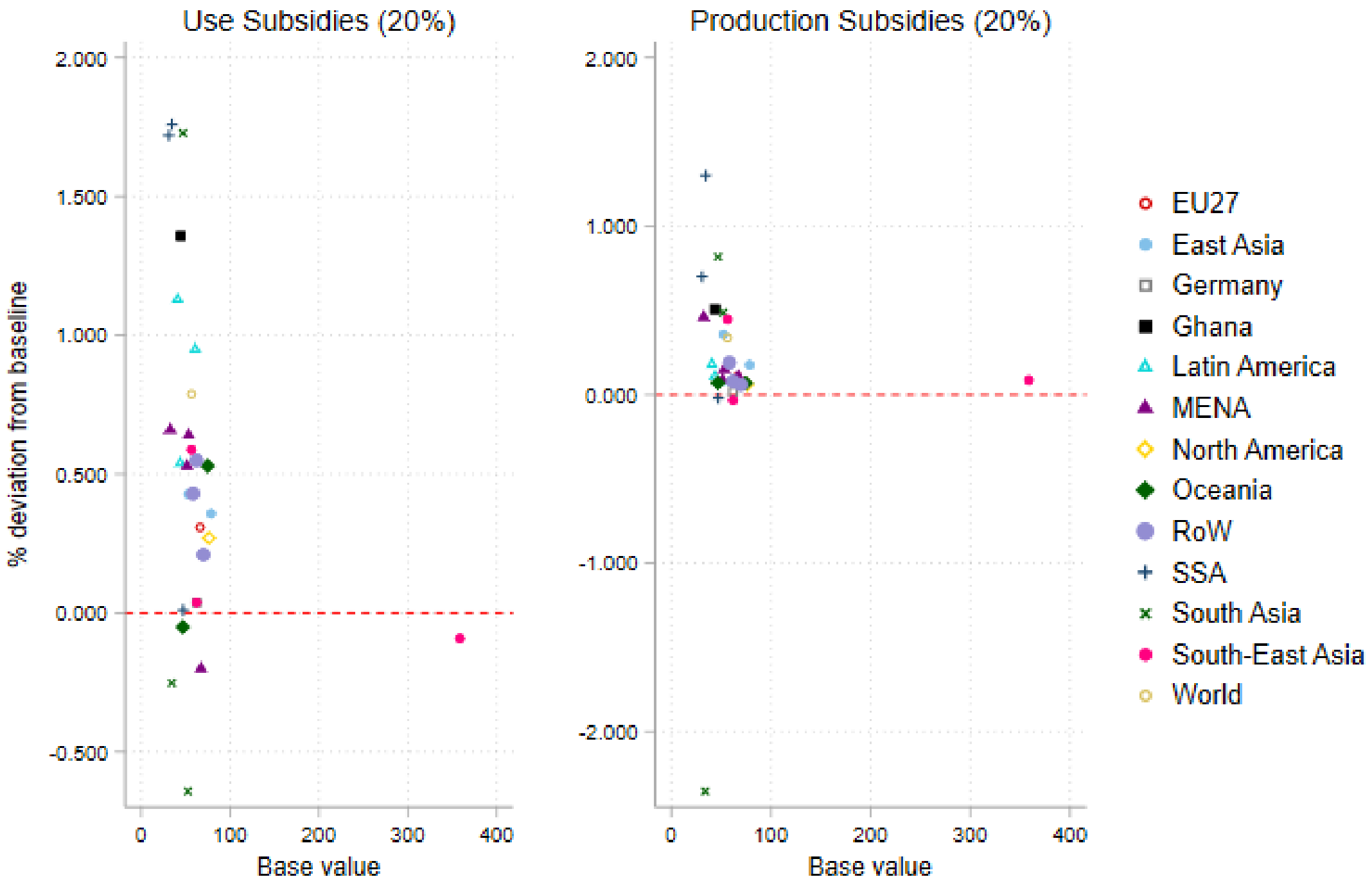

Notes: Own elaboration. The figure shows the results for scenarios with 20% subsidies on all legumes.

### 4.7 Welfare

Finally, welfare, as measured by utility changes, slightly increases in every scenario (see Figures 16 and D12). At the global level (see Table B3), only agricultural households show some welfare improvements, being similar under use or production subsidies and ranging between 0.1% and 0.4%. In Germany, there are virtually no welfare effects. In Ghana, all households are slightly better off, and the effects are very similar in both agricultural and non-agricultural households, and under either use or production subsidies: Welfare increases range between 0.1% and 3%. In Germany, there are virtually no effects.

**Figure 16. Welfare, By Region**

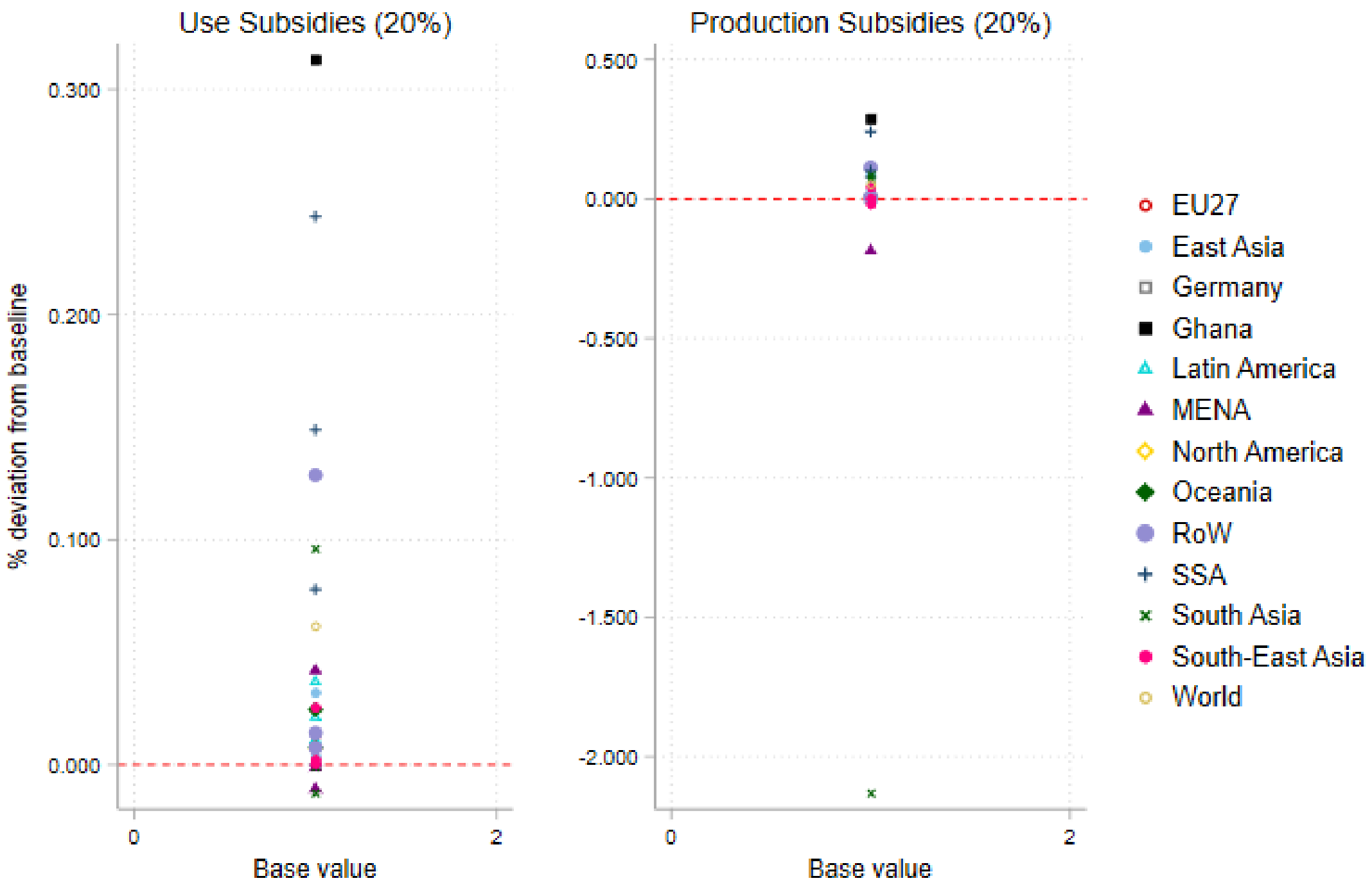

Notes: Own elaboration. The figure shows the results for scenarios with 20% subsidies on all legumes.

## 5. Discussion

Our findings show that production can effectively increase legume consumption and production. But implementing these subsidies globally faces feasibility barriers. Use subsidy targeting domestic final demand require a well-established tax system at the point of sale that allows for tax reduction/exemption for VAT or similar. This is especially challenging in the Global South, where a high share of legumes is marketed in more informal settings. In countries where existing agricultural subsidies already target specific crops, subsidizing legumes should be straightforward. Such policies are well established, for instance in Europe. However, adoption of such subsidies in richer countries could only exacerbate market asymmetries, since these more advanced economies have far more fiscal space to subsidize their farmers than developing countries. Use subsidies will drive up world market prices and thus increase costs to the consumer in countries where they are not introduced. Especially in countries specialized in the production of legumes for export, increasing world market prices could mean more serious, substantial environmental damage, mainly related to land use changes and deforestation (Pendrill et al., 2020). Production subsidies in the richer countries would have the opposite effects but could displace domestic production where they are not introduced, increasing dependency on imports.

Even more restrictive is the conflict between production subsidies and the legal framework of the World Trade Organization (WTO). Under the Agreement on Agriculture (AoA), incentives linked to production volume are classified under the "Amber Box," as they are explicitly considered to distort international trade and are subject to reduction commitments. In Europe, direct support under the CAP for protein-rich legumes falls under the "Amber Box", but the EU has a historical Aggregate Measure of Support (AMS) ceiling that is so high that it can absorb these payments without violating its international commitments. On the contrary, most developing countries do not have those historical margins. One alternative could be to invoke the *de minimus* rule (Article 6.4 of the AoA), which allows member countries to exclude small amounts of trade-distorting domestic support from their (AMS) calculations. Particularly, product-specific support is allowed up to a certain threshold, being a percentage of the total value of production of that specific crop: 5% in developed countries, 10% in developing countries, and 8.5% in China. Ultimately, to avoid litigation, any support mechanism could be designed as decoupled agri-environmental payments toward the WTO's "*Green Box*" - compensating producers for ecosystem services such as crop rotation rather than for crop yield - although it dilutes the direct stimulus on supply that a pure subsidy would seek to achieve.

# 6. Conclusions

The transition towards more sustainable food systems and nutritious diets requires carefully calibrated policy interventions; however, the empirical literature is still debating which mechanisms achieve these objectives whilst minimizing distortions in agricultural markets. This paper contributes to that literature by modelling, in a comparative manner, the systemic effects of two alternative approaches: final use subsidies versus production subsidies.

Previous studies based their analysis on highly aggregated agricultural products, focused primarily on Europe, employed partial-equilibrium analyses, and used highly aggregated product lists. We contribute to the literature by estimating the likely impacts of global subsidies on the production and consumption of legumes by using CGEBox, an open-source and open-access Computable General Equilibrium (CGE) modelling platform, with integrated accounting for land use, land cover, and energy. In addition to using a global model, we conducted the study with a high level of disaggregation of agricultural and food products and activities. This allows us to more accurately analyze the economic, environmental, and nutritional consequences of these policies. The analysis specifically focused on Ghana (as one example of an LMI country) and Germany (as one example of a HI country), two countries with different agroecological and socioeconomic conditions facing multiple food system issues. In terms of many conditions related to agricultural structures, food supply chains, dietary patterns, and nutrition and health outcomes, Germany is somewhat typical for other countries in Europe and North America, whereas Ghana is somewhat typical for Africa.

We found that use subsidies are more effective for reducing legume final demand prices and stimulating household intake of macronutrients, particularly proteins, without displacing the consumption of other foods. In contrast, production subsidies are more effective at boosting domestic supply, leading to an expansion in arable land used for legumes, which grows slightly at the expense of land used for livestock. Subsidizing final consumption leads to a decrease in intermediate demand for legumes, primarily affecting their use as animal feed. In contrast, production subsidies increase intermediate demand across all sectors, significantly benefiting egg, meat, and livestock production. The implementation of subsidies tends to increase international trade in legumes. The impact of these subsidies reveals profound regional disparities, influenced by land use and patterns of production and consumption. In producing regions such as Latin America and North America, where agricultural land use is high and competition for land from livestock is severe, we find that subsidies cannot effectively increase legume production. In contrast, in regions with a low initial proportion of land allocated to these crops, such as East Asia, subsidies prove highly effective in expanding supply and stimulating domestic demand. In this way, our research contributes to the previous literature with new evidence on which policy interventions could promote legume production and consumption and therefore the transition towards more sustainable food systems.

The findings of this study emphasize again that the choice of policy instrument must reflect its objectives. If the priority is to foster a healthier diet, policymakers should opt for use subsidies, which reduce consumer prices of the targeted products, increase their demand, and decrease consumption of less healthy alternatives. Conversely, if the main objective is to foster production activities with non-market benefits, production subsidies are preferable. To avoid loss of natural vegetation, subsidies for other crops could be reduced.

Furthermore, our findings highlight structural constraints on the implementation of subsidies for groups of production: the simultaneous promotion of various pulses exacerbates competition for arable land and limits the expansion of individual crops such as soybeans, particularly in land-constrained economies such as Germany. Finally, our study suggests no relevant welfare losses globally and even small welfare gains in some developing countries such as Ghana.

# References

Aguiar, A., Baldos, U., Chepeliev, M., Corong, E., & Simonato, T. (2025). The Global Trade Analysis Project (GTAP) Data Base: Version 12. *Journal of Global Economic Analysis*, *10*(2), 1-45.

Alandia G, Pulvento C, Sellami MH, Hoidal N, Anemone T, Nigussie E, Agüero JJ, Lavini A, Jacobsen SE (2020) Grain legumes may enhance high-quality food production in Europe. In: Hirich A, Choukr-Allah R, Ragab R (eds) Emerging research in alternative crops. Springer International Publishing, Cham, pp 25–53. https://doi.org/10.1007/978-3-319-90472-6_2

Akibode, C. S., & Maredia, M. K. (2012). Global and regional trends in production, trade and consumption of food legume crops (Staff Paper No. 2012-10). Michigan State University, Department of Agricultural, Food, and Resource Economics. https://doi.org/10.22004/ag.econ.136293

Balázs, B.; Kelemen, E.; Centofanti, T.; Vasconcelos, M.W.; Iannetta, P.P.M. (2021) Policy Interventions Promoting Sustainable Food- and Feed-Systems: A Delphi Study of Legume Production and Consumption. Sustainability 2021, 13, 7597. https://doi.org/10.3390/ su13147597.

Britz, W. (2022a). Disaggregating agro-food sectors in the GTAP data base. *Journal of Global Economic Analysis*, *7*(1), 44-75 Britz, W., & van der Mensbrugghe, D. (2018). CGEBox: A flexible, modular and extendable framework for CGE analysis in GAMS. *Journal of Global Economic Analysis*, *3*(2), 106-177.

Britz, W., (2022b). CGEBox –a flexible and modular toolkit for CGE modelling with a GUI. Version 2026, available at: https://www.ilr1.uni-bonn.de/en/research/research-groups/economic-modeling-of-agricultural-systems/media/cgebox_gui.pdf/@@download/file/cgebox_GUI.pdf

Britz, W. (2022). Estimating a global MAIDADS demand system considering demography, climate and norms. *Bio-Based and Applied Economics*, *10*(3), 219–238. https://doi.org/10.36253/bae-10488

Britz, W., & Roson, R. (2019). G-RDEM: A GTAP-Based Recursive Dynamic CGE Model for Long-Term Baseline Generation and Analysis. *Journal of Global Economic Analysis*, *4*(1), 50-96. https://doi.org/10.21642/JGEA.040103AF

Britz, W. & van der Mensbrugghe, D., 2018. CGEBox: a flexible, modular and extendable framework for CGE analysis in GAMS. J. Glob. Econ. Anal. 3 (2), 106–177.

Bues, A., S. Preiβel, M. Reckling, P. Zander, T. Kuhlman, K. Topp, C. Watson, K. Lindström, F.L. Stoddard & D. Murphy-Bokern, (2013). The environmental role of legumes in the new Common Agricultural Policy. European Parliament, Brussels, Policy Department B: Structural and Cohesion Policies, document IP/B/AGRI/IC/2012-067.

Campbell, B.M., Beare, D.J., Bennett, E.M., Hall-Spencer, J.M., Ingram, J.S., Jaramillo, F., Ortiz, R., Ramankutty, N., Sayer, J.A., Shindell, D., (2017). Agriculture production as a major driver of the Earth system exceeding planetary boundaries. Ecol. Soc. 22.

Chepeliev, M. (2020). Development of the non-CO2 GHG emissions database for the GTAP data base version 10A (GTAP Research Memorandum No. 32). Global Trade Analysis Project (GTAP), Purdue University. http://dx.doi.org/10.21642/gtap.rm32

Chepeliev, M. (2022). A Revised CO2 Emissions Database for GTAP

Curran, J., (2012). The nutritional value and health benefits of pulses in relation to obesity, diabetes, heart disease and cancer. Br. J. Nutr. 108, S1–S2.

Drinkwater L.E., Wagoner P. and Sarrantonio M. (1998) 'Legume-based cropping systems have reduced carbon and nitrogen losses', Nature, 396: 262–65.

European Commission (2020). Farm to Fork Strategy: for a Fair, Healthy and Environmentally-Friendly Food System.

FAO (2018). The Future of food and agriculture – Alternative pathways to 2050.

FAO (2025). *Food balance sheets 2010–2023*, FAOSTAT Analytical Brief No. 112. Food and Agriculture Organization of the United Nations. https://doi.org/10.4060/cd7162en

Fontagné, L., Guimbard, H., & Orefice, G. (2019). Product-level trade elasticities, CEPI working paper No. 2019-17, https://hal.archives-ouvertes.fr/hal-02444897/document

Heinrichs, J., Jouan, J., Pahmeyer, C., & Britz, W. (2021). Integrated assessment of legume production challenged by European policy interaction: A case-study approach from French and German dairy farms. Q Open, 2021, 1, 1–19, DOI: 10.1093/qopen/qoaa011.

Helming, J. & T. Kuhlman (2015). "Impact of a combined meat tax and vegetable protein subsidy on European agriculture," 148th Seminar, November 30-December 1, 2015, The Hague, The Netherlands 229271, European Association of Agricultural Economists.

Jensen E.S., Peoples M.B., Boddey R.M., Gresshoff P.M., Hauggaard-Nielsen H., Alves B.J.R. and Morrison M.J. (2012) 'Legumes for mitigation of climate change and the provision of feedstock for biofuels and biorefineries. A review', Agronomy for Sustainable Development, 32: 329–64.

Jouan J., A. Ridier & M. Carof, (2020). "Legume production and use in feed: Analysis of levers to improve protein self-sufficiency from foresight scenarios," Journal of Cleaner Production, Volume 274, 2020, 123085, ISSN 0959-6526, https://doi.org/10.1016/j.jclepro.2020.123085.

Kaimila, Y., Olotu, O. A., Clegg, M. E., Jackson, K. G., & Lovegrove, J. A. (2025). Pulse and legume consumption is associated with a more optimal nutrient intake and a higher EAT-Lancet index in a representative UK population. *European Journal of Nutrition*, *64*(3), 139.

Kebede E. (2020) Grain legumes production and productivity in Ethiopian smallholder agricultural system, contribution to livelihoods and the way forward. Cogent Food Agric 6:1722353. https://doi.org/10.1080/23311932.2020.1722353

Kuhlman, T., Helming, J., & Linderhof, V. (2017). Optimizing legume cropping: The policy questions. En D. Murphy-Bokern, F. Stoddard, y C. Watson (Eds.), Legumes in cropping systems (pp. 226–243). CABI.

Lee, H.-L., Hertel, T. W., Rose, S., & Avetisyan, M. (2009). An integrated global land use data base for CGE analysis of climate policy options. En T. W. Hertel, S. Rose, & R. S. J. Tol (Eds.), *Economic Analysis of Land Use in Global Climate Policy* (pp. 75–98). Routledge.

McDougall, R., & Golub, A. (2007). *GTAP-E: A revised energy-environmental version of the GTAP model* (GTAP Research Memorandum No. 15). Center for Global Trade Analysis, Purdue University. https://doi.org/10.21642/GTAP.RM15.

McDonald, S. and Thierfelder, K. (2014). Globe v2: A SAM Based Global CGE Model using GTAP Data

Michalis, E., Ragkos, A. & Malesios, C. Market-based or policy strategies to increase the cultivation of legumes for livestock? Effects on the cropping pattern and economic performance. Agric Econ 13, 3 (2025). https://doi.org/10.1186/s40100-024-00339-1

Notz I, Topp CF, Schuler J, Alves S, Gallardo LA, Dauber J, Haase T, Hargreaves PR, Hennessy M, Iantcheva A, Jeanneret P, Kay S, Recknagel J, Rittler L, Vasiljević M, Watson CA, Reckling M (2023) Transition to legume-supported farming in Europe through redesigning cropping systems. Agron Sustain Dev 43:12.

Pendrill, F., Persson, F. & Kastner, T. (2020). 'Deforestation risk embodied in production and consumption of agricultural and forestry commodities 2005-2017'. Chalmers University of Technology, Senckenberg Society for Nature Research, SEI, and Ceres Inc. DOI: 10.5281/zenodo.4250532.

Peoples M.B., Brockwell J., Herridge D.F., Rochester I.J., Alves B.J.R., Urquiaga S., Boddey R.M., Dakora F.D., Bhattarai S., Maskey S.L., Sampet C., Rerkasem B., Khan D.F., Hauggaard-Nielsen H.

and Jensen E.S. (2009) ‘The contributions of nitrogen-fixing crop legumes to the productivity of agricultural systems’, Symbiosis, 48: 1–17.

Peters, J. C. (2016). GTAP-E-Power: An Electricity-detailed Economy-wide Model. *Journal of Global Economic Analysis*, *1*(2), 156-187. https://doi.org/10.21642/JGEA.010204AF

Roson, R., & Britz, W. (2021). Simulating long run structural change with a dynamic general equilibrium model. *International Journal of Computational Economics and Econometrics*, *11*(4), 368–405. doi.org

Springmann, M., Freund, F. Options for reforming agricultural subsidies from health, climate, and economic perspectives. Nat Commun 13, 82 (2022). https://doi.org/10.1038/s41467-021-27645-2.

Steinfeld, H., Gerber, P., Wassenaar, T., Castel, V., Rosales, M., & De Haan, C. (2006). Livestock's long shadow. *FaO, Rome, 229*

Van der Mensbrugghe, D. (2018). The standard GTAP model in GAMS, version 7. *Journal of Global Economic Analysis, 3*(1), 1-83

van der Mensbrugghe, D., & Peters, J. C. (2020). *Volume preserving CES and CET formulations* (GTAP Working Paper No. 87). Center for Global Trade Analysis, Department of Agricultural Economics, Purdue University. https://www.gtap.agecon.purdue.edu/resources/res_display.asp?RecordID=6160

van Loon, M. & A. Seyyedmajid & Pronk, A. & Fodor, N. & I., Viorel & Kryvoshein, O. & Kryvobok, O. & Marrou, H. & Mihail, R. & Minguez, M. & Pulina, A. & Reckling, M. & Rittler, L. & Roggero, P. & Stoddard, F. & Topp, F. & van der Wel, J.& Watson, C. & Ittersum, M. (2023). Grain legume production in Europe for food, feed and meat substitution. Global Food Security. 39. 100723. 10.1016/j.gfs.2023.100723.

Willett, W., Rockstrom, J., Loken, B., Springmann, M., Lang, T., Vermeulen, S., Garnett, T., Tilman, D., DeClerck, F., Wood, A., (2019). Food in the Anthropocene: the EAT–Lancet Commission on healthy diets from sustainable food systems. Lancet 393, 447–492.

Williams, M., Stout, J., Roth, B., Cass, S., Papa, V., and Rees, B. (2014). Environmental implications of legume cropping. Legume Futures Report 3.7.

Wilts, R., & Britz, W. (2024). Distribution matters: Long-term quantification of the Sustainable Development Goals with household detail for different socio-economic pathways. *Global Transitions*, *6*, 173-186

Zander, P., Amjath-Babu, T. S., Preissel, S., Reckling, M., Bues, A., Schläfke, N., Kuhlman, T., Bachinger, J., Uthes, S., Stoddard, F. L., Murphy-Bokern, D., & Watson, C. A. (2016). Grain legume decline and potential recovery in European agriculture: A review. Agronomy for Sustainable Development, 36(2), 26.

# Appendix A. Database Description

## Table A1. List of regions/nations in the model

| Region Name | Geographical Reference | Income Level (WB classification) | Composition |
|---|---|---|---|
| EU27, HI | EU27 | High-income | Austria, Belgium, Bulgaria, Croatia, Cyprus, Czech Republic, Denmark, Estonia, Finland, France, Greece, Hungary, Ireland, Italy, Latvia, Lithuania, Luxembourg, Malta, Netherlands, Poland, Portugal, Romania, Slovakia, Slovenia, Spain, Sweden |
| Germany | Germany | High-income | Germany |
| Ghana | Ghana | Lower-Middle-Income | Ghana |
| Latin America, LMI | Latin America | Lower-Middle-Income | Bolivia, Venezuela, Honduras, Haiti |
| Latin America, UMI | Latin America | Upper-Middle-Income | Mexico, Argentina, Brazil, Colombia, Ecuador, Paraguay, Peru, Rest of South America, Guatemala, Nicaragua, El Salvador, Dominican Republic, Jamaica |
| Latin America, HI | Latin America | High-income | Chile, Uruguay, Costa Rica, Panama, Rest of Central America, Puerto Rico, Trinidad and Tobago, Caribbean |
| MENA, LI | MENA | Low-Income | Syria, Rest of Western Asia |
| MENA, LMI | MENA | Lower-Middle-Income | Lebanon, Palestine, Egypt, Morocco, Tunisia |
| MENA, UMI | MENA | Upper-Middle-Income | Iran Islamic Republic of, Iraq, Jordan, Turkey, Algeria, Rest of North Africa |
| MENA, HI | MENA | High-income | Bahrain, Israel, Kuwait, Oman, Qatar, Saudi Arabia, United Arab Emirates |
| North America, HI | North America | High-income | Canada, United States of America, Rest of North America |
| Oceania, UMI | Oceania | Upper-Middle-Income | Rest of Oceania |
| Oceania, HI | Oceania | High-income | Australia, New Zealand |
| Rest of Eastern Asia, HI | Rest of Eastern Asia | Upper-Middle-Income | China, Mongolia, Rest of East Asia |
| Rest of Eastern Asia, UMI | Rest of Eastern Asia | High-income | Hong Kong, Japan, Korea, Taiwan, Brunei Darussalam |
| Rest of Southern Asia, LI | Rest of Southern Asia | Low-Income | Afghanistan |
| Rest of Southern Asia, LMI | Rest of Southern Asia | Lower-Middle-Income | Bangladesh, India, Nepal, Pakistan, Rest of South Asia |
| Rest of Southern Asia, UMI | Rest of Southern Asia | Upper-Middle-Income | Sri Lanka |
| RoW, LMI | RoW | Lower-Middle-Income | Kyrgyztan, Tajikistan, Uzbekistan |
| RoW, UMI | RoW | Upper-Middle-Income | Albania, Serbia, Belarus, Ukraine, Rest of Eastern Europe, Kazakhstan, Rest of Former Soviet Union, Armenia, Azerbaijan, Georgia, Rest of the World |
| RoW, HI | RoW | High-income | United Kingdom, Switzerland, Norway, Rest of EFTA, Russian Federation, Rest of Europe |
| SSA, LI | SSA | Low-Income | Burkina Faso, Mali, Niger, Nigeria, Rest of Western Africa, Central African Republic, Chad, Democratic Republic of Congo, Burundi, Ethiopia, Madagascar, Malawi, Mozambique, Rwanda, Sudan, Uganda, Rest of Eastern Africa, Rest of South African Customs |
| SSA, LMI | SSA | Lower-Middle-Income | Benin, Cameroon, Cote d'Ivoire, Guinea, Mauritania, Senegal, Togo, Angola, Congo, Sao Tome and Principe, Comoros, Kenya, Tanzania, Zambia, Zimbabwe, Eswatini, Namibia |
| SSA, UMI | SSA | Upper-Middle-Income | Equatorial Guinea, Gabon, Mauritius, Botswana, South Africa |
| South-East Asia, LMI | South-East Asia | Lower-Middle-Income | Cambodia, Lao People's Democratic Republ, Rest of Southeast Asia |
| South-East Asia, UMI | South-East Asia | Upper-Middle-Income | Indonesia, Malaysia, Philippines, Thailand, Viet Nam |
| South-East Asia, HI | South-East Asia | High-income | Singapore |

Notes: own elaboration.

**Table A2. List of products in the model**

| Product key | Product name | Broad Category (GTAP product) |
|---|---|---|
| pdr-c | Paddy rice | Paddy rice |
| wht-c | Wheat | Wheat |
| barl-c | Barley | Cereal grains nec |
| maiz-c | Maiz | Cereal grains nec |
| sorg-c | Sorghum | Cereal grains nec |
| oat-c | Oats | Cereal grains nec |
| rye-c | Rye | Cereal grains nec |
| ocer-c | Other cereals | Cereal grains nec |
| soy-c | Soy bean | Oil seeds |
| palm-c | Palm oil fruit | Oil seeds |
| rape-c | Rape seed | Oil seeds |
| sunf-c | Sunflower seed | Oil seeds |
| grndN-c | Groundnuts | Oil seeds |
| oso-c | Other oilseeds | Oil seeds |
| olv-c | Olives | Vegetables, fruit, nuts |
| pota-c | Potatoes | Vegetables, fruit, nuts |
| cass-c | Cassava | Vegetables, fruit, nuts |
| rttb-c | Other roots and tubers | Vegetables, fruit, nuts |
| leg-c | Leguminosae | Vegetables, fruit, nuts |
| oveg-c | Other vegetables | Vegetables, fruit, nuts |
| toma-c | Tomatoes | Vegetables, fruit, nuts |
| appl-c | Apples | Vegetables, fruit, nuts |
| grap-c | Grapes | Vegetables, fruit, nuts |
| citr-c | Citrus fruits | Vegetables, fruit, nuts |
| banp-c | Bananas and plantains | Vegetables, fruit, nuts |
| ofru-c | Other fruits | Vegetables, fruit, nuts |
| v_fo-c | Rest of v_f | Vegetables, fruit, nuts |
| c_b-c | Sugar cane, sugar beet | Sugar cane, sugar beet |
| pfb-c | Plant-based fibers | Plant-based fibers |
| coco-c | Cocoa beans | Crops nec |
| teas-c | Teas | Crops nec |
| coff-c | Coffee beans | Crops nec |
| toba-c | Tobacco, raw | Crops nec |
| ocro-c | Other crops | Crops nec |
| rmk-c | Raw milk | Raw milk |
| cattle-c | Cattle for meat | Cattle,sheep,goats,horses |
| orum-c | Other ruminant for meat | Cattle,sheep,goats,horses |
| pig-c | Pigs | Animal products nec |
| Eggs-c | Eggs | Animal products nec |
| poul-c | Poultry for meat | Animal products nec |
| wol-c | Wool, silk-worm cocoons | Wool, silk-worm cocoons |
| oapo-c | Other animal products | Animal products nec |
| frs-c | Forestry | Forestry |
| fsh-c | Fishing | Fishing |
| coa-c | Coal | Coal |
| oil-c | Oil | Oil |
| gas-c | Gas | Gas |
| oxt-c | Non-metallic minerals mining | Non-metallic minerals mining |
| ctlMeat-c | Cattle meet | Bovine meat products |
| orumMeat-c | Other ruminant meat cuts | Bovine meat products |
| PigCut-c | Pig meat cuts | Meat products nec |
| othMeat-c | Other meat | Meat products nec |
| meatPrep-c | Meat preparations, sausages | Meat products nec |
| olivOil-c | Olive oil | Vegetable oils and fats |
| palmOil-c | Palm oil | Vegetable oils and fats |
| soyCake-c | Soybean cake | Vegetable oils and fats |
| soyOil-c | Soybean oil | Vegetable oils and fats |
| rapCake-c | Rape seed cake | Vegetable oils and fats |
| rapOil-c | Rape seed oil | Vegetable oils and fats |
| SunfCake-c | Sunflower seed cake | Vegetable oils and fats |
| SunfOil-c | Sunflower seed oil | Vegetable oils and fats |
| grndNCake-c | Groundnut cake | Vegetable oils and fats |
| grndNOil-c | Groundnut oil | Vegetable oils and fats |
| animFats-c | Animal fats | Vegetable oils and fats |
| volo-c | Other vegetable oils and cakes | Vegetable oils and fats |
| butt-c | Butter | Dairy products |
| ches-c | Cheese | Dairy products |
| smp-c | Skimmed milk powder | Dairy products |
| wmp-c | Whole milk powder | Dairy products |
| whey-c | Whey | Dairy products |
| yogh-c | Yoghurt | Dairy products |
| butm-c | Buttermilk | Dairy products |
| skim-c | Skimmed milk | Dairy products |
| crem-c | Creme | Dairy products |
| icec-c | Ice creme | Dairy products |
| case-c | Casein | Dairy products |
| milo-c | Other dairy | Dairy products |
| pcr-c | Processed rice | Processed rice |
| sgr-c | Sugar | Sugar |
| ofd_juice-c | Juices | Food products nec |
| ofd_v_f-c | Food processing, fruits and veg based | Food products nec |
| ofd_flour-c | Cereals flours | Food products nec |
| ofd_byMill-c | By-products of cereal flours (bran, germ etc.) | Food products nec |
| bread_pastry-c | Bread and pastry | Food products nec |
| coff_tea-c | Coffee, teas | Food products nec |
| ofd_cere-c | Processed food, cereals based other | Food products nec |
| ofd_rest-c | Food processing rest | Food products nec |
| wine-c | Wine | Beverages and tobacco products |
| beer-c | Beer | Beverages and tobacco products |
| OthBevTob-c | Other beverages and tobacco products | Beverages and tobacco products |
| tex-c | Textiles | Textiles |
| wap-c | Wearing apparel | Wearing apparel |
| lea-c | Leather products | Leather products |
| lum-c | Wood products | Wood products |
| ppp-c | Paper products, publishing | Paper products, publishing |
| p_c-c | Petroleum, coal products | Petroleum, coal products |
| chm-c | Chemicals | Chemicals |
| bph-c | Basic pharmaceutical products | Basic pharmaceutical products |
| rpp-c | Rubber and plastics | Rubber and plastics |
| nmm-c | Mineral products nec | Mineral products nec |
| i_s-c | Iron and steel | Iron and steel |
| nfm-c | Metals nec | Metals nec |
| fmp-c | Metal products | Metal products |
| ele-c | Computer,electronic,optical prd | Computer,electronic,optical prd |
| eeq-c | Electrical equipment | Electrical equipment |
| ome-c | Machinery and equipment nec | Machinery and equipment nec |
| mvh-c | Motor vehicles and parts | Motor vehicles and parts |
| otn-c | Transport equipment nec | Transport equipment nec |
| omf-c | Manufactures nec | Manufactures nec |
| tnd-c | Electricity transm. and dist. | Electricity transm. and dist. |
| ely-c | Electricity | Electricity |
| gdt-c | Gas manufacture, distribution | Gas manufacture, distribution |
| wtr-c | Water | Water |
| cns-c | Construction | Construction |
| trd-c | Trade | Trade |
| afs-c | Accommodation,Food,service act | Accommodation,Food,service act |
| otp-c | Transport nec | Transport nec |
| wtp-c | Water transport | Water transport |
| atp-c | Air transport | Air transport |
| whs-c | Warehousing, support activities | Warehousing, support activities |
| cmn-c | Communication | Communication |
| ofi-c | Financial services nec | Financial services nec |
| ins-c | Insurance | Insurance |
| rsa-c | Real estate activities | Real estate activities |
| obs-c | Business services nec | Business services nec |
| ros-c | Recreational and other serv | Recreational and other serv |
| osg-c | Public Administration, defense | Public Administration, defense |
| edu-c | Education | Education |
| hht-c | Human health, Social work act | Human health, Social work act |
| dwe-c | Dwellings | Dwellings |

Notes: own elaboration.

**Table A3. List of activities in the model**

| Sector Key | Sector Description |
|---|---|
| pdr_irr-a | Paddy rice, irrigated |
| pdr_rai-a | Paddy rice, rainfed |
| wht_irr-a | Wheat, irrigated |
| wht_rai-a, | Wheat, rainfed |
| sorg_irr-a | Sorghum irrigated |
| sorg_rai-a | Sorghum, rainfed |
| barl_irr-a | Barley, irrigated |
| barl_rai-a | Barley, rainfed |
| rye_irr-a | Rye, irrigated |
| rye_rai-a | Rye, rainfed |
| oat_irr-a | Oats, irrigated |
| oat_rai-a | Oats, rainfed |
| maiz_irr-a | Maize, irrigated |
| maiz_rai-a | Maize, rainfed |
| ocer_irr-a | Other cereals, irrigated |
| ocer_rai-a | Other cereals, rainfed |
| soy_irr-a | Soy bean, irrigated |
| soy_rai-a | soy bean, rainfed |
| palm_irr-a | Palm, irrigated |
| palm_rai-a | Palm, rainfed |
| rape_irr-a | Rape seed, irrigated |
| rape_rai-a | Rape seed, rainfed |
| sunf_irr-a | Sunflower seed, irrigated |
| sunf_rai-a | Sunflower seed, rainfed |
| grndn_rai-a | Ground nut, rainfed |
| grndn_irr-a | Ground nut, irrigated |
| oso_irr-a | Other oilseeds, irrigated |
| oso_rai-a | Other oilseeds, rainfed |
| pota_irr-a | Potatoes, irrigated |
| pota_rai-a | Potatoes, rainfed |
| cass_irr-a | Cassava, irrigated |
| cass_rai-a | Cassava, rainfed |
| rttb_irr-a | Other roots and tubers, irrigated |
| rttb_rai-a | Other roots and tubers, rainfed |
| olv_irr-a | Olives, irrigated |
| olv_rai-a | Olives, rainfed |
| leg_irr-a | Leguminosae, irrigated |
| leg_rai-a | Leguminosae, rainfed |
| toma_irr-a | Tomatoes, irrigated |
| toma_rai-a | Tomatoes, rainfed |
| oveg_irr-a | Other vegetable, irrigated |
| oveg_rai-a | Other vegetables, rainfed |
| citr_irr-a | Citrus fruits, irrigated |
| citr_rai-a | Citrus fruits, rainfed |
| banp_irr-a | Bananas and plantains, irrigated |
| banp_rai-a | Bananas and plantains, rainfed |
| appl_irr-a | Apples, irrigated |
| appl_rai-a | Apples, rainfed |
| grap_irr-a | Grapes, irrigated |
| grap_rai-a | Grapes, rainfed |
| ofru_irr-a | Other fruits, irrigated |
| ofru_rai-a | Other fruits, rainfed |
| v_fo_irr-a | Rest of v_f, irrigated |
| v_fo_rai-a | Rest of v_f, rainfed |
| c_b_irr-a | Sugar cane and beet, irrigated |
| c_b_rai-a | Sugar cane and beet, rainfed |
| pfb_irr-a | Plant-based fibers, irrigated |
| pfb_rai-a | Plant-based fibers, rainfed |
| coco_irr-a | Cocoa beans, irrigated |
| coco_rai-a | Cocoa beans, rainfed |
| teas_irr-a | Teas, irrigated |
| teas_rai-a | Teas, rainfed |
| coff_irr-a | Coffee, irrigated |
| coff_rai-a | Coffee, rainfed |
| toba_irr-a | Tobacco, irrigated |
| toba_rai-a | Tobacco, rainfed |
| ocro_irr-a | Other crops, irrigated |
| ocro_rai-a | Other crops, rainfed |
| pcr-a | Processed rice |
| rmk-a | Raw milk |
| cattle-a | Cattle for meat |
| orum-a | Other ruminant for meat |
| pig-a | Pigs |
| Eggs-a | Eggs |
| poul-a | Poultry for meat |
| wol-a | Wool, silk-worm cocoons |
| oapo-a | Other animal products |
| ctlMeat-a | Cattle meat |
| orumMeat-a | Other ruminant meat |
| PigCut-a | Pig meat cuts |
| othMeat-a | Other meat |
| meatPrep-a | Meat preparations, sausages |
| frs-a | Forestry |
| fsh-a | Fishing |
| coa-a | Coal |
| oil-a | Oil |
| gas-a | Gas |
| oxt-a | Non-metallic minerals mining |
| olivOil-a | Olive oil |
| palmOil-a | Palm oil |
| soyCake-a | Soybean cake |
| soyOil-a | Soybean oil |
| rapCake-a | Rape seed cake |
| rapOil-a | Rape seed oil |
| SunfCake-a | Sunflower seed cake |
| SunfOil-a | Sunflower seed oil |
| grndNCake-a | Groundnut cake |
| grndNOil-a | Groundnut oil |
| animFats-a | Animal fats |
| volo-a | Other vegetable oils and cakes |
| butt-a | Butter |
| ches-a | Cheese |
| smp-a | Skimmed milk powder |
| wmp-a | Whole milk powder |
| whey-a | Whey |
| yogh-a | Yoghurt |
| butm-a | Buttermilk |
| skim-a | Skimmed milk |
| crem-a | Creme |
| icec-a | Ice creme |
| case-a | Casein |
| milo-a | Other dairy |
| sgr-a | Sugar |
| ofd_juice-a | Juices |
| based ofd_v_f-a | Food processing, fruits and vegs |
| ofd_flour-a | Cereal flours |
| ofd_byMill-a | By-products of cereal flours (bran, |
| bread_pastry-a | Bread and pastry |
| coff_tea-a | Coffee, teas |
| ofd_cere-a | Processed food, cereals based other |
| ofd_rest-a | Food processing rest |
| wine-a | Wine |
| beer-a | Beer |
| OthBevTob-a | Other beverages and tobacco products |
| tex-a | Textiles |
| wap-a | Wearing apparel |
| lea-a | Leather products |
| lum-a | Wood products |
| ppp-a | Paper products, publishing |
| fmp-a | Metal products |
| mvh-a | Motor vehicles and parts |
| otn-a | Transport equipment nec |
| omf-a | Manufactures nec |
| p_c-a | Petroleum, coal products |
| chm-a | Chemicals |
| bph-a | Basic pharmaceutical products |
| rpp-a | Rubber and plastics |
| nmm-a | Mineral products nec |
| i_s-a | Iron and steel |
| nfm-a | Metals nec |
| ele-a | Computer,electronic,optical prd |
| eeq-a | Electrical equipment |
| ome-a | Machinery and equipment nec |
| nuclearbl-a | Nuclear power |
| coalbl-a | Coal power baseload |
| gasbl-a | Gas power baseload |
| windbl-a | Wind power |
| hydrobl-a | Hydro power baseload |
| oilbl-a | Oil power baseload |
| otherbl-a | Other baseload |
| gasp-a | Gas power peakload |
| hydrop-a | Hydro power peakload |
| oilp-a | Oil power peakload |
| solarp-a | Solar power |
| tnd-a | Electricity transm. and dist. |
| gdt-a | Gas manufacture, distribution |
| wtr-a | Water |
| cns-a | Construction |
| trd-a | Trade |
| afs-a | Accommodation,Food,service act |
| otp-a | Transport nec |
| wtp-a | Water transport |
| atp-a | Air transport |
| whs-a | Warehousing, support activities |
| cmn-a | Communication |
| ofi-a | Financial services nec |
| ins-a | Insurance |
| rsa-a | Real estate activities |
| obs-a | Business services nec |
| ros-a | Recreational and other serv |
| osg-a | Public Administration, defense |
| edu-a | Education |
| hht-a | Human health, Social work act |
| dwe-a | Dwellings |

Notes: own elaboration.

# Appendix B. Results: Germany, Ghana and World

## Table B1. Product-Level Results, Selected Regions and Scenarios (% deviation from Base)

| Dimension | Region | Product | Base Value | Use Subsidy | | | | | | | | Production Subsidy | | | | | | | |
|---|---|---|---|---|---|---|---|---|---|---|---|---|---|---|---|---|---|---|---|
| | | | | 10% Subsidy | | | | 20% Subsidy | | | | 10% Subsidy | | | | 20% Subsidy | | | |
| | | | | Soybeans | Groundnut | Leguminosae | All Legumes | Soybeans | Groundnut | Leguminosae | All Legumes | Soybeans | Groundnut | Leguminosae | All Legumes | Soybeans | Groundnut | Leguminosae | All Legumes |
| Consumer Prices | Germany | Soy bean | 1.090 | -7.3 | 0.0 | 0.2 | -7.1 | -14.6 | 0.0 | 0.5 | -14.2 | -7.9 | 0.0 | 0.3 | -4.8 | -16.0 | 0.0 | 0.6 | -9.7 |
| Consumer Prices | Germany | Other Leguminosae | 1.091 | 0.0 | 0.0 | -6.7 | -6.7 | 0.0 | 0.0 | -13.3 | -13.3 | 0.0 | 0.0 | -3.3 | -2.3 | 0.0 | 0.0 | -6.7 | -4.6 |
| Consumer Prices | Germany | Groundnut | 1.091 | 0.0 | 0.0 | 0.1 | 0.1 | 0.0 | 0.0 | 0.2 | 0.2 | -0.1 | -0.1 | 0.1 | -4.6 | -0.1 | -0.2 | 0.3 | -14.0 |
| Consumer Prices | Ghana | Soy bean | 1.000 | -5.7 | 0.0 | 0.3 | -5.4 | -11.5 | 0.0 | 0.7 | -10.9 | -4.9 | 0.0 | 0.3 | -1.3 | -9.9 | 0.0 | 0.6 | -2.6 |
| Consumer Prices | Ghana | Other Leguminosae | 1.000 | 0.1 | 0.0 | -6.2 | -6.0 | 0.3 | 0.0 | -12.4 | -12.1 | 0.2 | 0.0 | -3.5 | -1.3 | 0.4 | 0.0 | -7.0 | -2.6 |
| Consumer Prices | Ghana | Groundnut | 1.000 | 0.1 | -9.7 | 0.3 | -9.4 | 0.2 | -19.3 | 0.6 | -18.7 | 0.0 | -0.8 | 0.2 | -3.4 | 0.1 | -1.6 | 0.5 | -6.7 |
| Consumer Prices | World | Soy bean | 1.209 | -1.7 | 0.0 | 0.1 | -1.6 | -3.6 | 0.0 | 0.2 | -3.4 | -1.0 | 0.0 | 0.1 | -0.9 | -1.7 | 0.0 | 0.2 | -1.9 |
| Consumer Prices | World | Other Leguminosae | 0.981 | 0.0 | 0.0 | -7.1 | -7.1 | 0.0 | 0.1 | -14.3 | -14.2 | 0.2 | 0.0 | -3.6 | -3.2 | 0.4 | 0.1 | -7.3 | -6.7 |
| Consumer Prices | World | Groundnut | 1.002 | 0.0 | -9.3 | 0.2 | -9.1 | 0.0 | -18.6 | 0.4 | -18.3 | 0.0 | -5.6 | 0.2 | -6.1 | 0.1 | -11.4 | 0.5 | -12.3 |
| Producer Prices | Germany | Soy bean | 1.000 | 2.1 | 0.0 | 0.2 | 2.3 | 4.6 | 0.0 | 0.5 | 5.1 | -7.9 | 0.0 | 0.3 | -4.8 | -16.0 | 0.0 | 0.6 | -9.7 |
| Producer Prices | Germany | Other Leguminosae | 1.000 | 0.0 | 0.0 | 1.9 | 1.9 | 0.1 | 0.0 | 4.3 | 4.4 | 0.0 | 0.0 | -3.4 | -3.2 | 0.0 | 0.0 | -6.7 | -6.5 |
| Producer Prices | Germany | Groundnut | 0.000 | 0.0 | 0.0 | 0.0 | 0.0 | 0.0 | 0.0 | 0.0 | 0.0 | 0.0 | 0.0 | 0.0 | 0.0 | 0.0 | 0.0 | 0.0 | 0.0 |
| Producer Prices | Ghana | Soy bean | 1.000 | 4.8 | 0.0 | 0.3 | 5.1 | 10.6 | 0.0 | 0.7 | 11.4 | -4.9 | 0.0 | 0.3 | -1.3 | -9.9 | 0.0 | 0.6 | -2.6 |
| Producer Prices | Ghana | Other Leguminosae | 1.000 | 0.1 | 0.0 | 4.3 | 4.4 | 0.3 | 0.0 | 9.5 | 9.9 | 0.2 | 0.0 | -3.5 | -1.3 | 0.4 | 0.0 | -7.0 | -2.6 |
| Producer Prices | Ghana | Groundnut | 1.000 | 0.1 | 0.4 | 0.3 | 0.7 | 0.2 | 0.9 | 0.6 | 1.6 | 0.0 | -0.8 | 0.2 | -3.4 | 0.1 | -1.6 | 0.5 | -6.7 |
| Producer Prices | World | Soy bean | 1.000 | 0.1 | 0.0 | 0.1 | 0.2 | 0.1 | 0.0 | 0.3 | 0.4 | -1.5 | 0.0 | 0.1 | -1.1 | -3.3 | -0.1 | 0.2 | -2.4 |
| Producer Prices | World | Other Leguminosae | 1.000 | 0.0 | 0.0 | 3.3 | 3.3 | 0.0 | 0.1 | 7.4 | 7.5 | 0.2 | 0.0 | -3.5 | -3.1 | 0.6 | 0.1 | -7.0 | -6.5 |
| Producer Prices | World | Groundnut | 1.000 | 0.0 | 0.5 | 0.1 | 0.6 | 0.0 | 1.1 | 0.4 | 1.4 | 0.0 | -2.9 | 0.2 | -4.9 | 0.1 | -6.0 | 0.3 | -10.0 |
| Output | Germany | Soy bean | 0.040 | 6.1 | 0.0 | -0.1 | 5.9 | 13.2 | 0.0 | -0.3 | 12.8 | 6.7 | 0.0 | -0.2 | 4.0 | 14.6 | 0.0 | -0.4 | 8.3 |
| Output | Germany | Other Leguminosae | 0.290 | 0.0 | 0.0 | 2.8 | 2.7 | -0.1 | 0.0 | 6.2 | 6.1 | 0.0 | 0.0 | 4.0 | 5.2 | -0.1 | 0.0 | 8.4 | 10.9 |
| Output | Germany | Groundnut | 0.000 | 0.0 | 0.0 | 0.0 | 0.0 | 0.0 | 0.0 | 0.0 | 0.0 | 0.0 | 0.0 | 0.0 | 0.0 | 0.0 | 0.0 | 0.0 | 0.0 |
| Output | Ghana | Soy bean | 0.080 | 4.2 | 0.0 | -0.1 | 4.1 | 9.0 | 0.0 | -0.2 | 8.8 | 4.4 | 0.0 | -0.1 | 1.2 | 9.4 | 0.0 | -0.1 | 2.4 |
| Output | Ghana | Other Leguminosae | 0.200 | 0.0 | 0.0 | 4.8 | 4.7 | -0.1 | 0.0 | 10.6 | 10.5 | 0.0 | 0.0 | 3.7 | 1.6 | -0.1 | 0.1 | 7.9 | 3.2 |
| Output | Ghana | Groundnut | 0.430 | -0.1 | 0.6 | -0.1 | 0.3 | -0.3 | 1.3 | -0.3 | 0.7 | -0.2 | -1.2 | -0.1 | 0.7 | -0.5 | -2.4 | -0.2 | 1.2 |
| Output | World | Soy bean | 168.890 | 0.2 | -0.1 | 0.0 | 0.1 | 0.4 | -0.2 | -0.1 | 0.2 | 1.5 | -0.1 | 0.0 | 0.6 | 3.2 | -0.2 | -0.1 | 1.3 |
| Output | World | Other Leguminosae | 76.900 | 0.0 | 0.0 | 4.6 | 4.6 | 0.0 | -0.1 | 10.4 | 10.3 | -0.2 | 0.0 | 4.0 | 3.6 | -0.4 | -0.1 | 8.4 | 7.7 |
| Output | World | Groundnut | 46.380 | -0.1 | 2.4 | -0.1 | 2.2 | -0.1 | 5.2 | -0.1 | 4.9 | -0.6 | 2.4 | -0.1 | 4.8 | -1.4 | 5.1 | -0.1 | 10.4 |
| Household Demand | Germany | Soy bean | 0.040 | 6.1 | 0.0 | -0.1 | 5.9 | 13.2 | 0.0 | -0.3 | 12.8 | 6.7 | 0.0 | -0.2 | 4.0 | 14.6 | 0.0 | -0.4 | 8.3 |
| Household Demand | Germany | Other Leguminosae | 0.172 | 0.0 | 0.0 | 9.2 | 9.1 | 0.0 | 0.0 | 20.2 | 20.2 | 0.0 | 0.0 | 4.4 | 3.0 | 0.0 | 0.0 | 9.2 | 6.2 |
| Household Demand | Germany | Groundnut | 0.002 | -0.3 | 0.0 | 0.0 | -0.4 | -0.7 | 0.0 | -0.1 | -0.8 | -0.3 | 0.1 | -0.1 | 5.1 | -0.6 | 0.3 | -0.2 | 17.5 |
| Household Demand | Ghana | Soy bean | 0.077 | 4.9 | 0.0 | 0.0 | 4.8 | 10.5 | 0.0 | -0.1 | 10.3 | 4.2 | 0.0 | 0.0 | 1.3 | 9.0 | 0.0 | -0.1 | 2.7 |
| Household Demand | Ghana | Other Leguminosae | 0.133 | 0.0 | 0.0 | 8.2 | 8.1 | -0.1 | 0.0 | 18.1 | 17.9 | 0.0 | 0.0 | 4.6 | 2.0 | -0.1 | 0.1 | 9.8 | 4.1 |
| Household Demand | Ghana | Groundnut | 0.000 | -0.1 | 11.8 | -0.1 | 11.7 | -0.1 | 26.6 | -0.1 | 26.3 | 0.0 | 0.9 | 0.0 | 4.2 | 0.0 | 1.8 | -0.1 | 8.8 |
| Household Demand | World | Soy bean | 32.431 | 1.3 | -0.5 | -0.1 | 0.7 | 2.8 | -1.1 | -0.2 | 1.5 | 0.8 | -0.1 | -0.1 | 0.3 | 1.2 | -0.2 | -0.1 | 0.7 |
| Household Demand | World | Other Leguminosae | 43.870 | 0.0 | -0.1 | 9.5 | 9.4 | 0.0 | -0.1 | 21.1 | 20.9 | -0.2 | 0.0 | 4.6 | 4.2 | -0.4 | -0.1 | 9.8 | 9.1 |
| Household Demand | World | Groundnut | 14.974 | -0.2 | 7.7 | -0.1 | 7.3 | -0.5 | 17.1 | -0.3 | 16.2 | -0.1 | 4.0 | -0.1 | 4.6 | -0.2 | 8.5 | -0.2 | 9.9 |
| Intermediate Demand | Germany | Soy bean | 2.110 | 0.0 | 0.0 | 0.0 | 0.0 | -0.1 | 0.0 | -0.1 | -0.1 | 0.1 | 0.0 | 0.0 | 0.4 | 0.3 | 0.0 | 0.0 | 0.8 |
| Intermediate Demand | Germany | Other Leguminosae | 0.280 | 0.0 | 0.0 | -2.2 | -2.2 | 0.0 | 0.0 | -4.8 | -4.8 | 0.0 | 0.0 | 3.4 | 2.7 | 0.0 | 0.0 | 7.1 | 5.6 |
| Intermediate Demand | Germany | Groundnut | 0.030 | 0.0 | 0.0 | 0.0 | 0.0 | 0.0 | 0.0 | 0.0 | 0.0 | 0.0 | 0.0 | 0.0 | 0.6 | 0.0 | 0.0 | 0.0 | 2.4 |
| Intermediate Demand | Ghana | Soy bean | 0.010 | -2.9 | 0.0 | -0.5 | -3.4 | -6.1 | 0.0 | -1.2 | -7.2 | 3.7 | -0.1 | -0.6 | 2.2 | 7.9 | -0.1 | -1.2 | 4.9 |
| Intermediate Demand | Ghana | Other Leguminosae | 0.070 | 0.0 | 0.0 | -1.8 | -1.8 | 0.1 | 0.0 | -3.8 | -3.8 | 0.0 | 0.0 | 1.8 | 0.9 | 0.1 | 0.0 | 3.8 | 1.8 |
| Intermediate Demand | Ghana | Groundnut | 0.340 | -0.1 | 0.1 | -0.1 | -0.1 | -0.1 | 0.2 | -0.2 | -0.2 | -0.1 | -0.2 | -0.1 | 1.4 | -0.3 | -0.4 | -0.2 | 2.7 |
| Intermediate Demand | World | Soy bean | 148.560 | 0.0 | 0.0 | 0.0 | 0.0 | -0.1 | 0.1 | 0.0 | 0.0 | 1.4 | -0.1 | 0.0 | 0.6 | 3.2 | -0.2 | -0.1 | 1.2 |
| Intermediate Demand | World | Other Leguminosae | 34.230 | 0.0 | 0.0 | -1.6 | -1.6 | 0.0 | 0.0 | -3.4 | -3.4 | -0.2 | 0.0 | 3.0 | 2.6 | -0.4 | 0.0 | 6.2 | 5.5 |
| Intermediate Demand | World | Groundnut | 31.550 | 0.0 | -0.2 | 0.0 | -0.2 | 0.0 | -0.4 | 0.0 | -0.4 | -0.9 | 1.7 | -0.1 | 4.9 | -2.0 | 3.4 | -0.1 | 10.5 |
| Exports | Germany | Soy bean | 0.000 | 1.7 | -0.1 | 0.2 | 1.8 | 3.7 | -0.2 | 0.6 | 4.0 | 35.1 | -0.1 | 0.0 | 10.7 | 89.0 | -0.2 | 0.1 | 23.1 |
| Exports | Germany | Other Leguminosae | 0.110 | -0.1 | 0.0 | 5.6 | 5.5 | -0.1 | 0.0 | 12.5 | 12.4 | -0.1 | 0.0 | 4.6 | 7.0 | -0.2 | 0.0 | 9.5 | 14.8 |
| Exports | Germany | Groundnut | 0.000 | 0.0 | 0.0 | 0.0 | 0.0 | 0.0 | 0.0 | 0.0 | 0.0 | 0.0 | 0.0 | 0.0 | 0.0 | 0.0 | 0.0 | 0.0 | 0.0 |
| Exports | Ghana | Soy bean | 0.000 | -17.9 | -0.3 | -0.4 | -18.6 | -34.7 | -0.6 | -0.9 | -35.8 | 22.8 | -0.8 | -0.3 | -1.1 | 53.4 | -1.6 | -0.6 | -2.3 |
| Exports | Ghana | Other Leguminosae | 0.000 | -0.5 | 0.0 | -3.9 | -4.4 | -1.0 | 0.0 | -7.9 | -8.9 | -0.4 | 0.1 | 5.3 | -0.9 | -0.9 | 0.2 | 11.4 | -2.3 |
| Exports | Ghana | Groundnut | 0.090 | -0.4 | 2.5 | -0.3 | 1.8 | -0.8 | 5.5 | -0.7 | 3.9 | -0.5 | -5.2 | -0.2 | -2.0 | -1.0 | -10.4 | -0.4 | -4.8 |
| Exports | World | Soy bean | 90.750 | 0.2 | -0.1 | 0.0 | 0.1 | 0.4 | -0.2 | 0.0 | 0.2 | -1.3 | -0.1 | 0.0 | -0.7 | -2.8 | -0.2 | 0.0 | -1.4 |
| Exports | World | Other Leguminosae | 13.670 | 0.0 | 0.0 | 2.3 | 2.3 | 0.0 | 0.0 | 5.2 | 5.2 | 0.1 | 0.0 | 2.1 | 0.9 | 0.2 | 0.0 | 4.5 | 2.1 |
| Exports | World | Groundnut | 2.050 | -0.2 | 0.7 | -0.2 | 0.3 | -0.4 | 1.5 | -0.5 | 0.6 | -0.4 | -0.5 | -0.2 | 3.4 | -0.9 | -0.8 | -0.5 | 8.6 |
| Imports | Germany | Soy bean | 2.120 | 0.0 | 0.0 | 0.0 | 0.0 | -0.1 | 0.0 | -0.1 | -0.1 | 0.1 | 0.0 | 0.0 | 0.4 | 0.3 | 0.0 | 0.0 | 0.8 |
| Imports | Germany | Other Leguminosae | 0.290 | 0.0 | 0.0 | 2.6 | 2.6 | 0.0 | 0.0 | 5.8 | 5.8 | 0.0 | 0.0 | 3.9 | 2.1 | 0.0 | 0.0 | 8.3 | 4.4 |
| Imports | Germany | Groundnut | 0.030 | 0.0 | 0.0 | 0.0 | 0.0 | 0.0 | 0.0 | 0.0 | -0.1 | 0.0 | 0.0 | 0.0 | 0.9 | -0.1 | 0.0 | 0.0 | 3.3 |
| Imports | Ghana | Soy bean | 0.010 | -0.1 | 0.0 | -0.6 | -0.7 | -0.3 | 0.0 | -1.3 | -1.6 | 1.8 | -0.1 | -0.6 | 4.5 | 3.5 | -0.2 | -1.3 | 9.7 |
| Imports | Ghana | Other Leguminosae | 0.010 | 0.3 | 0.0 | -0.6 | -0.4 | 0.6 | 0.0 | -1.3 | -0.8 | 0.3 | 0.0 | 1.0 | 2.3 | 0.6 | 0.0 | 2.1 | 4.7 |
| Imports | Ghana | Groundnut | 0.000 | 0.1 | -0.5 | -0.2 | -0.6 | 0.3 | -1.2 | -0.4 | -1.4 | 0.3 | 3.2 | -0.2 | 5.9 | 0.6 | 7.8 | -0.4 | 12.3 |
| Imports | World | Soy bean | 103.360 | 0.2 | -0.1 | 0.0 | 0.1 | 0.5 | -0.2 | 0.0 | 0.3 | -1.3 | -0.1 | 0.0 | -0.7 | -2.7 | -0.2 | 0.0 | -1.4 |
| Imports | World | Other Leguminosae | 15.440 | 0.0 | 0.0 | 2.2 | 2.2 | 0.0 | 0.0 | 5.1 | 5.0 | 0.1 | 0.0 | 2.2 | 0.9 | 0.2 | 0.0 | 4.6 | 2.1 |
| Imports | World | Groundnut | 2.270 | -0.2 | 0.7 | -0.2 | 0.3 | -0.4 | 1.6 | -0.5 | 0.6 | -0.4 | -0.6 | -0.2 | 3.1 | -0.9 | -1.0 | -0.5 | 7.1 |
| Nutrient Intake, Calories | Germany | Soy bean | 1.010 | 5.5 | 0.0 | -0.1 | 5.4 | 12.0 | 0.0 | -0.3 | 11.7 | 6.1 | 0.0 | -0.2 | 3.7 | 13.4 | 0.0 | -0.4 | 7.7 |
| Nutrient Intake, Calories | Germany | Other Leguminosae | 3.340 | 0.0 | 0.0 | 8.5 | 8.5 | 0.0 | 0.0 | 18.8 | 18.8 | 0.0 | 0.0 | 4.5 | 3.1 | 0.0 | 0.0 | 9.5 | 6.4 |
| Nutrient Intake, Calories | Germany | Groundnut | 0.050 | -0.3 | 0.0 | 0.0 | -0.4 | -0.7 | 0.0 | -0.1 | -0.8 | -0.3 | 0.1 | -0.1 | 5.1 | -0.6 | 0.3 | -0.2 | 17.5 |
| Nutrient Intake, Calories | Germany | All Commodities | 2079.620 | 0.0 | 0.0 | 0.0 | 0.0 | 0.0 | 0.0 | 0.0 | 0.0 | 0.0 | 0.0 | 0.0 | 0.0 | 0.0 | 0.0 | 0.0 | 0.0 |
| Nutrient Intake, Calories | Ghana | Soy bean | 0.300 | 4.9 | 0.0 | -0.1 | 4.8 | 10.5 | 0.0 | -0.1 | 10.3 | 4.2 | 0.0 | 0.0 | 1.4 | 9.0 | 0.1 | -0.1 | 2.7 |
| Nutrient Intake, Calories | Ghana | Other Leguminosae | 48.280 | 0.0 | 0.0 | 8.2 | 8.1 | -0.1 | -0.1 | 18.1 | 17.9 | 0.0 | 0.0 | 4.6 | 2.0 | -0.1 | 0.1 | 9.8 | 4.1 |
| Nutrient Intake, Calories | Ghana | Groundnut | 0.000 | 0.0 | 0.0 | 0.0 | 0.0 | 0.0 | 0.0 | 0.0 | 0.0 | 0.0 | 0.0 | 0.0 | 0.0 | 0.0 | 0.0 | 0.0 | 0.0 |
| Nutrient Intake, Calories | Ghana | All Commodities | 1800.570 | 0.1 | 0.0 | 0.3 | 0.3 | 0.1 | 0.0 | 0.6 | 0.7 | 0.1 | 0.0 | 0.2 | 0.2 | 0.1 | 0.0 | 0.4 | 0.4 |
| Nutrient Intake, Calories | World | Soy bean | 10.420 | 0.9 | -0.5 | -0.1 | 0.4 | 2.0 | -1.1 | -0.2 | 0.8 | 0.5 | -0.1 | 0.0 | 0.2 | 0.8 | -0.1 | -0.1 | 0.4 |
| Nutrient Intake, Calories | World | Other Leguminosae | 36.640 | 0.0 | -0.1 | 8.9 | 8.8 | -0.1 | -0.2 | 19.7 | 19.4 | -0.1 | -0.1 | 5.0 | 4.0 | -0.2 | -0.1 | 10.7 | 8.7 |
| Nutrient Intake, Calories | World | Groundnut | 2.150 | -0.1 | 8.9 | -0.1 | 8.6 | -0.3 | 19.9 | -0.3 | 19.3 | -0.1 | 3.3 | -0.1 | 4.9 | -0.2 | 7.0 | -0.2 | 10.5 |
| Nutrient Intake, Calories | World | All Commodities | 1822.480 | 0.0 | 0.0 | 0.2 | 0.2 | 0.0 | 0.0 | 0.3 | 0.3 | 0.0 | 0.0 | 0.1 | 0.1 | 0.0 | 0.0 | 0.2 | 0.2 |
| Nutrient Intake, Proteins | Germany | Soy bean | 0.100 | 5.5 | 0.0 | -0.1 | 5.4 | 12.0 | 0.0 | -0.3 | 11.7 | 6.1 | 0.0 | -0.2 | 3.7 | 13.4 | 0.0 | -0.4 | 7.7 |
| Nutrient Intake, Proteins | Germany | Other Leguminosae | 0.220 | 0.0 | 0.0 | 8.5 | 8.5 | 0.0 | 0.0 | 18.8 | 18.8 | 0.0 | 0.0 | 4.5 | 3.1 | 0.0 | 0.0 | 9.5 | 6.4 |
| Nutrient Intake, Proteins | Germany | Groundnut | 0.000 | -0.3 | 0.0 | 0.0 | -0.4 | -0.7 | 0.0 | -0.1 | -0.8 | -0.3 | 0.1 | -0.1 | 5.1 | -0.6 | 0.3 | -0.2 | 17.5 |
| Nutrient Intake, Proteins | Germany | All Commodities | 61.590 | 0.0 | 0.0 | 0.0 | 0.0 | 0.0 | 0.0 | 0.0 | 0.0 | 0.0 | 0.0 | 0.0 | 0.0 | 0.0 | 0.0 | 0.0 | 0.0 |
| Nutrient Intake, Proteins | Ghana | Soy bean | 0.020 | 4.9 | 0.0 | -0.1 | 4.8 | 10.5 | 0.0 | -0.1 | 10.3 | 4.2 | 0.0 | 0.0 | 1.4 | 9.0 | 0.1 | -0.1 | 2.7 |
| Nutrient Intake, Proteins | Ghana | Other Leguminosae | 2.840 | 0.0 | 0.0 | 8.2 | 8.1 | -0.1 | -0.1 | 18.1 | 17.9 | 0.0 | 0.0 | 4.6 | 2.0 | -0.1 | 0.1 | 9.8 | 4.1 |
| Nutrient Intake, Proteins | Ghana | Groundnut | 0.000 | 0.0 | 0.0 | 0.0 | 0.0 | 0.0 | 0.0 | 0.0 | 0.0 | 0.0 | 0.0 | 0.0 | 0.0 | 0.0 | 0.0 | 0.0 | 0.0 |
| Nutrient Intake, Proteins | Ghana | All Commodities | 43.520 | 0.1 | 0.0 | 0.6 | 0.6 | 0.1 | 0.0 | 1.3 | 1.4 | 0.1 | 0.0 | 0.4 | 0.3 | 0.1 | 0.0 | 0.8 | 0.5 |
| Nutrient Intake, Proteins | World | Soy bean | 0.770 | 1.6 | -0.5 | -0.1 | 1.0 | 3.5 | -1.0 | -0.2 | 2.3 | 0.6 | -0.1 | -0.1 | 0.2 | 0.9 | -0.1 | -0.1 | 0.5 |
| Nutrient Intake, Proteins | World | Other Leguminosae | 2.440 | 0.0 | -0.1 | 8.9 | 8.8 | -0.1 | -0.2 | 19.7 | 19.5 | -0.1 | -0.1 | 5.0 | 4.0 | -0.2 | -0.1 | 10.7 | 8.6 |
| Nutrient Intake, Proteins | World | Groundnut | 0.090 | -0.2 | 8.4 | -0.1 | 8.1 | -0.3 | 18.7 | -0.3 | 18.0 | -0.1 | 3.6 | -0.1 | 5.1 | -0.2 | 7.7 | -0.2 | 10.9 |
| Nutrient Intake, Proteins | World | All Commodities | 55.970 | 0.0 | 0.0 | 0.3 | 0.4 | 0.0 | 0.0 | 0.8 | 0.8 | 0.0 | 0.0 | 0.2 | 0.2 | 0.0 | 0.0 | 0.4 | 0.3 |
| Nutrient Intake, Fats | Germany | Soy bean | 0.040 | 5.5 | 0.0 | -0.1 | 5.4 | 12.0 | 0.0 | -0.3 | 11.7 | 6.1 | 0.0 | -0.2 | 3.7 | 13.4 | 0.0 | -0.4 | 7.7 |
| Nutrient Intake, Fats | Germany | Other Leguminosae | 0.010 | 0.0 | 0.0 | 8.5 | 8.5 | 0.0 | 0.0 | 18.8 | 18.8 | 0.0 | 0.0 | 4.5 | 3.1 | 0.0 | 0.0 | 9.5 | 6.4 |
| Nutrient Intake, Fats | Germany | Groundnut | 0.000 | -0.3 | 0.0 | 0.0 | -0.4 | -0.7 | 0.0 | -0.1 | -0.8 | -0.3 | 0.1 | -0.1 | 5.1 | -0.6 | 0.3 | -0.2 | 17.5 |
| Nutrient Intake, Fats | Germany | All Commodities | 92.810 | 0.0 | 0.0 | 0.0 | 0.0 | 0.0 | 0.0 | 0.0 | 0.0 | 0.0 | 0.0 | 0.0 | 0.0 | 0.0 | 0.0 | 0.0 | 0.0 |
| Nutrient Intake, Fats | Ghana | Soy bean | 0.020 | 4.9 | 0.0 | -0.1 | 4.8 | 10.5 | 0.0 | -0.1 | 10.3 | 4.2 | 0.0 | 0.0 | 1.4 | 9.0 | 0.1 | -0.1 | 2.7 |
| Nutrient Intake, Fats | Ghana | Other Leguminosae | 0.250 | 0.0 | 0.0 | 8.2 | 8.1 | -0.1 | -0.1 | 18.1 | 17.9 | 0.0 | 0.0 | 4.6 | 2.0 | -0.1 | 0.1 | 9.8 | 4.1 |
| Nutrient Intake, Fats | Ghana | Groundnut | 0.000 | 0.0 | 0.0 | 0.0 | 0.0 | 0.0 | 0.0 | 0.0 | 0.0 | 0.0 | 0.0 | 0.0 | 0.0 | 0.0 | 0.0 | 0.0 | 0.0 |
| Nutrient Intake, Fats | Ghana | All Commodities | 32.270 | 0.1 | 0.0 | 0.1 | 0.1 | 0.1 | 0.0 | 0.2 | 0.3 | 0.1 | 0.0 | 0.1 | 0.2 | 0.1 | 0.0 | 0.2 | 0.3 |
| Nutrient Intake, Fats | World | Soy bean | 0.480 | 0.6 | -0.5 | -0.1 | 0.0 | 1.2 | -1.1 | -0.2 | 0.0 | 0.4 | -0.1 | 0.0 | 0.1 | 0.7 | -0.1 | -0.1 | 0.3 |
| Nutrient Intake, Fats | World | Other Leguminosae | 0.200 | 0.0 | -0.1 | 8.9 | 8.8 | -0.1 | -0.2 | 19.9 | 19.6 | -0.1 | -0.1 | 5.3 | 4.2 | -0.1 | -0.1 | 11.2 | 8.9 |
| Nutrient Intake, Fats | World | Groundnut | 0.160 | -0.1 | 9.1 | -0.1 | 8.9 | -0.2 | 20.5 | -0.3 | 19.9 | -0.1 | 3.1 | -0.1 | 4.8 | -0.3 | 6.6 | -0.2 | 10.3 |
| Nutrient Intake, Fats | World | All Commodities | 51.830 | 0.0 | 0.0 | 0.0 | 0.0 | 0.0 | 0.0 | 0.0 | 0.0 | 0.0 | 0.0 | 0.0 | 0.0 | 0.0 | 0.0 | 0.0 | 0.1 |

Own elaboration.

## Table B2. Land Cover and Land Use Changes, Selected Regions and Scenarios (% deviation from Base)

| Land Cover / Land Use | Region | Land | Base Value | Use Subsidy | | | | | | | | Production Subsidy | | | | | | | |
|---|---|---|---|---|---|---|---|---|---|---|---|---|---|---|---|---|---|---|---|
| | | | | 10% Subsidy | | | | 20% Subsidy | | | | 10% Subsidy | | | | 20% Subsidy | | | |
| | | | | Soybeans | Groundnut | Leguminosae | All Legumes | Soybeans | Groundnut | Leguminosae | All Legumes | Soybeans | Groundnut | Leguminosae | All Legumes | Soybeans | Groundnut | Leguminosae | All Legumes |
| land cover | Germany | Forest | 10,957 | 0.0 | 0.0 | 0.0 | 0.0 | 0.0 | 0.0 | -0.1 | -0.1 | 0.0 | 0.0 | 0.0 | 0.0 | 0.0 | 0.0 | -0.1 | -0.1 |
| land cover | Germany | PastureLand | 4,619 | 0.0 | 0.0 | -0.1 | -0.2 | -0.1 | 0.0 | -0.3 | -0.4 | 0.0 | 0.0 | -0.2 | -0.2 | -0.1 | 0.0 | -0.4 | -0.5 |
| land cover | Germany | cropLand | 11,967 | 0.0 | 0.0 | 0.1 | 0.1 | 0.0 | 0.0 | 0.2 | 0.2 | 0.0 | 0.0 | 0.1 | 0.1 | 0.0 | 0.0 | 0.2 | 0.3 |
| land cover | Ghana | Forest | 6,487 | 0.0 | 0.0 | 0.0 | 0.0 | 0.0 | 0.0 | -0.1 | -0.1 | 0.0 | 0.0 | 0.0 | 0.0 | 0.0 | 0.0 | 0.0 | -0.1 |
| land cover | Ghana | PastureLand | 4,851 | 0.0 | 0.0 | 0.0 | 0.0 | 0.0 | 0.0 | 0.0 | 0.0 | 0.0 | 0.0 | 0.0 | 0.0 | 0.0 | 0.0 | 0.0 | 0.1 |
| land cover | Ghana | cropLand | 6,360 | 0.0 | 0.0 | 0.0 | 0.0 | 0.0 | 0.0 | 0.1 | 0.1 | 0.0 | 0.0 | 0.0 | 0.0 | 0.0 | 0.0 | 0.1 | 0.0 |
| land cover | World | Forest | 1,599,473 | 0.0 | 0.0 | 0.0 | 0.0 | 0.0 | 0.0 | -0.1 | -0.1 | 0.0 | 0.0 | 0.0 | 0.0 | 0.0 | 0.0 | -0.1 | -0.1 |
| land cover | World | PastureLand | 3,166,371 | 0.0 | 0.0 | -0.1 | -0.1 | 0.0 | 0.0 | -0.1 | -0.2 | 0.0 | 0.0 | -0.1 | -0.1 | 0.0 | 0.0 | -0.1 | -0.1 |
| land cover | World | cropLand | 1,581,407 | 0.0 | 0.0 | 0.2 | 0.2 | 0.0 | 0.0 | 0.4 | 0.4 | 0.0 | 0.0 | 0.1 | 0.1 | 0.1 | 0.0 | 0.3 | 0.3 |
| land use | Germany | Ground nut, irrigated | 0 | 0.0 | 0.0 | 0.0 | 0.0 | 0.0 | 0.0 | 0.0 | 0.0 | 0.0 | 0.0 | 0.0 | 0.0 | 0.0 | 0.0 | 0.0 | 0.0 |
| land use | Germany | Ground nut, rainfed | 0 | 0.0 | 0.0 | 0.0 | 0.0 | 0.0 | 0.0 | 0.0 | 0.0 | 0.0 | 0.0 | 0.0 | 0.0 | 0.0 | 0.0 | 0.0 | 0.0 |
| land use | Germany | Leguminosae, irrigated | 2 | 0.0 | 0.0 | 2.1 | 2.1 | -0.1 | 0.0 | 4.7 | 4.6 | 0.0 | 0.0 | 3.1 | 4.0 | -0.1 | 0.0 | 6.4 | 8.4 |
| land use | Germany | Leguminosae, rainfed | 127 | -0.1 | 0.0 | 2.5 | 2.4 | -0.1 | 0.0 | 5.5 | 5.4 | -0.1 | 0.0 | 3.6 | 4.7 | -0.1 | 0.0 | 7.5 | 9.8 |
| land use | Germany | Soy bean, irrigated | 0 | 0.0 | 0.0 | 0.0 | 0.0 | 0.0 | 0.0 | 0.0 | 0.0 | 0.0 | 0.0 | 0.0 | 0.0 | 0.0 | 0.0 | 0.0 | 0.0 |
| land use | Germany | soy bean, rainfed | 47 | 5.2 | 0.0 | -0.2 | 5.0 | 11.2 | 0.0 | -0.5 | 10.7 | 5.7 | 0.0 | -0.3 | 3.2 | 12.5 | 0.0 | -0.6 | 6.7 |
| land use | Ghana | Ground nut, irrigated | 124 | -0.1 | 0.5 | -0.2 | 0.3 | -0.3 | 1.2 | -0.4 | 0.5 | -0.2 | -1.1 | -0.1 | 0.6 | -0.4 | -2.2 | -0.3 | 1.1 |
| land use | Ghana | Ground nut, rainfed | 124 | -0.1 | 0.5 | -0.2 | 0.3 | -0.3 | 1.2 | -0.4 | 0.5 | -0.2 | -1.1 | -0.1 | 0.6 | -0.4 | -2.2 | -0.3 | 1.1 |
| land use | Ghana | Leguminosae, irrigated | 0 | 0.0 | 0.0 | 0.0 | 0.0 | 0.0 | 0.0 | 0.0 | 0.0 | 0.0 | 0.0 | 0.0 | 0.0 | 0.0 | 0.0 | 0.0 | 0.0 |
| land use | Ghana | Leguminosae, rainfed | 54 | 0.0 | 0.0 | 4.5 | 4.5 | -0.1 | 0.0 | 10.0 | 9.9 | 0.0 | 0.0 | 3.5 | 1.5 | -0.1 | 0.1 | 7.5 | 3.0 |
| land use | Ghana | Soy bean, irrigated | 17 | 3.9 | 0.0 | -0.1 | 3.9 | 8.5 | 0.0 | -0.2 | 8.3 | 4.1 | 0.0 | -0.1 | 1.1 | 8.8 | 0.0 | -0.1 | 2.2 |
| land use | Ghana | soy bean, rainfed | 66 | 5.0 | 0.0 | -0.1 | 4.8 | 10.9 | 0.0 | -0.3 | 10.5 | 5.2 | 0.0 | -0.1 | 1.3 | 11.3 | 0.1 | -0.2 | 2.8 |
| land use | World | Ground nut, irrigated | 14,534 | -0.1 | 2.0 | -0.3 | 1.6 | -0.1 | 4.4 | -0.7 | 3.5 | -0.3 | 1.6 | -0.3 | 3.7 | -0.6 | 3.5 | -0.7 | 7.7 |
| land use | World | Ground nut, rainfed | 14,534 | -0.1 | 2.0 | -0.3 | 1.6 | -0.1 | 4.4 | -0.7 | 3.5 | -0.3 | 1.6 | -0.3 | 3.7 | -0.6 | 3.5 | -0.7 | 7.7 |
| land use | World | Leguminosae, irrigated | 3,248 | 0.0 | -0.1 | 4.3 | 4.2 | -0.1 | -0.3 | 9.6 | 9.2 | -0.2 | -0.1 | 5.0 | 3.7 | -0.3 | -0.3 | 10.9 | 8.0 |
| land use | World | Leguminosae, rainfed | 28,262 | 0.0 | -0.1 | 4.3 | 4.2 | 0.0 | -0.2 | 9.4 | 9.2 | -0.1 | -0.1 | 3.9 | 2.8 | -0.1 | -0.2 | 8.2 | 6.0 |
| land use | World | Soy bean, irrigated | 3,299 | 0.4 | -0.1 | -0.3 | -0.1 | 0.8 | -0.3 | -0.8 | -0.2 | 6.2 | -0.2 | -0.3 | 3.3 | 13.1 | -0.5 | -0.7 | 6.9 |
| land use | World | soy bean, rainfed | 111,167 | 0.2 | -0.1 | -0.2 | -0.1 | 0.5 | -0.2 | -0.5 | -0.1 | 0.7 | -0.1 | -0.2 | 0.7 | 1.3 | -0.2 | -0.4 | 1.4 |

Own elaboration.

## Table B3. Utility Changes, Selected Regions and Scenarios (% deviation from Base)

| Region | Institution | Base Value | Use Subsidy | | | | | | | | Production Subsidy | | | | | | | |
|---|---|---|---|---|---|---|---|---|---|---|---|---|---|---|---|---|---|---|
| | | | 10% Subsidy | | | | 20% Subsidy | | | | 10% Subsidy | | | | 20% Subsidy | | | |
| | | | Soybeans | Groundnut | Leguminosae | All Legumes | Soybeans | Groundnut | Leguminosae | All Legumes | Soybeans | Groundnut | Leguminosae | All Legumes | Soybeans | Groundnut | Leguminosae | All Legumes |
| Germany | Households | 1 | 0 | 0.0 | 0.0 | 0.0 | 0.0 | 0.0 | 0.0 | 0.0 | 0.0 | 0.0 | 0.0 | 0.0 | 0.0 | 0.0 | 0.0 | 0.0 |
| Germany | Agricultural | 1 | 0 | 0.0 | 0.0 | 0.0 | 0.0 | 0.0 | 0.1 | 0.1 | 0.0 | 0.0 | 0.0 | 0.0 | 0.0 | 0.0 | 0.1 | 0.1 |
| Germany | Non-Agricultural | 1 | 0 | 0.0 | 0.0 | 0.0 | 0.0 | 0.0 | 0.0 | 0.0 | 0.0 | 0.0 | 0.0 | 0.0 | 0.0 | 0.0 | 0.0 | 0.0 |
| Ghana | Households | 1 | 0 | 0.0 | 0.1 | 0.1 | 0.1 | 0.0 | 0.2 | 0.3 | 0.1 | 0.0 | 0.1 | 0.1 | 0.1 | 0.0 | 0.2 | 0.3 |
| Ghana | Agricultural | 1 | 0 | 0.0 | 0.1 | 0.2 | 0.1 | 0.0 | 0.2 | 0.4 | 0.0 | 0.0 | 0.1 | 0.1 | 0.1 | -0.1 | 0.2 | 0.2 |
| Ghana | Non-Agricultural | 1 | 0 | 0.0 | 0.1 | 0.1 | 0.1 | 0.0 | 0.2 | 0.3 | 0.1 | 0.0 | 0.1 | 0.2 | 0.1 | 0.0 | 0.2 | 0.3 |
| World | Households | 1 | 0 | 0.0 | 0.0 | 0.0 | 0.0 | 0.0 | 0.0 | 0.1 | 0.0 | 0.0 | 0.0 | 0.0 | 0.0 | 0.0 | 0.0 | 0.1 |
| World | Agricultural | 1 | 0 | 0.0 | 0.2 | 0.2 | 0.0 | 0.1 | 0.4 | 0.5 | 0.1 | 0.0 | 0.2 | 0.2 | 0.1 | 0.1 | 0.4 | 0.5 |
| World | Non-Agricultural | 1 | 0 | 0.0 | 0.0 | 0.0 | 0.0 | 0.0 | 0.0 | 0.0 | 0.0 | 0.0 | 0.0 | 0.0 | 0.0 | 0.0 | 0.0 | 0.0 |

Own elaboration

# Appendix C. Additional results: All Regions

## Table C1. Selected Results, 20% Subsidy

| Region/Country | Dimension | Soybeans | | | Other Leguminosae | | | Groundnut | | | All products | | |
|---|---|---|---|---|---|---|---|---|---|---|---|---|---|
| | | Base Value | Use Subsidy (% deviation) | Production Subsidy (% deviation) | Base Value | Use Subsidy (% deviation) | Production Subsidy (% deviation) | Base Value | Use Subsidy (% deviation) | Production Subsidy (% deviation) | Base Value | Use Subsidy (% deviation) | Production Subsidy (% deviation) |
| EU27, HI | Output | 1.9 | -0.5 | 16.0 | 3.3 | 8.4 | 7.0 | 0.0 | 0.0 | 0.0 | 0.0 | 0.0 | 0.0 |
| EU27, HI | Household Demand | 0.1 | 12.0 | 5.1 | 2.8 | 18.4 | 5.6 | 0.0 | -1.0 | 28.5 | 0.0 | 0.0 | 0.0 |
| EU27, HI | Nutrient Intake, Calories | 0.1 | 7.7 | 4.7 | 19.6 | 17.1 | 5.7 | 0.0 | -1.0 | 28.5 | 2038.8 | 0.1 | 0.1 |
| EU27, HI | Nutrient Intake, Proteins | 0.0 | 7.7 | 4.7 | 1.4 | 17.1 | 5.7 | 0.0 | -1.0 | 28.5 | 65.7 | 0.3 | 0.1 |
| EU27, HI | Nutrient Intake, Fats | 0.0 | 7.7 | 4.7 | 0.1 | 17.1 | 5.7 | 0.0 | -1.0 | 28.5 | 85.7 | 0.0 | 0.0 |
| East Asia, HI | Output | 20.2 | -0.4 | 12.1 | 31.2 | 9.7 | 7.7 | 22.6 | 3.8 | 12.4 | 0.0 | 0.0 | 0.0 |
| East Asia, HI | Household Demand | 26.4 | -1.4 | 0.1 | 13.8 | 24.3 | 9.9 | 3.7 | 27.1 | 10.0 | 0.0 | 0.0 | 0.0 |
| East Asia, HI | Nutrient Intake, Calories | 50.3 | -1.4 | 0.1 | 24.2 | 24.0 | 9.9 | 5.3 | 27.0 | 10.0 | 2107.4 | 0.2 | 0.1 |
| East Asia, HI | Nutrient Intake, Proteins | 3.3 | -1.4 | 0.1 | 1.6 | 24.0 | 9.9 | 0.2 | 27.0 | 10.0 | 78.1 | 0.4 | 0.2 |
| East Asia, HI | Nutrient Intake, Fats | 2.5 | -1.4 | 0.1 | 0.1 | 24.0 | 9.9 | 0.4 | 27.0 | 10.0 | 54.4 | 0.1 | 0.1 |
| East Asia, UMI | Output | 0.1 | 5.7 | 96.5 | 0.3 | 13.5 | 81.3 | 0.0 | 1.3 | 148.1 | 0.0 | 0.0 | 0.0 |
| East Asia, UMI | Household Demand | 1.0 | 13.7 | 4.3 | 0.4 | 18.6 | 60.2 | 0.0 | 16.9 | 51.4 | 0.0 | 0.0 | 0.0 |
| East Asia, UMI | Nutrient Intake, Calories | 14.5 | 13.7 | 2.7 | 4.5 | 17.9 | 56.5 | 0.0 | 0.0 | 0.0 | 1450.9 | 0.1 | 0.2 |
| East Asia, UMI | Nutrient Intake, Proteins | 1.5 | 13.7 | 2.7 | 0.3 | 17.9 | 56.5 | 0.0 | 0.0 | 0.0 | 52.1 | 0.4 | 0.4 |
| East Asia, UMI | Nutrient Intake, Fats | 0.5 | 13.7 | 2.7 | 0.0 | 17.9 | 56.5 | 0.0 | 0.0 | 0.0 | 52.0 | 0.1 | 0.1 |
| Germany | Output | 0.0 | 12.8 | 8.3 | 0.3 | 6.1 | 10.9 | 0.0 | 0.0 | 0.0 | 0.0 | 0.0 | 0.0 |
| Germany | Household Demand | 0.0 | 12.8 | 8.3 | 0.2 | 20.2 | 6.2 | 0.0 | -0.8 | 17.5 | 0.0 | 0.0 | 0.0 |
| Germany | Nutrient Intake, Calories | 1.0 | 11.7 | 7.7 | 3.3 | 18.8 | 6.4 | 0.1 | -0.8 | 17.5 | 2079.6 | 0.0 | 0.0 |
| Germany | Nutrient Intake, Proteins | 0.1 | 11.7 | 7.7 | 0.2 | 18.8 | 6.4 | 0.0 | -0.8 | 17.5 | 61.6 | 0.0 | 0.0 |
| Germany | Nutrient Intake, Fats | 0.0 | 11.7 | 7.7 | 0.0 | 18.8 | 6.4 | 0.0 | -0.8 | 17.5 | 92.8 | 0.0 | 0.0 |
| Ghana | Output | 0.1 | 8.8 | 2.4 | 0.2 | 10.5 | 3.2 | 0.4 | 0.7 | 1.2 | 0.0 | 0.0 | 0.0 |
| Ghana | Household Demand | 0.1 | 10.3 | 2.7 | 0.1 | 17.9 | 4.1 | 0.0 | 26.3 | 8.8 | 0.0 | 0.0 | 0.0 |
| Ghana | Nutrient Intake, Calories | 0.3 | 10.3 | 2.7 | 48.3 | 17.9 | 4.1 | 0.0 | 0.0 | 0.0 | 1800.6 | 0.7 | 0.4 |
| Ghana | Nutrient Intake, Proteins | 0.0 | 10.3 | 2.7 | 2.8 | 17.9 | 4.1 | 0.0 | 0.0 | 0.0 | 43.5 | 1.4 | 0.5 |
| Ghana | Nutrient Intake, Fats | 0.0 | 10.3 | 2.7 | 0.3 | 17.9 | 4.1 | 0.0 | 0.0 | 0.0 | 32.3 | 0.3 | 0.3 |
| Latin America, HI | Output | 0.5 | -0.5 | 8.2 | 0.2 | 12.4 | 4.3 | 0.0 | 7.0 | 4.7 | 0.0 | 0.0 | 0.0 |
| Latin America, HI | Household Demand | 0.0 | 13.8 | 1.4 | 0.2 | 19.2 | 3.5 | 0.0 | 15.6 | 1.3 | 0.0 | 0.0 | 0.0 |
| Latin America, HI | Nutrient Intake, Calories | 0.6 | 13.8 | 1.4 | 18.5 | 18.2 | 3.6 | 0.3 | 15.6 | 1.3 | 1352.0 | 0.2 | 0.1 |
| Latin America, HI | Nutrient Intake, Proteins | 0.1 | 13.8 | 1.4 | 1.4 | 18.2 | 3.6 | 0.0 | 15.6 | 1.3 | 43.7 | 0.5 | 0.1 |
| Latin America, HI | Nutrient Intake, Fats | 0.0 | 13.8 | 1.4 | 0.1 | 18.2 | 3.6 | 0.0 | 15.6 | 1.3 | 43.0 | 0.0 | 0.0 |
| Latin America, LMI | Output | 1.0 | -0.3 | 3.4 | 0.3 | 10.8 | 1.4 | 0.0 | 3.2 | 2.8 | 0.0 | 0.0 | 0.0 |
| Latin America, LMI | Household Demand | 0.0 | 15.6 | 1.6 | 0.2 | 19.2 | 2.7 | 0.0 | 16.2 | 2.0 | 0.0 | 0.0 | 0.0 |
| Latin America, LMI | Nutrient Intake, Calories | 0.3 | 15.6 | 1.6 | 31.8 | 19.0 | 2.7 | 0.8 | 13.3 | 2.1 | 1533.9 | 0.4 | 0.1 |
| Latin America, LMI | Nutrient Intake, Proteins | 0.0 | 15.6 | 1.6 | 2.5 | 19.0 | 2.7 | 0.0 | 13.3 | 2.1 | 40.8 | 1.1 | 0.2 |
| Latin America, LMI | Nutrient Intake, Fats | 0.0 | 15.6 | 1.6 | 0.1 | 19.0 | 2.7 | 0.1 | 13.3 | 2.1 | 45.6 | 0.0 | 0.0 |
| Latin America, UMI | Output | 73.8 | 0.3 | -1.7 | 3.4 | 10.0 | -0.4 | 2.1 | 0.6 | 0.5 | 0.0 | 0.0 | 0.0 |
| Latin America, UMI | Household Demand | 0.0 | 17.9 | 0.7 | 2.2 | 19.4 | 1.5 | 0.0 | 24.7 | 0.8 | 0.0 | 0.0 | 0.0 |
| Latin America, UMI | Nutrient Intake, Calories | 0.1 | 8.5 | 0.7 | 44.9 | 18.9 | 1.5 | 0.0 | 22.5 | 0.8 | 1955.0 | 0.4 | 0.0 |
| Latin America, UMI | Nutrient Intake, Proteins | 0.0 | 8.5 | 0.7 | 3.2 | 18.9 | 1.5 | 0.0 | 22.5 | 0.8 | 60.6 | 1.0 | 0.1 |
| Latin America, UMI | Nutrient Intake, Fats | 0.0 | 8.5 | 0.7 | 0.2 | 18.9 | 1.5 | 0.0 | 22.5 | 0.8 | 69.6 | 0.0 | 0.0 |
| MENA, HI | Output | 0.0 | 0.0 | 0.0 | 0.1 | 0.9 | 63.5 | 0.1 | 0.7 | 354.2 | 0.0 | 0.0 | 0.0 |
| MENA, HI | Household Demand | 0.0 | -0.1 | 1.0 | 0.4 | -6.6 | 4.5 | 0.0 | -2.0 | 15.0 | 0.0 | 0.0 | 0.0 |
| MENA, HI | Nutrient Intake, Calories | 0.0 | 0.0 | 0.0 | 20.7 | -6.8 | 4.7 | 0.0 | -1.0 | 57.5 | 2064.0 | -0.1 | 0.0 |
| MENA, HI | Nutrient Intake, Proteins | 0.0 | 0.0 | 0.0 | 1.7 | -6.8 | 4.7 | 0.0 | -1.0 | 57.5 | 67.2 | -0.2 | 0.1 |
| MENA, HI | Nutrient Intake, Fats | 0.0 | 0.0 | 0.0 | 0.1 | -6.8 | 4.7 | 0.0 | -1.0 | 57.5 | 70.1 | -0.1 | 0.0 |
| MENA, LI | Output | 0.0 | 1.2 | 51.7 | 0.2 | 19.9 | 28.1 | 0.1 | 1.0 | 68.0 | 0.0 | 0.0 | 0.0 |
| MENA, LI | Household Demand | 0.0 | 17.4 | 11.8 | 0.3 | 22.0 | 15.3 | 0.0 | 25.7 | 31.1 | 0.0 | 0.0 | 0.0 |
| MENA, LI | Nutrient Intake, Calories | 0.0 | 16.4 | 11.8 | 14.9 | 21.9 | 15.3 | 0.1 | 20.3 | 28.8 | 1136.1 | 0.2 | 0.1 |
| MENA, LI | Nutrient Intake, Proteins | 0.0 | 16.4 | 11.8 | 1.3 | 21.9 | 15.3 | 0.0 | 20.3 | 28.8 | 32.5 | 0.7 | 0.5 |
| MENA, LI | Nutrient Intake, Fats | 0.0 | 16.4 | 11.8 | 0.1 | 21.9 | 15.3 | 0.0 | 20.3 | 28.8 | 29.0 | -0.1 | -0.1 |
| MENA, LMI | Output | 0.9 | 0.3 | 0.6 | 0.8 | 6.4 | -0.7 | 0.9 | -0.1 | -0.8 | 0.0 | 0.0 | 0.0 |
| MENA, LMI | Household Demand | 0.0 | -0.2 | 1.2 | 0.2 | 24.2 | 3.2 | 0.0 | 27.4 | 2.9 | 0.0 | 0.0 | 0.0 |
| MENA, LMI | Nutrient Intake, Calories | 0.0 | -0.2 | 1.2 | 13.3 | 23.7 | 3.3 | 0.0 | 11.9 | 2.3 | 1710.2 | 0.2 | 0.0 |
| MENA, LMI | Nutrient Intake, Proteins | 0.0 | -0.2 | 1.2 | 1.2 | 23.7 | 3.3 | 0.0 | 11.9 | 2.3 | 51.4 | 0.5 | 0.1 |
| MENA, LMI | Nutrient Intake, Fats | 0.0 | -0.2 | 1.2 | 0.1 | 23.7 | 3.3 | 0.0 | 11.9 | 2.3 | 36.1 | 0.0 | 0.1 |
| MENA, UMI | Output | 0.2 | -0.1 | 11.8 | 1.9 | 12.6 | 6.3 | 0.4 | 4.1 | 11.6 | 0.0 | 0.0 | 0.0 |
| MENA, UMI | Household Demand | 0.0 | -0.5 | 0.9 | 0.7 | 28.3 | 5.5 | 0.0 | 27.3 | 9.2 | 0.0 | 0.0 | 0.0 |
| MENA, UMI | Nutrient Intake, Calories | 0.0 | -0.2 | 0.8 | 16.8 | 25.9 | 5.7 | 0.4 | 18.8 | 9.1 | 1830.6 | 0.2 | 0.1 |
| MENA, UMI | Nutrient Intake, Proteins | 0.0 | -0.2 | 0.8 | 1.5 | 25.9 | 5.7 | 0.0 | 18.8 | 9.1 | 53.1 | 0.6 | 0.2 |
| MENA, UMI | Nutrient Intake, Fats | 0.0 | -0.2 | 0.8 | 0.1 | 25.9 | 5.7 | 0.0 | 18.8 | 9.1 | 58.7 | 0.0 | 0.0 |
| North America, HI | Output | 60.1 | 0.1 | -0.8 | 5.4 | 7.8 | -0.9 | 2.9 | 0.3 | -2.2 | 0.0 | 0.0 | 0.0 |
| North America, HI | Household Demand | 0.0 | 19.1 | 0.8 | 2.2 | 23.3 | 2.2 | 0.0 | 26.9 | 0.9 | 0.0 | 0.0 | 0.0 |
| North America, HI | Nutrient Intake, Calories | 0.3 | 3.9 | 0.6 | 17.7 | 22.8 | 2.2 | 0.1 | 18.3 | 0.8 | 2423.3 | 0.1 | 0.1 |
| North America, HI | Nutrient Intake, Proteins | 0.0 | 3.9 | 0.6 | 1.2 | 22.8 | 2.2 | 0.0 | 18.3 | 0.8 | 76.3 | 0.3 | 0.1 |
| North America, HI | Nutrient Intake, Fats | 0.0 | 3.9 | 0.6 | 0.1 | 22.8 | 2.2 | 0.0 | 18.3 | 0.8 | 113.7 | -0.1 | 0.1 |
| Oceania, HI | Output | 0.0 | -0.1 | 1.5 | 2.9 | 12.0 | -1.6 | 0.0 | 0.2 | 1.6 | 0.0 | 0.0 | 0.0 |
| Oceania, HI | Household Demand | 0.0 | 17.4 | 1.1 | 0.3 | 24.1 | 2.9 | 0.0 | 24.8 | 2.2 | 0.0 | 0.0 | 0.0 |
| Oceania, HI | Nutrient Intake, Calories | 0.0 | 16.9 | 1.1 | 29.5 | 23.8 | 2.9 | 0.0 | 16.6 | 2.2 | 2148.5 | 0.2 | 0.0 |
| Oceania, HI | Nutrient Intake, Proteins | 0.0 | 16.9 | 1.1 | 2.0 | 23.8 | 2.9 | 0.0 | 16.6 | 2.2 | 74.8 | 0.5 | 0.1 |
| Oceania, HI | Nutrient Intake, Fats | 0.0 | 16.9 | 1.1 | 0.2 | 23.8 | 2.9 | 0.0 | 16.6 | 2.2 | 93.9 | -0.1 | 0.0 |
| Oceania, UMI | Output | 0.0 | 0.0 | 0.0 | 0.0 | 8.5 | 56.4 | 0.0 | 10.5 | 31.3 | 0.0 | 0.0 | 0.0 |
| Oceania, UMI | Household Demand | 0.0 | 0.0 | 0.0 | 0.0 | -6.0 | 6.6 | 0.0 | 13.8 | 18.6 | 0.0 | 0.0 | 0.0 |
| Oceania, UMI | Nutrient Intake, Calories | 0.0 | 0.0 | 0.0 | 4.8 | -6.0 | 6.6 | 0.7 | 13.7 | 17.6 | 1446.8 | 0.0 | 0.0 |
| Oceania, UMI | Nutrient Intake, Proteins | 0.0 | 0.0 | 0.0 | 0.4 | -6.0 | 6.6 | 0.0 | 13.7 | 17.6 | 46.8 | -0.1 | 0.1 |
| Oceania, UMI | Nutrient Intake, Fats | 0.0 | 0.0 | 0.0 | 0.1 | -6.0 | 6.6 | 0.1 | 13.7 | 17.6 | 46.1 | 0.0 | 0.0 |
| RoW, HI | Output | 1.4 | -0.4 | 10.3 | 2.7 | 4.8 | 2.4 | 0.0 | 0.0 | 0.0 | 0.0 | 0.0 | 0.0 |
| RoW, HI | Household Demand | 0.4 | 15.2 | 0.6 | 0.6 | 22.0 | 4.2 | 0.2 | -3.4 | 3.3 | 0.0 | 0.0 | 0.0 |
| RoW, HI | Nutrient Intake, Calories | 0.4 | 13.1 | 0.6 | 14.1 | 17.6 | 4.7 | 0.2 | -3.4 | 3.3 | 2048.5 | 0.1 | 0.0 |
| RoW, HI | Nutrient Intake, Proteins | 0.0 | 13.1 | 0.6 | 1.0 | 17.6 | 4.7 | 0.0 | -3.4 | 3.3 | 69.9 | 0.2 | 0.1 |
| RoW, HI | Nutrient Intake, Fats | 0.0 | 13.1 | 0.6 | 0.1 | 17.6 | 4.7 | 0.0 | -3.4 | 3.3 | 77.3 | 0.0 | 0.0 |
| RoW, LMI | Output | 0.0 | 0.2 | 8.5 | 0.2 | 5.7 | 9.1 | 0.1 | 14.5 | 9.5 | 0.0 | 0.0 | 0.0 |
| RoW, LMI | Household Demand | 0.0 | 17.4 | 4.2 | 0.1 | 25.5 | 5.8 | 0.1 | 19.7 | 9.7 | 0.0 | 0.0 | 0.0 |
| RoW, LMI | Nutrient Intake, Calories | 0.0 | 0.0 | 0.0 | 13.1 | 22.3 | 5.9 | 0.6 | 19.7 | 9.7 | 1792.3 | 0.3 | 0.1 |
| RoW, LMI | Nutrient Intake, Proteins | 0.0 | 0.0 | 0.0 | 0.9 | 22.3 | 5.9 | 0.0 | 19.7 | 9.7 | 58.4 | 0.4 | 0.2 |
| RoW, LMI | Nutrient Intake, Fats | 0.0 | 0.0 | 0.0 | 0.1 | 22.3 | 5.9 | 0.1 | 19.7 | 9.7 | 63.2 | 0.1 | 0.1 |
| RoW, UMI | Output | 1.1 | 0.4 | 5.0 | 0.5 | 7.4 | 0.9 | 0.0 | 0.0 | 0.0 | 0.0 | 0.0 | 0.0 |
| RoW, UMI | Household Demand | 0.0 | 18.7 | 2.2 | 0.2 | 21.0 | 2.8 | 0.0 | -2.8 | 9.0 | 0.0 | 0.0 | 0.0 |
| RoW, UMI | Nutrient Intake, Calories | 0.2 | 17.8 | 2.2 | 26.5 | 19.8 | 2.9 | 0.1 | -2.8 | 9.1 | 1886.7 | 0.3 | 0.0 |
| RoW, UMI | Nutrient Intake, Proteins | 0.0 | 17.8 | 2.2 | 1.8 | 19.8 | 2.9 | 0.0 | -2.8 | 9.1 | 62.1 | 0.6 | 0.1 |
| RoW, UMI | Nutrient Intake, Fats | 0.0 | 17.8 | 2.2 | 0.1 | 19.8 | 2.9 | 0.0 | -2.8 | 9.1 | 63.8 | 0.0 | 0.0 |
| SSA, LI | Output | 0.3 | 4.9 | 6.7 | 5.0 | 10.5 | 6.0 | 2.6 | 1.2 | 10.5 | 0.0 | 0.0 | 0.0 |
| SSA, LI | Household Demand | 0.1 | 15.0 | 5.5 | 3.8 | 16.1 | 6.1 | 0.2 | 24.4 | 12.1 | 0.0 | 0.0 | 0.0 |
| SSA, LI | Nutrient Intake, Calories | 0.5 | 15.0 | 5.5 | 54.5 | 16.1 | 6.1 | 0.5 | 24.3 | 12.1 | 1213.6 | 0.7 | 0.3 |
| SSA, LI | Nutrient Intake, Proteins | 0.1 | 15.0 | 5.5 | 3.4 | 16.1 | 6.1 | 0.0 | 24.3 | 12.1 | 30.8 | 1.7 | 0.7 |
| SSA, LI | Nutrient Intake, Fats | 0.0 | 15.0 | 5.5 | 0.2 | 16.1 | 6.1 | 0.0 | 24.3 | 12.1 | 25.5 | 0.1 | 0.1 |
| SSA, LMI | Output | 0.1 | 0.1 | 19.2 | 2.0 | 11.3 | 12.6 | 3.0 | 11.7 | 11.5 | 0.0 | 0.0 | 0.0 |
| SSA, LMI | Household Demand | 0.0 | -4.8 | 2.4 | 1.8 | 15.5 | 10.9 | 2.9 | 12.2 | 11.0 | 0.0 | 0.0 | 0.0 |
| SSA, LMI | Nutrient Intake, Calories | 0.1 | -4.8 | 2.4 | 57.6 | 15.4 | 10.9 | 12.3 | 12.2 | 11.0 | 1417.5 | 0.6 | 0.5 |
| SSA, LMI | Nutrient Intake, Proteins | 0.0 | -4.8 | 2.4 | 3.8 | 15.4 | 10.9 | 0.6 | 12.2 | 11.0 | 34.5 | 1.8 | 1.3 |
| SSA, LMI | Nutrient Intake, Fats | 0.0 | -4.8 | 2.4 | 0.3 | 15.4 | 10.9 | 0.9 | 12.2 | 11.0 | 32.9 | 0.3 | 0.4 |
| SSA, UMI | Output | 1.3 | 7.9 | 6.8 | 0.1 | 6.3 | 7.6 | 0.1 | 5.7 | 17.0 | 0.0 | 0.0 | 0.0 |
| SSA, UMI | Household Demand | 0.9 | 10.7 | 7.2 | 0.0 | 19.7 | 8.4 | 0.0 | 11.6 | 10.6 | 0.0 | 0.0 | 0.0 |
| SSA, UMI | Nutrient Intake, Calories | 1.9 | 10.7 | 7.2 | 5.6 | 19.0 | 8.5 | 0.0 | 8.9 | 12.3 | 1538.4 | 0.0 | 0.0 |
| SSA, UMI | Nutrient Intake, Proteins | 0.2 | 10.7 | 7.2 | 0.4 | 19.0 | 8.5 | 0.0 | 8.9 | 12.3 | 47.1 | 0.0 | 0.0 |
| SSA, UMI | Nutrient Intake, Fats | 0.1 | 10.7 | 7.2 | 0.0 | 19.0 | 8.5 | 0.0 | 8.9 | 12.3 | 52.4 | 0.0 | 0.0 |
| South Asia, LI | Output | 0.0 | 2.0 | 469.9 | 0.0 | 6.2 | 225.6 | 0.0 | 0.0 | 0.0 | 0.0 | 0.0 | 0.0 |
| South Asia, LI | Household Demand | 0.0 | 0.1 | -3.3 | 0.0 | -5.4 | 2.6 | 0.0 | -1.0 | 6.3 | 0.0 | 0.0 | 0.0 |
| South Asia, LI | Nutrient Intake, Calories | 1.7 | 0.1 | -3.3 | 24.1 | -5.4 | 2.6 | 1.3 | -1.0 | 6.3 | 1235.4 | -0.1 | -2.5 |
| South Asia, LI | Nutrient Intake, Proteins | 0.1 | 0.1 | -3.3 | 1.7 | -5.4 | 2.6 | 0.1 | -1.0 | 6.3 | 33.6 | -0.3 | -2.4 |
| South Asia, LI | Nutrient Intake, Fats | 0.1 | 0.1 | -3.3 | 0.2 | -5.4 | 2.6 | 0.1 | -1.0 | 6.3 | 23.9 | 0.0 | -2.5 |
| South Asia, LMI | Output | 5.8 | 0.7 | 10.2 | 14.2 | 13.2 | 11.8 | 10.3 | 9.4 | 9.0 | 0.0 | 0.0 | 0.0 |
| South Asia, LMI | Household Demand | 1.4 | 11.5 | 3.6 | 12.2 | 20.8 | 9.9 | 7.6 | 12.7 | 9.2 | 0.0 | 0.0 | 0.0 |
| South Asia, LMI | Nutrient Intake, Calories | 0.4 | 11.4 | 3.6 | 65.1 | 20.8 | 9.9 | 2.4 | 12.7 | 9.2 | 1688.2 | 0.7 | 0.4 |
| South Asia, LMI | Nutrient Intake, Proteins | 0.0 | 11.4 | 3.6 | 4.3 | 20.8 | 9.9 | 0.1 | 12.7 | 9.2 | 46.3 | 1.7 | 0.8 |
| South Asia, LMI | Nutrient Intake, Fats | 0.0 | 11.4 | 3.6 | 0.5 | 20.8 | 9.9 | 0.2 | 12.7 | 9.2 | 39.0 | 0.1 | 0.1 |
| South Asia, UMI | Output | 0.0 | 0.6 | 13.4 | 0.0 | 3.0 | 30.7 | 0.1 | 0.3 | 8.3 | 0.0 | 0.0 | 0.0 |
| South Asia, UMI | Household Demand | 0.0 | 18.4 | 18.0 | 0.2 | -5.0 | 3.4 | 0.0 | 25.6 | 22.5 | 0.0 | 0.0 | 0.0 |
| South Asia, UMI | Nutrient Intake, Calories | 0.0 | 0.0 | 0.0 | 112.7 | -5.0 | 3.5 | 0.0 | 0.0 | 0.0 | 2221.0 | -0.2 | 0.2 |
| South Asia, UMI | Nutrient Intake, Proteins | 0.0 | 0.0 | 0.0 | 7.0 | -5.0 | 3.5 | 0.0 | 0.0 | 0.0 | 51.8 | -0.6 | 0.5 |
| South Asia, UMI | Nutrient Intake, Fats | 0.0 | 0.0 | 0.0 | 0.8 | -5.0 | 3.5 | 0.0 | 0.0 | 0.0 | 37.0 | 0.1 | 0.1 |
| South-East Asia, HI | Output | 0.0 | 0.0 | 0.0 | 0.0 | 8.9 | 239.4 | 0.0 | 0.0 | 0.0 | 0.0 | 0.0 | 0.0 |
| South-East Asia, HI | Household Demand | 0.0 | -0.4 | 1.3 | 0.0 | -6.9 | 8.5 | 0.0 | -3.1 | 31.8 | 0.0 | 0.0 | 0.0 |
| South-East Asia, HI | Nutrient Intake, Calories | 0.0 | 0.0 | 0.0 | 66.9 | -6.9 | 8.5 | 0.0 | 0.0 | 0.0 | 10389.0 | -0.1 | 0.1 |
| South-East Asia, HI | Nutrient Intake, Proteins | 0.0 | 0.0 | 0.0 | 3.5 | -6.9 | 8.5 | 0.0 | 0.0 | 0.0 | 358.0 | -0.1 | 0.1 |
| South-East Asia, HI | Nutrient Intake, Fats | 0.0 | 0.0 | 0.0 | 0.4 | -6.9 | 8.5 | 0.0 | 0.0 | 0.0 | 240.6 | 0.0 | 0.0 |
| South-East Asia, LMI | Output | 0.0 | 0.5 | -1.0 | 0.6 | 6.3 | -3.8 | 0.3 | 0.9 | -8.3 | 0.0 | 0.0 | 0.0 |
| South-East Asia, LMI | Household Demand | 0.0 | 18.8 | 0.8 | 0.0 | 0.0 | 0.0 | 0.0 | 26.7 | 4.2 | 0.0 | 0.0 | 0.0 |
| South-East Asia, LMI | Nutrient Intake, Calories | 1.5 | 18.4 | 0.7 | 0.1 | -8.9 | 5.7 | 0.0 | 0.0 | 0.0 | 2294.0 | 0.0 | 0.0 |
| South-East Asia, LMI | Nutrient Intake, Proteins | 0.1 | 18.4 | 0.7 | 0.0 | -8.9 | 5.7 | 0.0 | 0.0 | 0.0 | 61.5 | 0.0 | 0.0 |
| South-East Asia, LMI | Nutrient Intake, Fats | 0.0 | 18.4 | 0.7 | 0.0 | -8.9 | 5.7 | 0.0 | 0.0 | 0.0 | 34.3 | 0.0 | 0.0 |
| South-East Asia, UMI | Output | 0.1 | 6.2 | 58.6 | 1.2 | 13.5 | 42.2 | 0.3 | 5.0 | 34.0 | 0.0 | 0.0 | 0.0 |
| South-East Asia, UMI | Household Demand | 1.8 | 17.6 | 1.0 | 1.0 | 20.7 | 33.0 | 0.1 | 19.7 | 34.6 | 0.0 | 0.0 | 0.0 |
| South-East Asia, UMI | Nutrient Intake, Calories | 10.9 | 15.6 | 1.0 | 13.5 | 19.2 | 31.2 | 0.6 | 19.7 | 34.6 | 2034.9 | 0.1 | 0.2 |
| South-East Asia, UMI | Nutrient Intake, Proteins | 1.6 | 15.6 | 1.0 | 0.8 | 19.2 | 31.2 | 0.0 | 19.7 | 34.6 | 55.9 | 0.6 | 0.5 |
| South-East Asia, UMI | Nutrient Intake, Fats | 0.3 | 15.6 | 1.0 | 0.1 | 19.2 | 31.2 | 0.0 | 19.7 | 34.6 | 43.1 | 0.0 | 0.0 |
| World | Output | 168.9 | 0.2 | 1.3 | 76.9 | 10.3 | 7.7 | 46.4 | 4.9 | 10.4 | 0.0 | 0.0 | 0.0 |
| World | Household Demand | 32.4 | 1.5 | 0.7 | 43.9 | 20.9 | 9.1 | 15.0 | 16.2 | 9.9 | 0.0 | 0.0 | 0.0 |
| World | Nutrient Intake, Calories | 10.4 | 0.8 | 0.4 | 36.6 | 19.4 | 8.7 | 2.2 | 19.3 | 10.5 | 1822.5 | 0.3 | 0.2 |
| World | Nutrient Intake, Proteins | 0.8 | 2.3 | 0.5 | 2.4 | 19.5 | 8.6 | 0.1 | 18.0 | 10.9 | 56.0 | 0.8 | 0.3 |
| World | Nutrient Intake, Fats | 0.5 | 0.0 | 0.3 | 0.2 | 19.6 | 8.9 | 0.2 | 19.9 | 10.3 | 51.8 | 0.0 | 0.1 |

Notes: Own elaboration

**Figure C1. Prices of Legumes (% deviation from baseline)**
**Panel A: Producer Prices**

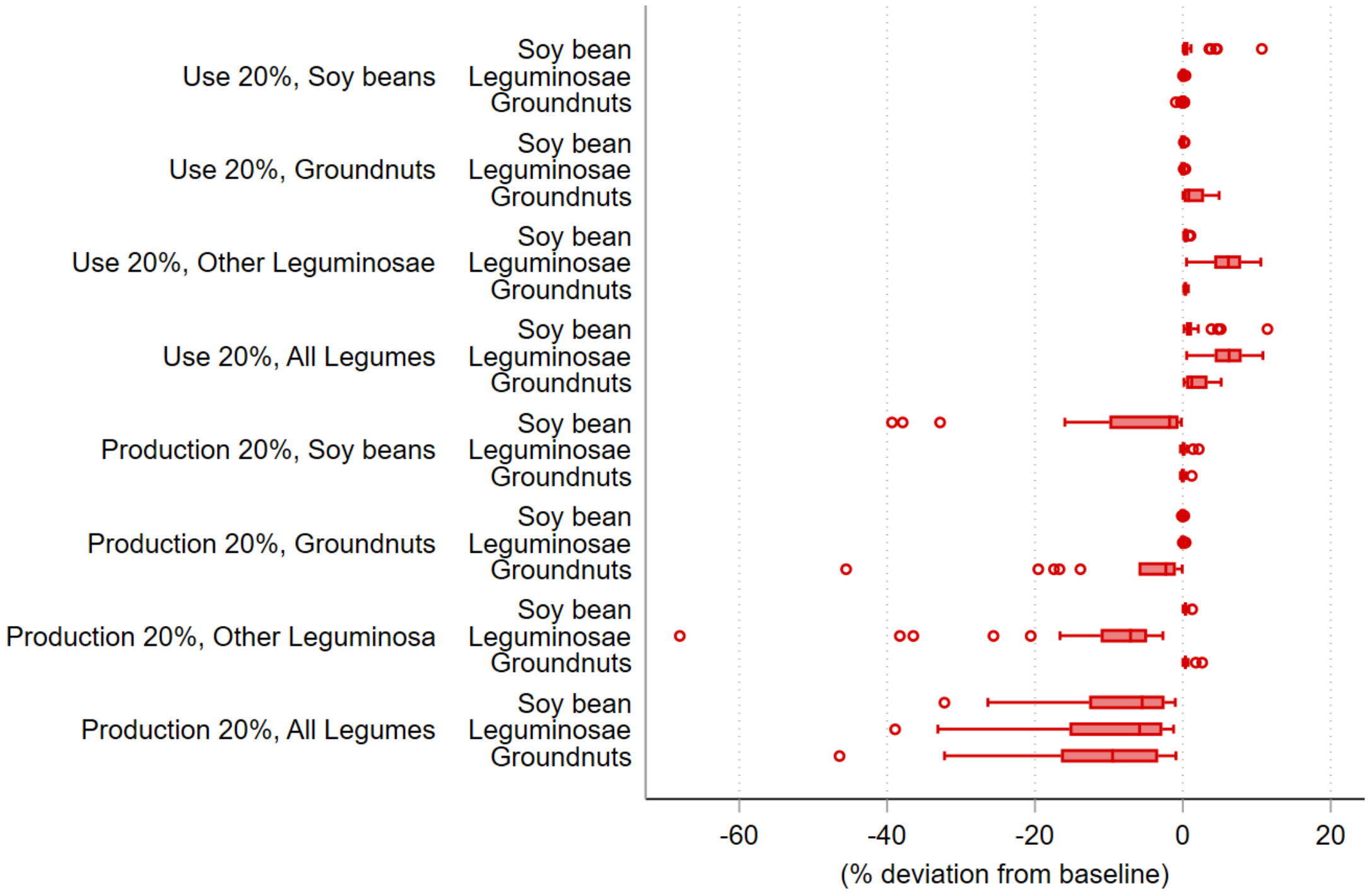

**Panel B: Consumer Prices**

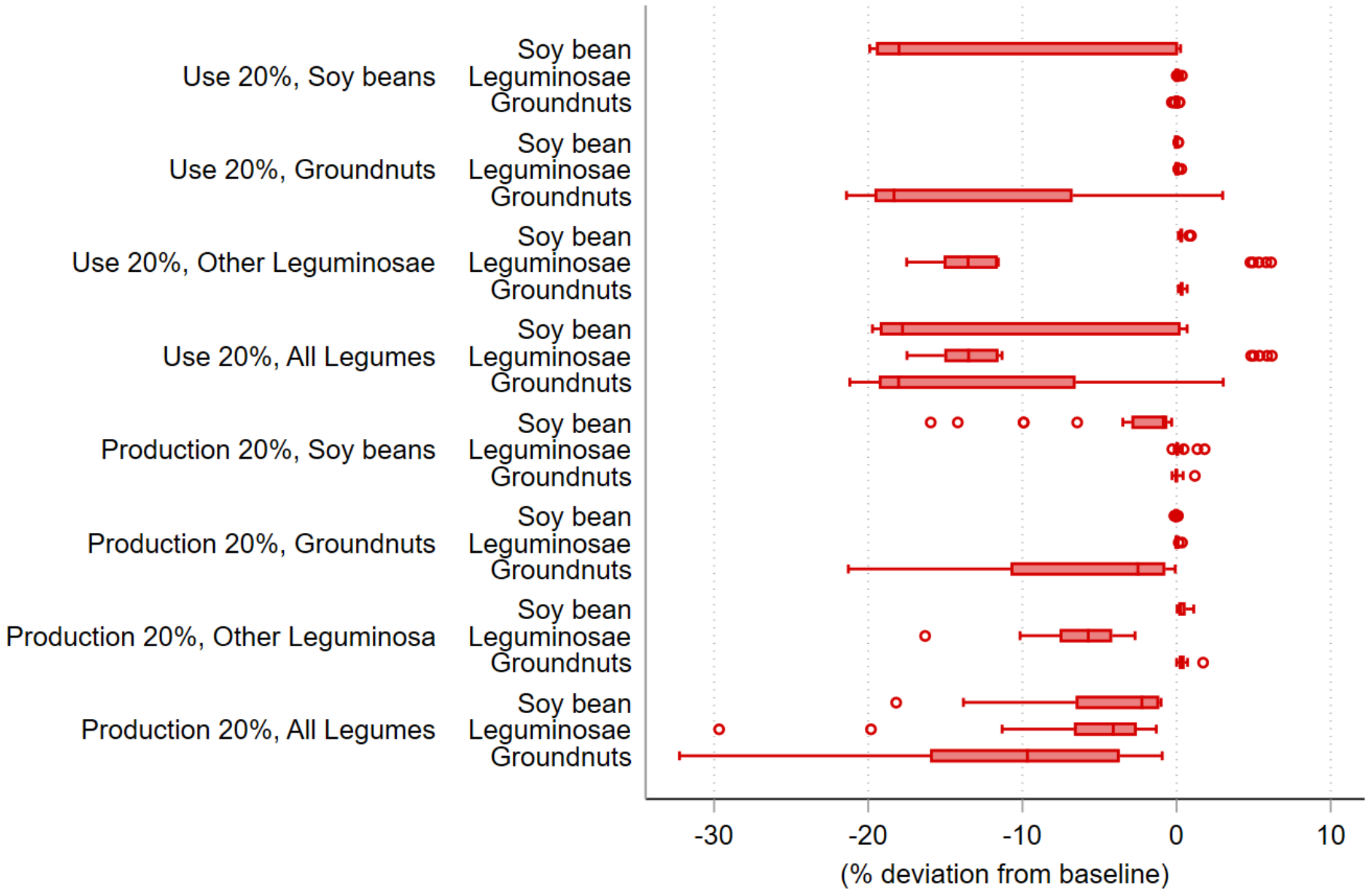

Notes: Own elaboration.

**Figure C2. Production of Legumes (% deviation from baseline)**

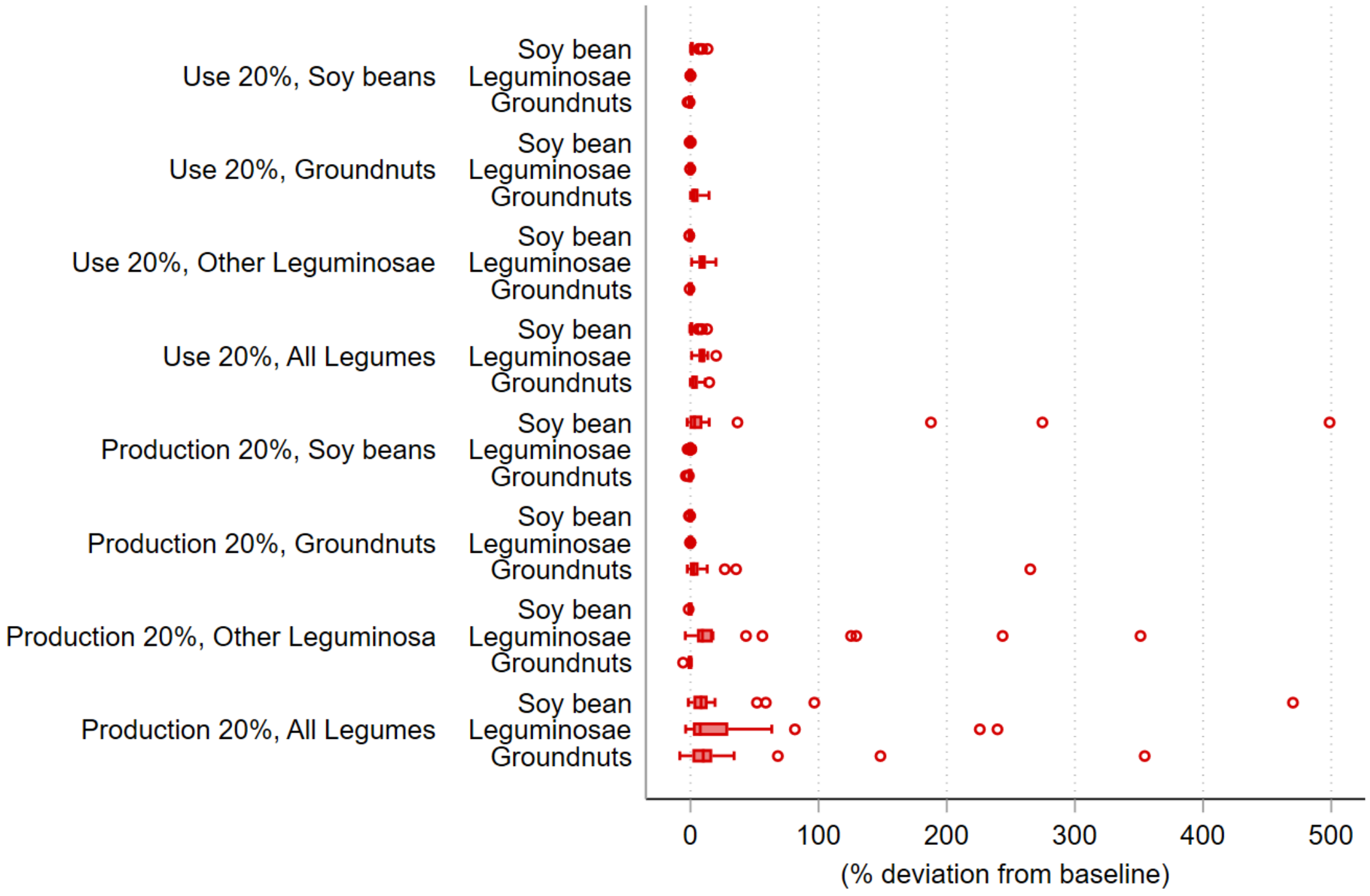

Notes: Own elaboration.

**Figure C3. Land Use for Legume Cultivation (% deviation from baseline)**

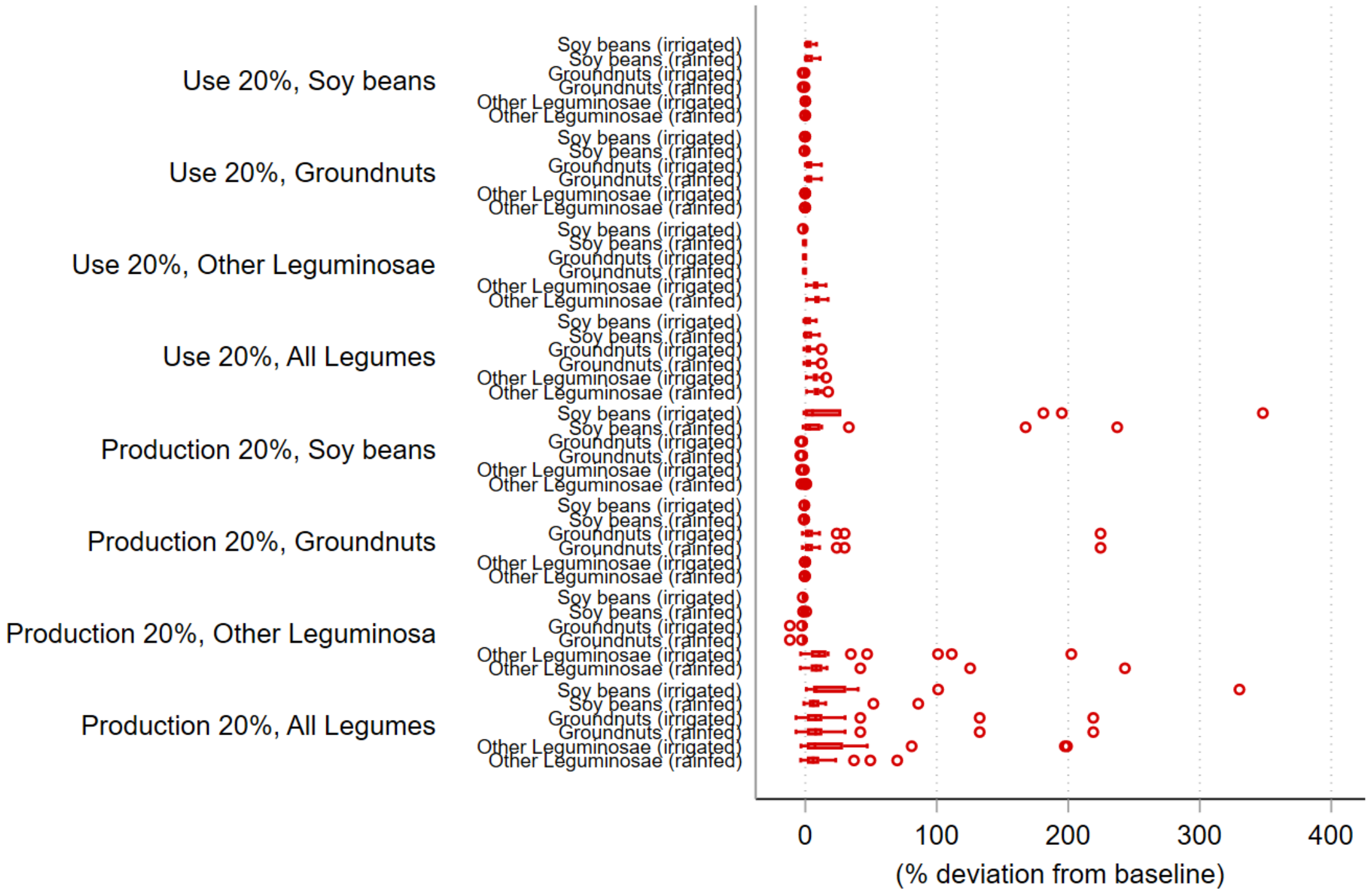

Notes: Own elaboration.

### Figure C4. Land Cover changes (% deviation from baseline)

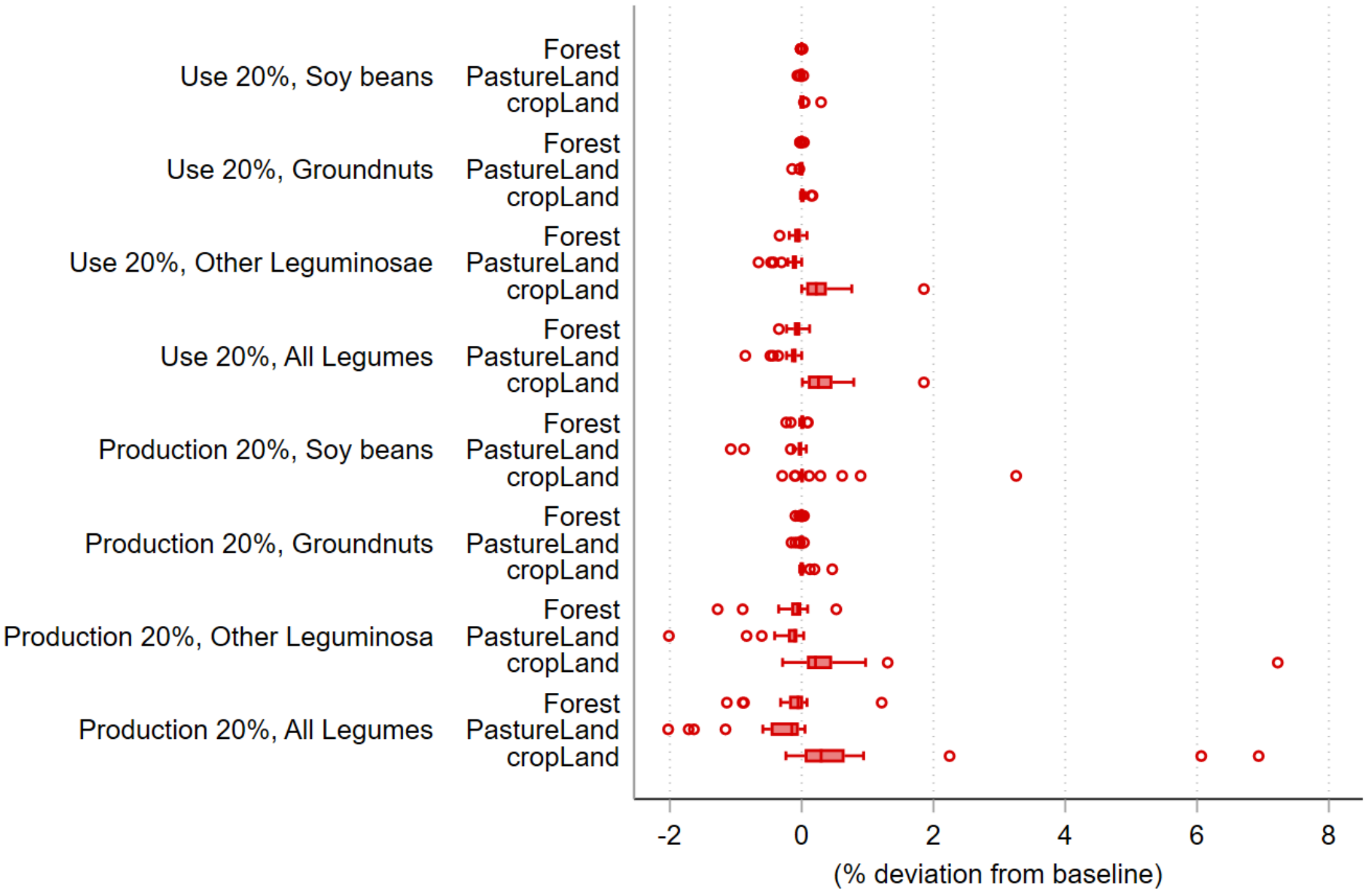

Notes: Own elaboration.

### Figure C5. Household Demand of Legumes (% deviation from baseline)

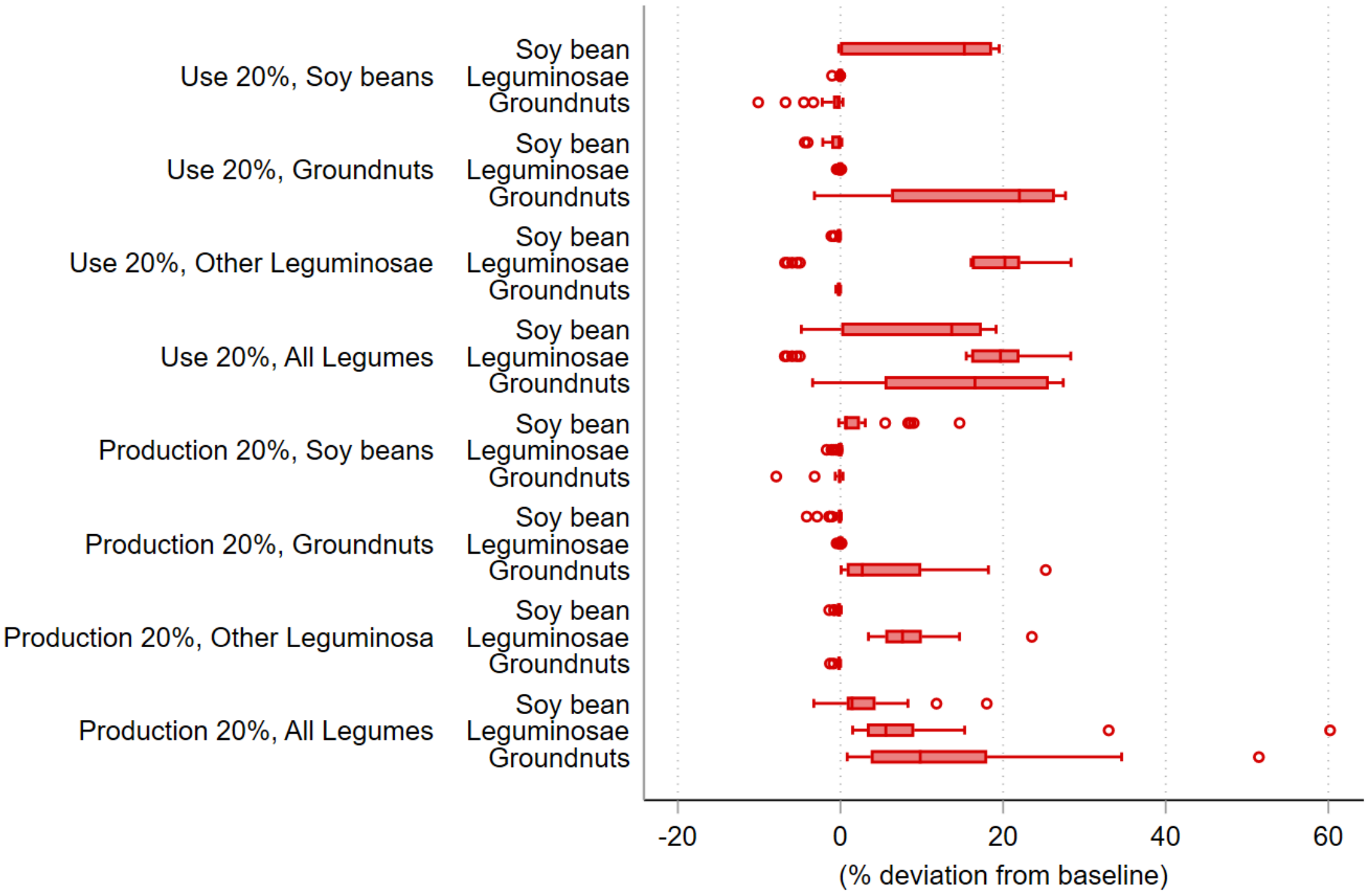

Notes: Own elaboration.

**Figure C6. Intermediate Demand of Legumes (% deviation from baseline)**

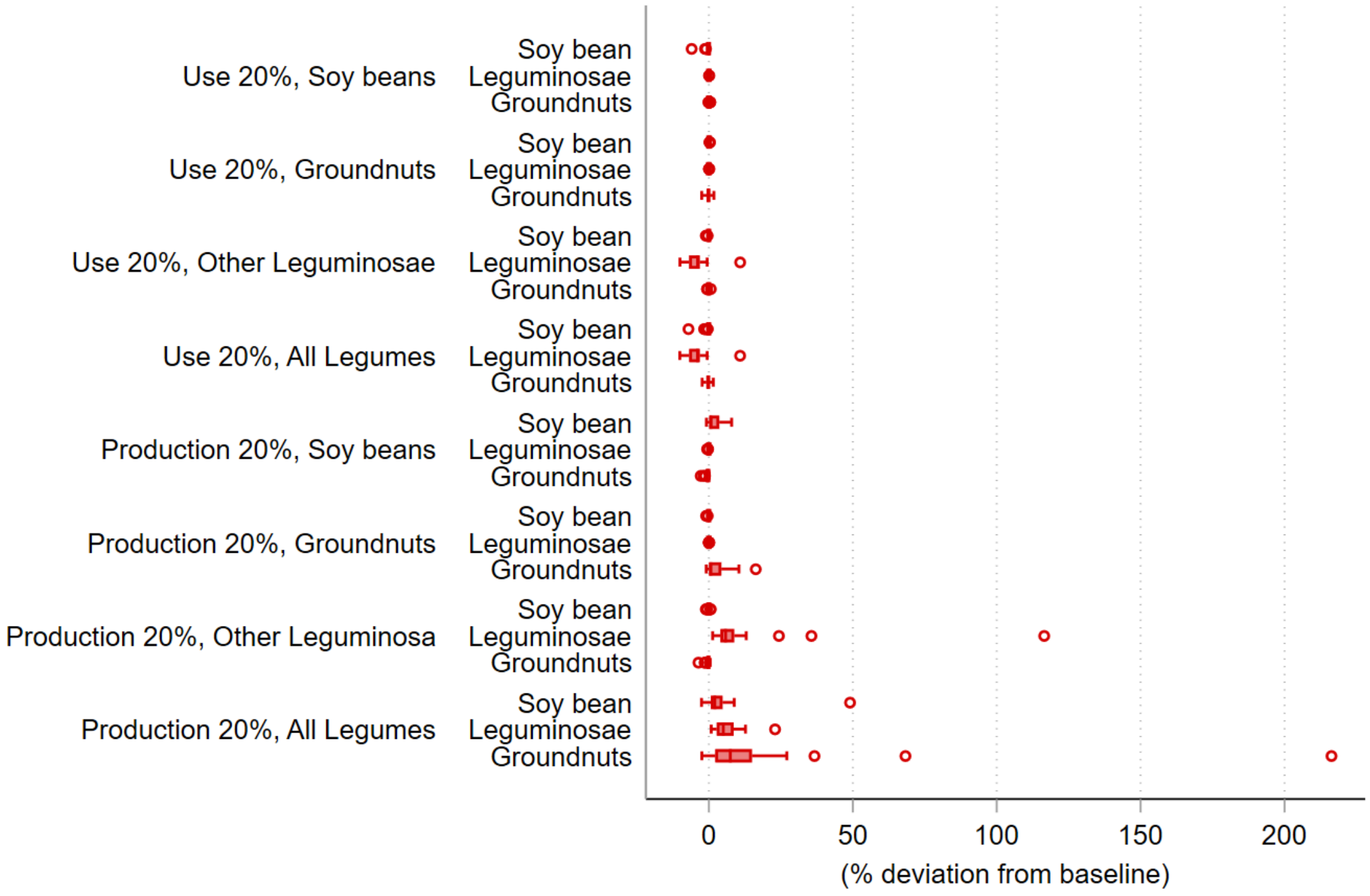

Notes: Own elaboration.

**Figure C7. Intermediate Demand of Legumes, by Sector (% deviation from baseline)**

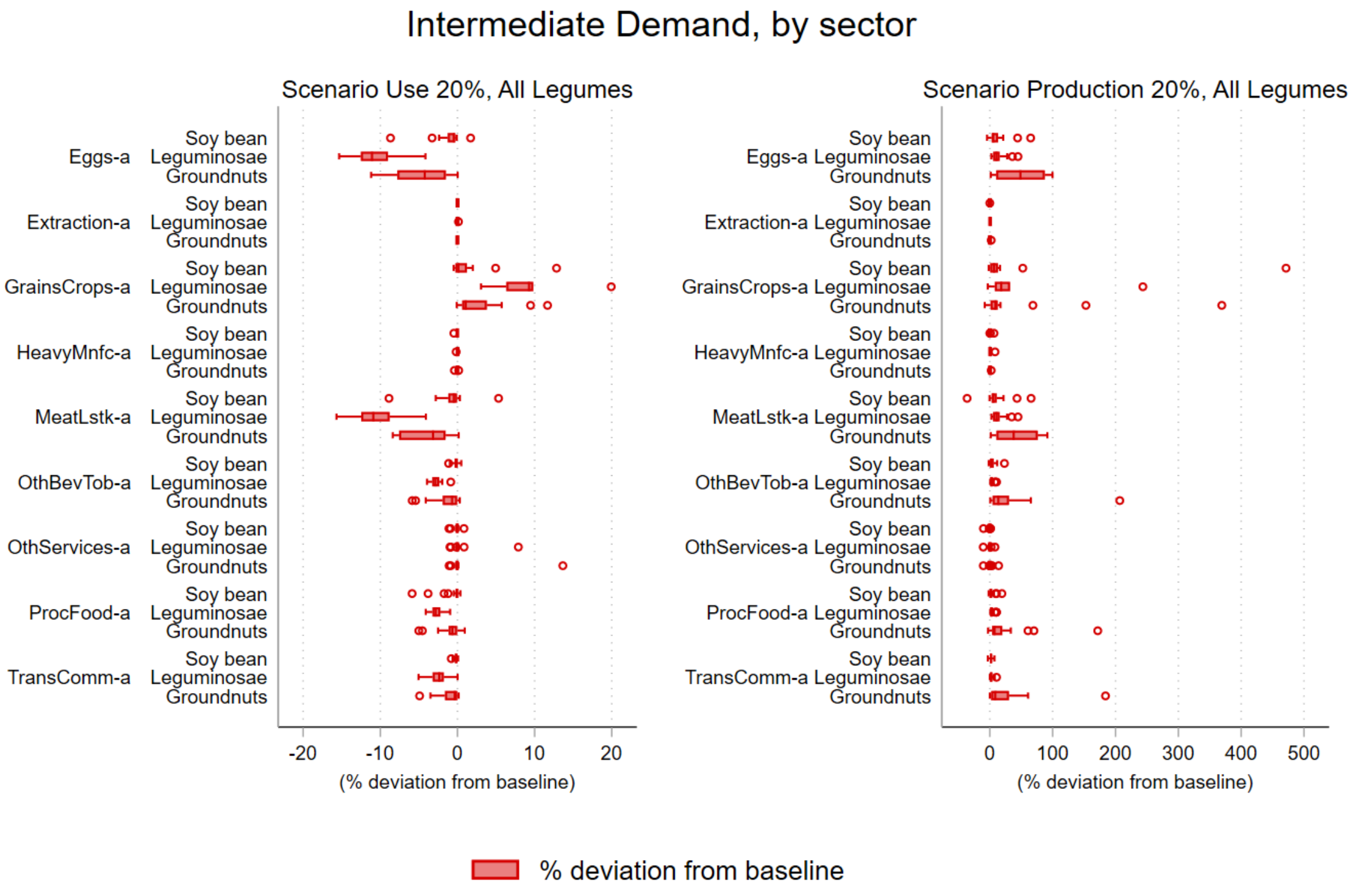

Notes: Own elaboration.

## Figure C8. Exports of Legumes (% deviation from baseline)

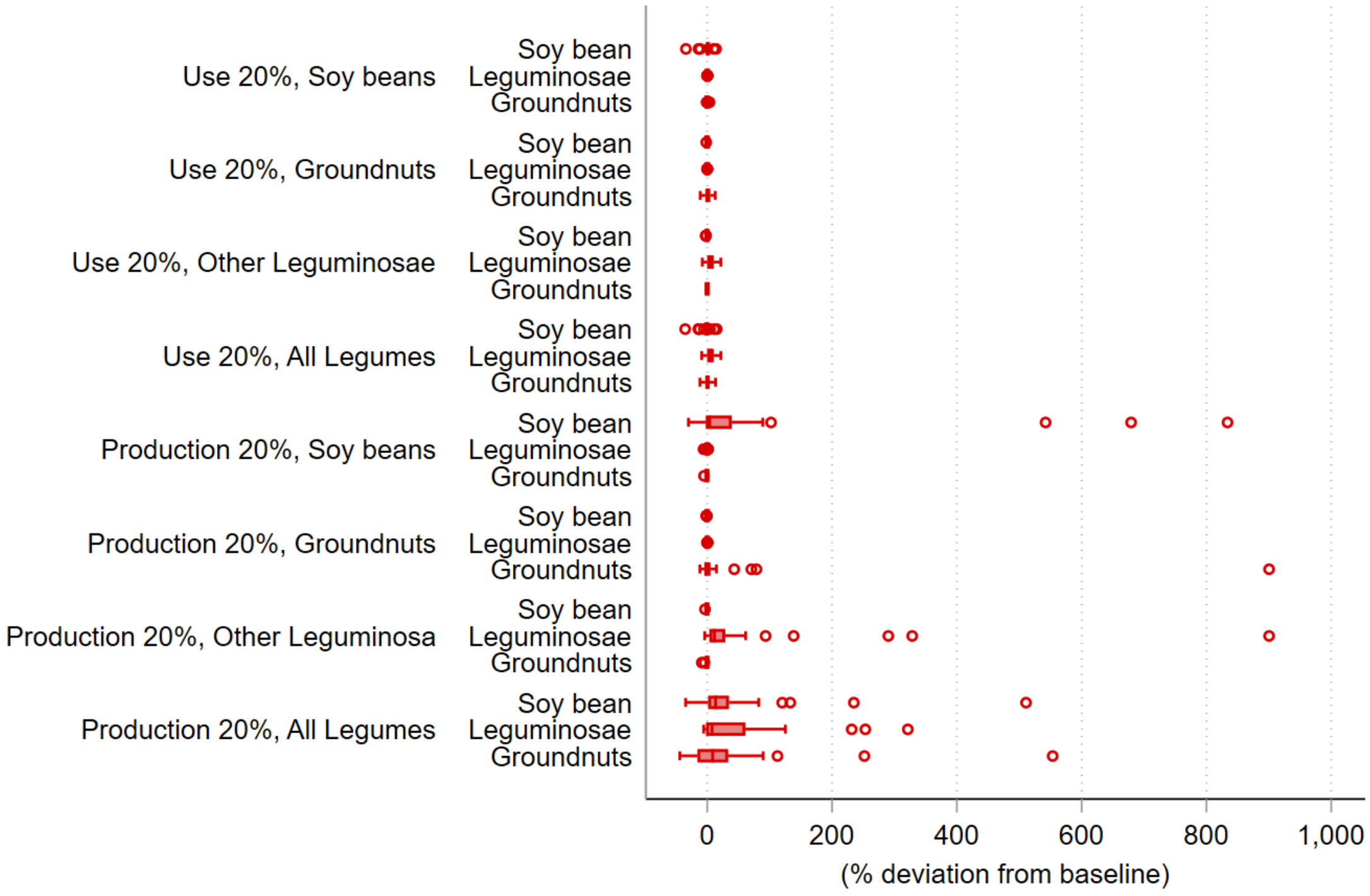

Notes: Own elaboration.

## Figure C9. Imports of Legumes (% deviation from baseline)

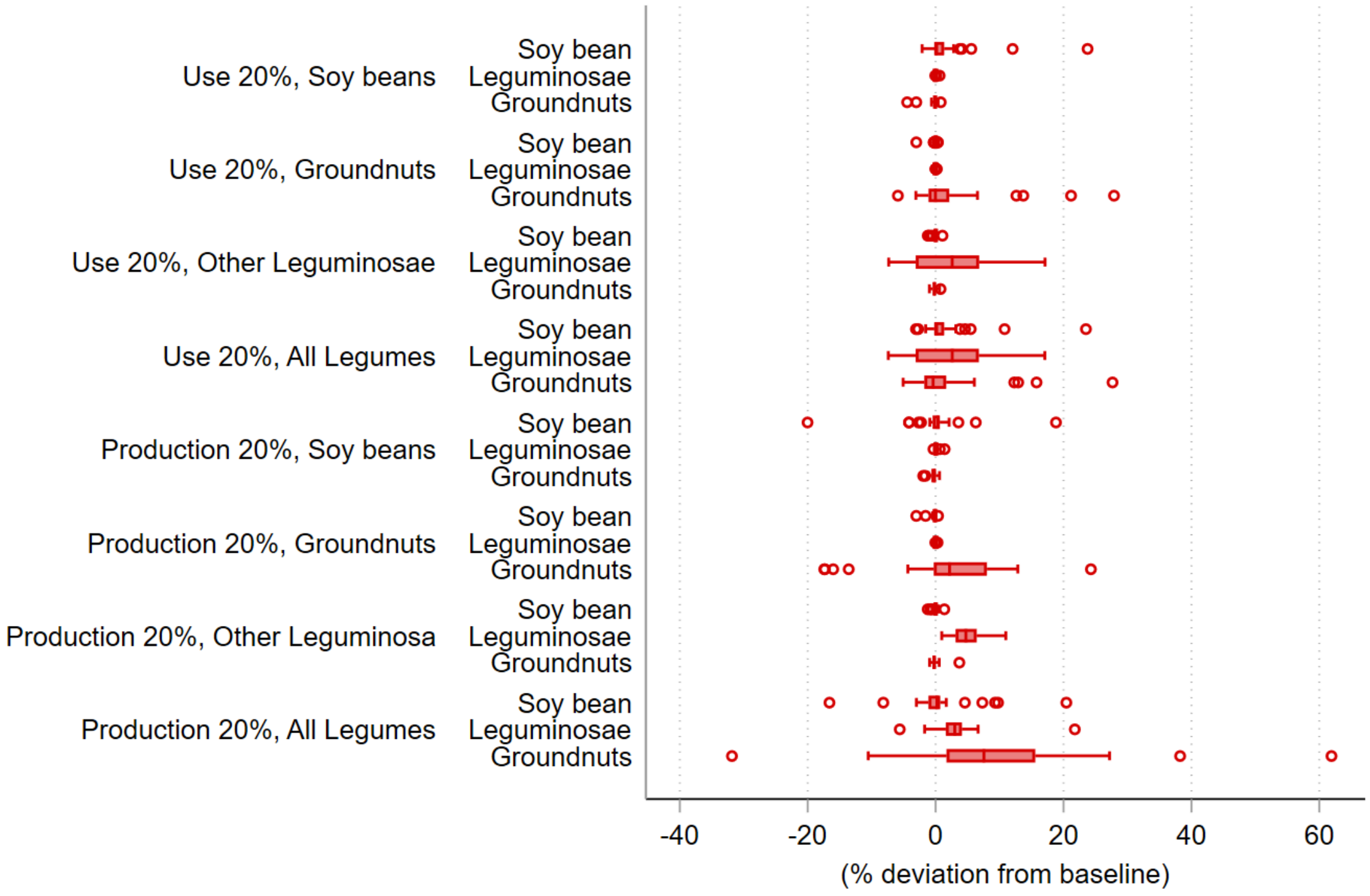

Notes: Own elaboration.

**Figure C10. Nutrient Intake, All Households (% deviation from baseline)**
**Panel A: Calories**

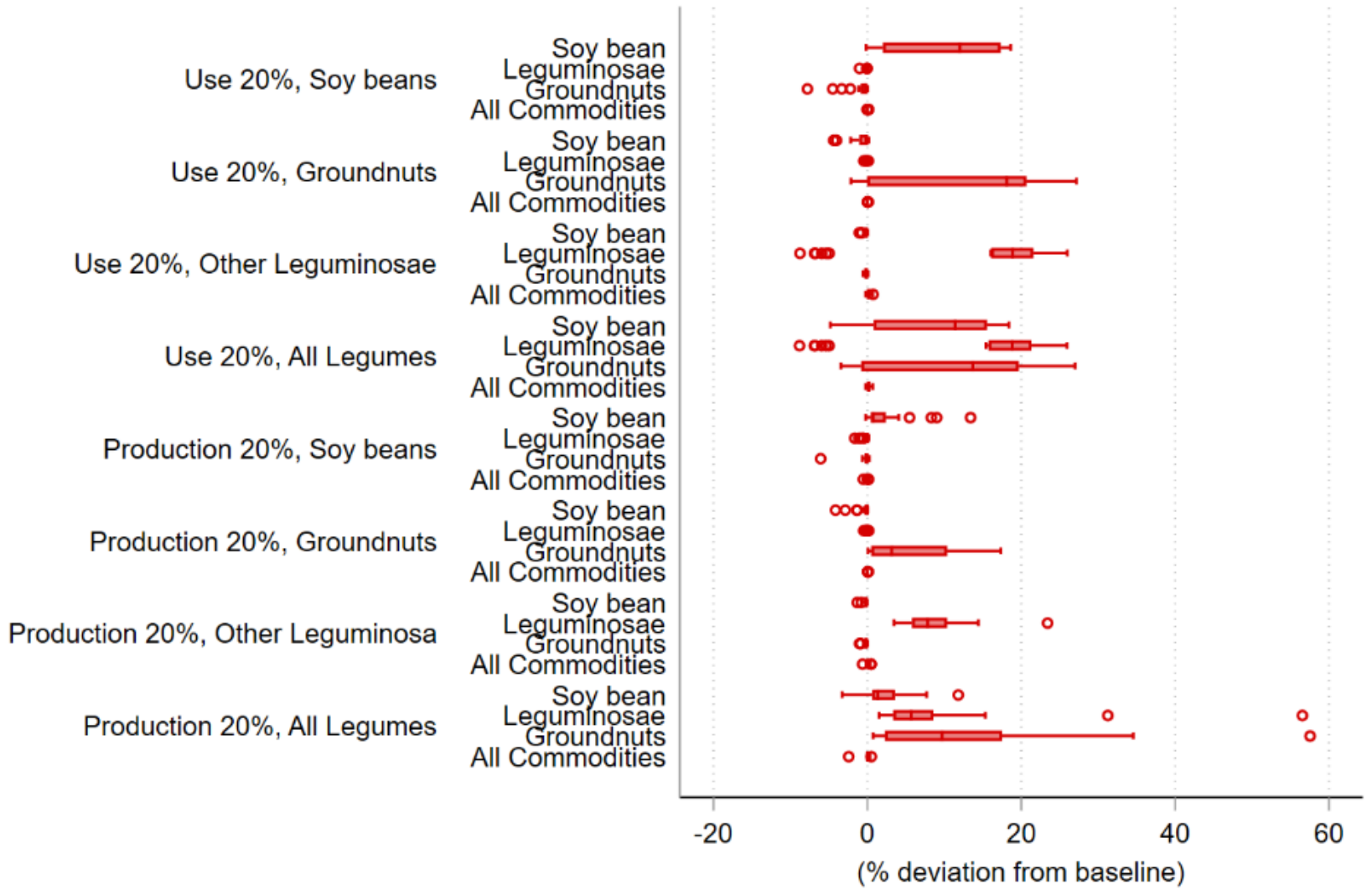

**Panel B: Proteins**

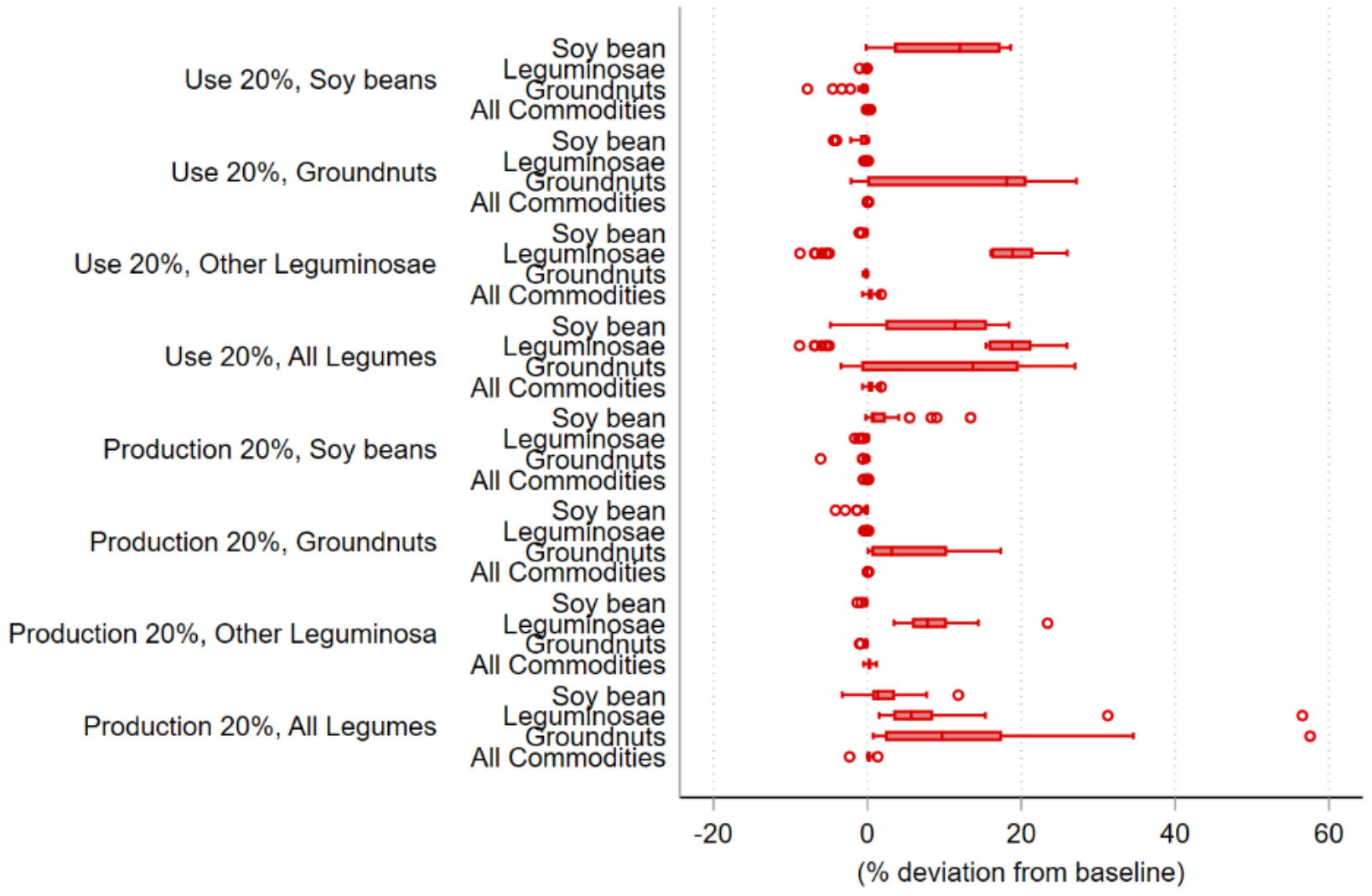

**Panel C: Fats**

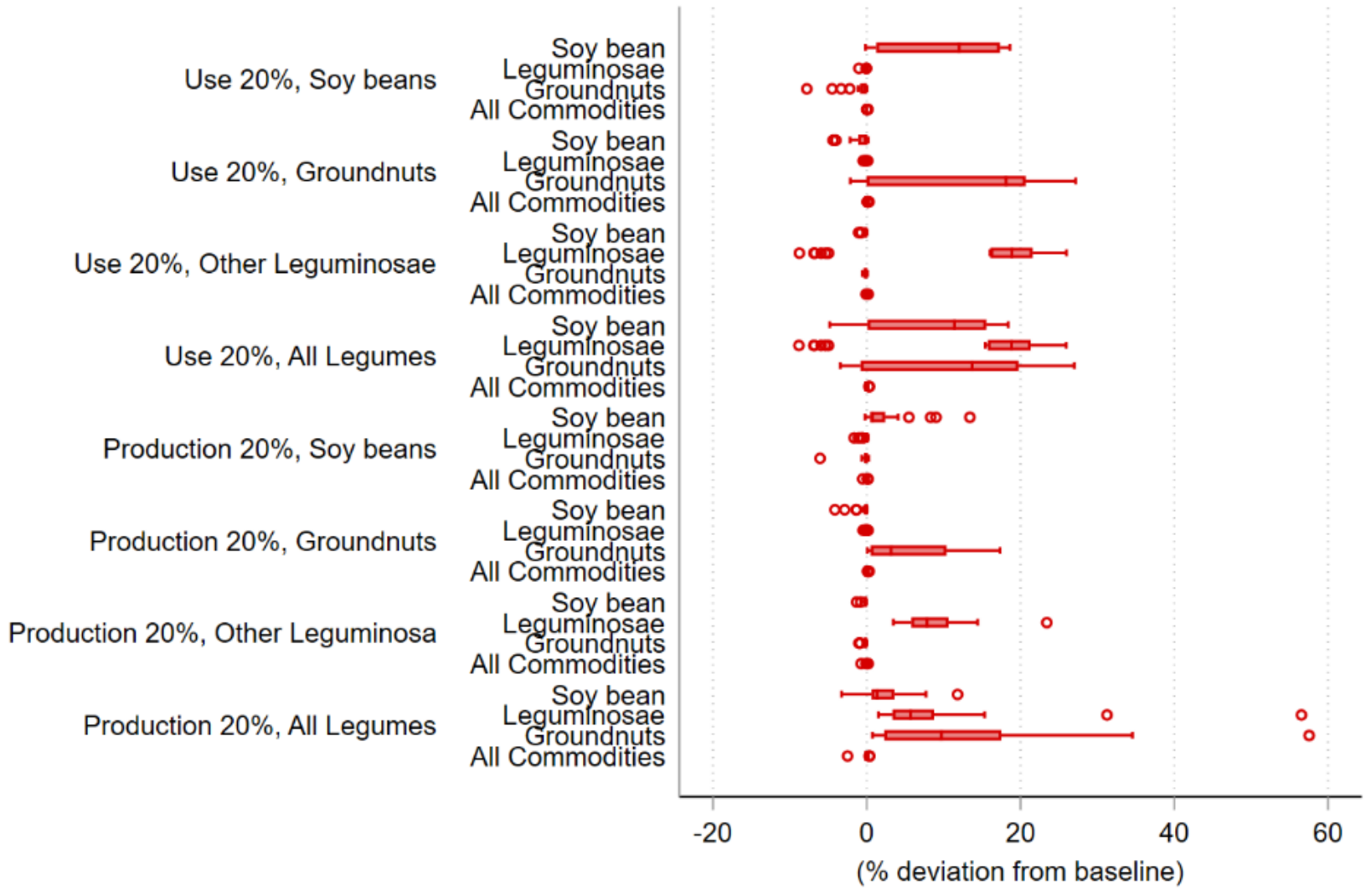

Notes: Own elaboration.

**Figure C11. GHG changes (% deviation from baseline)**

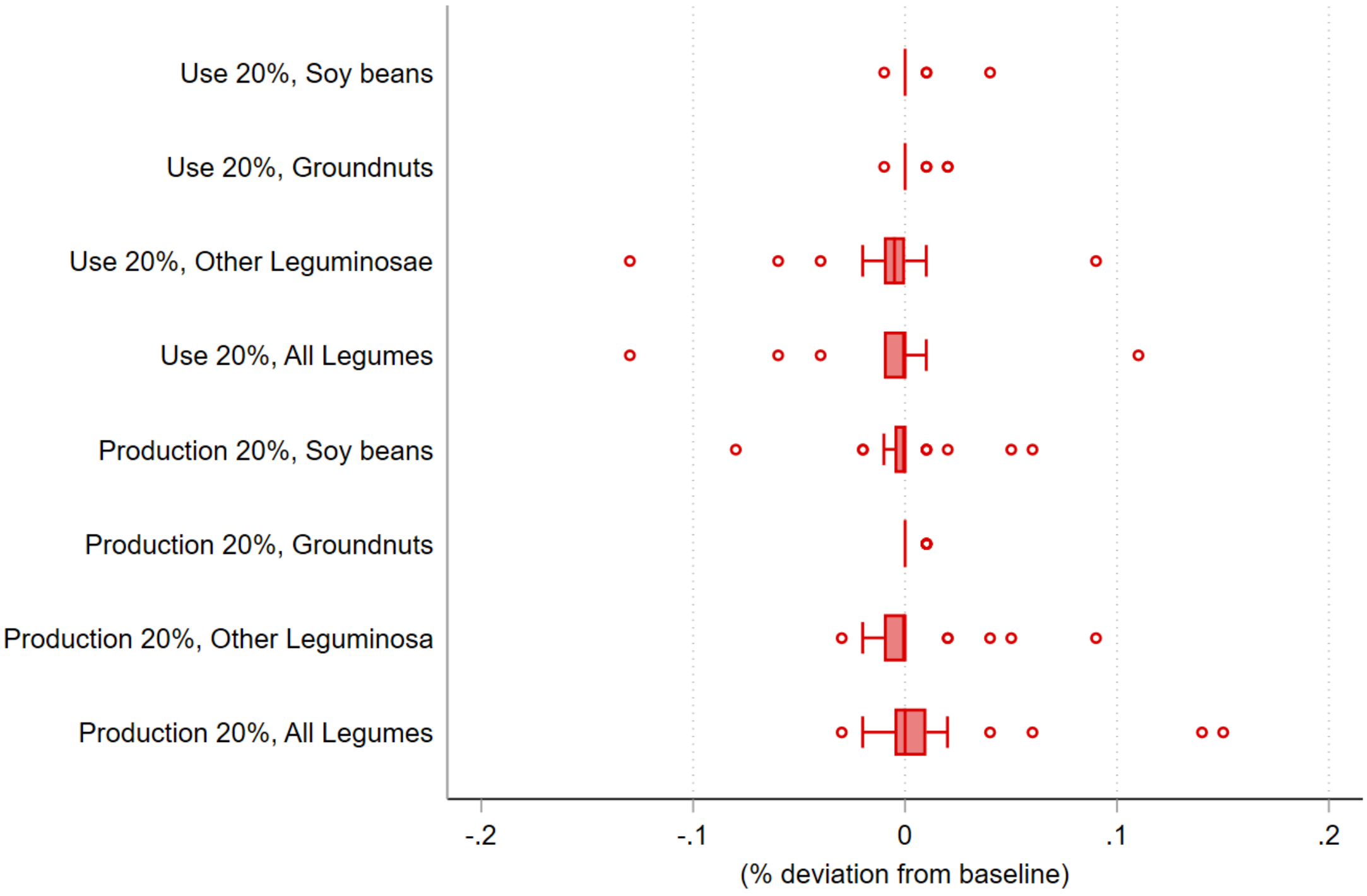

Notes: Own elaboration.

**Figure C12. Utility (% deviation from baseline)**

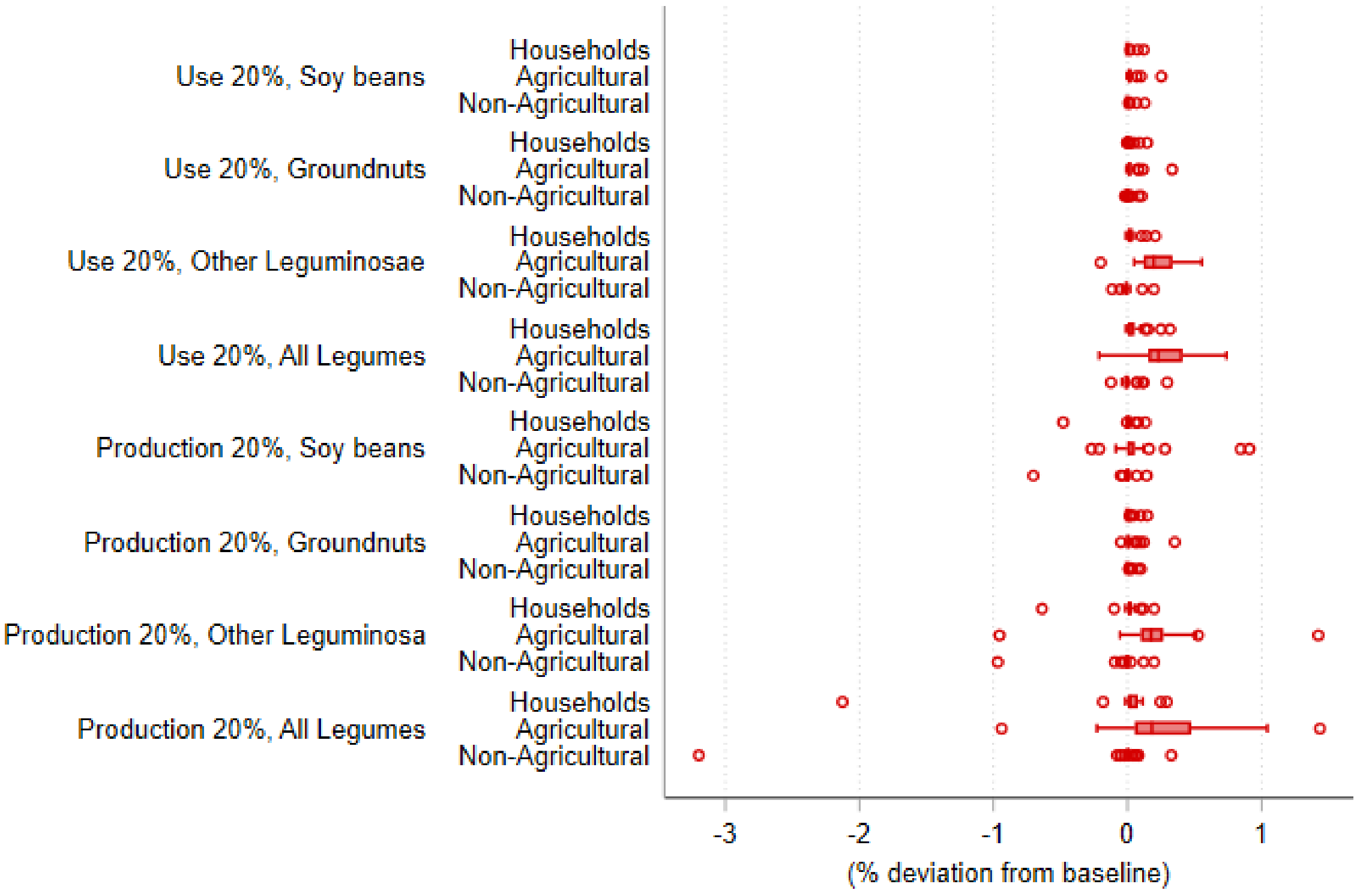

Notes: Own elaboration.